\documentclass[a4paper,10pt,notitlepage]{article}
\usepackage{amsfonts,amssymb,amsmath,amsthm,hyperref,colonequals}
\usepackage{caption}
\usepackage{cite}
\usepackage{tabularx}
\usepackage{graphicx}
\usepackage{shuffle}
\usepackage[nice]{nicefrac} 
\usepackage{url}

\usepackage{color}

\newcommand{\ointctrclockwise}{\oint}
\theoremstyle{definition}
\newcommand{\beqa}{\begin{eqnarray}}
\newcommand{\eeqa}{\end{eqnarray}}
\newcommand{\beq}{\begin{equation}}
\newcommand{\eeq}{\end{equation}}

\newcommand{\ft}{\mathfrak{t}}\newcommand{\fq}{\mathfrak{q}}
\newcommand{\fx}{\mathsf{x}}

\newcommand{\fQ}{\mathbf{\mathcal{Q}}}

\newcommand{\fA}{\mathsf{A}}
\newcommand{\fR}{\mathsf{R}}
\newcommand{\fw}{\mathsf{w}}

\newcommand{\calZ}{\mathcal{Z}}
\newcommand{\calM}{\mathcal{M}}

\newcommand{\tA}{\textsf{A}}

\newcommand{\tg}{\textsf{g}}

\newcommand{\ba}{\textbf{a}}\newcommand{\bA}{\textbf{A}}

\def\balpha{\boldsymbol{\alpha}}

\newcommand{\SU}{\text{SU}}
\newcommand{\bsl}{\text{sl}}

\newcommand{\form}[2]{\ensuremath{\left( #1, #2\right)}}

\newcommand{\Up}{\Upsilon_q}
\newcommand{\tN}{\textsf{N}^{\beta}}

\newmuskip\pFqskip
\mathchardef\pFcomma=\mathcode`,

\newcommand*\pFq[5]{%
   \begingroup
   \begingroup\lccode`~=`,
     \lowercase{\endgroup\def~}{\pFcomma\mkern\pFqskip}%
   \mathcode`,=\string"8000
   {}_{#1}\Phi_{#2}\biggl[\genfrac..{0pt}{}{#3}{#4};#5\biggr]%
   \endgroup
}

\begin{document}

\thispagestyle{empty}
\setcounter{page}{0}
\begin{flushright}\footnotesize
\texttt{DESY 14-236}\\
\texttt{HU-Mathematik-35-2014}\\
\texttt{HU-EP-14/57}\\
\vspace{0.5cm}
\end{flushright}
\setcounter{footnote}{0}

\begin{center}
{\huge{
\textbf{Toda 3-Point \vspace{0.5cm} Functions \\ From Topological Strings II
}
}}
\vspace{15mm}

{\sc 
Mikhail Isachenkov$^{a}$, Vladimir Mitev$^{b}$,   Elli Pomoni$^{a,c}$ }\\[5mm]

{\it $^a$DESY Hamburg, Theory Group, \\
Notkestrasse 85, D--22607 Hamburg, Germany
}\\[5mm]

{\it $^b$Institut f\"ur Mathematik und Institut f\"ur Physik,\\ Humboldt-Universit\"at zu Berlin\\
IRIS Haus, Zum Gro{\ss}en Windkanal 6,  12489 Berlin, Germany
}\\[5mm]

{\it $^c$Physics Division, National Technical University of Athens,\\
15780 Zografou Campus, Athens, Greece
}\\[5mm]

\texttt{mikhail.isachenkov@desy.de}\\
\texttt{mitev@math.hu-berlin.de}\\
\texttt{elli.pomoni@desy.de}\\[25mm]

\textbf{Abstract}\\[2mm]
\end{center}
In \cite{Mitev:2014isa} we proposed a formula for the 3-point structure constants of generic primary fields in the Toda field theory, derived using topological strings and the AGT-W correspondence from the partition functions of the non-Lagrangian $T_N$ theories on $S^4$. In this article, we obtain from it the well-known formula by Fateev and Litvinov and show that the degeneration on a first level of one of the three primary fields on the Toda side corresponds to a particular Higgsing of the $T_N$ theories.


\newpage
\setcounter{page}{1}


\tableofcontents

\bigskip

\section{Introduction}

Two-dimensional  conformal field theories (CFTs) include many models with important physical applications and  have provided a rich playground for exact solutions of CFTs. Even though extensive methods have been developed for rational CFTs throughout the years \cite{DiFrancesco:1997nk},  non-rational 2D CFTs are much less understood. So far, all non-rational CFTs that have been solved are versions of Liouville. 
A CFT is solved when its two and three point correlation functions are obtained and a crucial step in doing this for the Liouville CFT was the proposal  of the 3-point function by Dorn-Otto-Zamolodchikov-Zamolodchikov (DOZZ) \cite{Dorn:1994xn,Zamolodchikov:1995aa} based on insightful and powerful consistency checks. This proposal was  rigorously derived by Teschner \cite{Teschner:2001rv} who showed that the DOZZ 3-point function is a solution of the crossing symmetry equation.

The next natural step is to study multifield non-rational CFTs, a prototype of which is the Toda CFT. Obtaining the 3-point functions of the Toda CFT is a long-standing problem in mathematical physics. Attacking this problem purely by using 2D CFT techniques is a notoriously difficult task and results exist only for particular specializations of the external momenta. The state of the art can be found in the works of Fateev and Litvinov \cite{Fateev:2005gs,Fateev:2007ab,Fateev:2008bm}, who obtained the 3-point functions of primary operators if one of them is {\it appropriately degenerate}. 

In a previous publication \cite{Mitev:2014isa}, we presented a formula for the 3-point functions of three \textit{arbitrary} primaries of the Toda CFT. 
Our formula \eqref{eq:3pointfunctionsastopologicalstrings} was obtained  using techniques of a very different nature than \cite{Fateev:2005gs,Fateev:2007ab,Fateev:2008bm}, namely  topological stings, 5-brane web physics and the AGT-W correspondence. 
The purpose of the present paper is to push forward the program of further understanding and checking  it. 
We begin with  \eqref{eq:3pointfunctionsastopologicalstrings}, 
  specialize appropriately one of the external momenta 
  and obtain the formula of Fateev-Litvinov \cite{Fateev:2005gs} after a direct calculation, thus presenting a highly non-trivial check of our proposal.
Specializing means that
the Verma module for the primary field has a null-vector descendant at level one. In the rest of the paper, we will refer to them as {\it semi-degenerate}\footnote{A representation of \textbf{W}$_N$ can contain a null vector at some level higher than one. Such representations are called semi-degenerate as well. The 3-point functions containing one primary belonging to such representation and two generic ones will not be considered in the present paper.}, as opposed to the completely degenerate ones, containing $N-1$ linearly independent null-vectors. 
Furthermore, we believe that the techniques of  \cite{Mitev:2014isa} will provide the solution not only for the 3-points functions of $\textbf{W}_N$ primaries, but also for those involving descendent fields.  We leave this for a future work.

    The quirks of our formula for the 3-point functions  \eqref{eq:3pointfunctionsastopologicalstrings}  stem from the strategy employed in  \cite{Mitev:2014isa} to derive it. A key element was the AGT-W correspondence \cite{Alday:2009aq,Wyllard:2009hg}, which is a relation between 4D $\mathcal{N}=2$ $\SU(N)$ quiver gauge theories and the 2D $\textbf{W}_N$ Toda CFT. Specifically, upon an appropriate identification of the parameters, the correlation functions of the 2D Toda CFT are equal to the partition functions of the corresponding 4D $\mathcal{N}=2$ gauge theories. 
The conformal blocks of the 2D CFTs are given by the instanton partition functions of Nekrasov \cite{Alday:2009aq,Wyllard:2009hg}, while the 3-point structure constants  are  obtained by the  partition functions of the $T_N$ superconformal theories \cite{Kozcaz:2010af,Bao:2013pwa}. The $T_N$ theories have no Lagrangian description and thus their  partition functions were unknown until recently \cite{Mitev:2014isa,Bao:2013pwa, Hayashi:2013qwa}. The sole exception was the \textbf{W}$_2\equiv \bf{Vir}$ case, {\it i.e.} the Liouville case, whose 3-point structure constants are given by the famous DOZZ formula \cite{Dorn:1994xn,Zamolodchikov:1995aa} and equal to the partition function of  four free
hypermultiplets \cite{Gomis:2011pf,Bao:2013pwa}.

We were able to bypass the fact that  the $T_N$ theories have no known Lagrangian description by using a
 {\it generalized} version of  AGT-W: a relation between 5D gauge theories compactified on $S^1$ and  2D $q$-deformed Liouville/Toda CFT
 \cite{Awata:2009ur,Awata:2010yy,Schiappa:2009cc,Mironov:2011dk,Itoyama:2013mca,Bao:2011rc,Nieri:2013yra,Bao:2013pwa,Nieri:2013vba,Aganagic:2013tta,Aganagic:2014oia,
 Taki:2014fva,Dijkgraaf:2009pc,Cheng:2010yw,Tan:2013xba}, where the circumference  $\beta$ of the $S^1$ corresponds to the deformation parameter $q=e^{-\beta}$ of the CFT. In 5D,  the partition functions can be computed not only using localization, which requires a Lagrangian, but also by using
 the  powerful tool of topological strings \cite{Iqbal:2012xm}.
Employing this technology, we calculated in  \cite{Bao:2013pwa} (see also \cite{Hayashi:2013qwa})  the partition functions of the 5D $T_N$ theories and  suggested that they should be interpreted as the 3-point structure constants of the $q$-deformed Toda.   Subsequently,  we showed in \cite{Mitev:2014isa}  how to take the 4D limit, corresponding to $\beta \rightarrow 0$ or equivalently to $q\rightarrow 1$, thus obtaining the partition function \eqref{eq:def4DlimitcalZ} of the 4D $T_N$ theories. 
We want to stress that taking this limit is a tricky business, as the expression  \eqref{eq:3pointfunctionsastopologicalstrings}  includes non-trivial multiple sums and integrals.  This  is the reason why we will always work with the $q$-deformed formulas and take the limit only at the end.

This article is organized as follows. After briefly reminding  the reader of  the essentials of Toda CFTs, we recall the formula by Fateev and Litvinov for a {\it special class} of 3-point functions of Toda primaries, as well as its straightforward generalization to the conjectural $q$-deformed Toda theory. We then conclude section~\ref{sec:FL} by quoting our general proposal for {\it generic} 3-point functions of Toda primaries. To spell out the details of it, we will need some basics of the AGT dictionary collected in section~\ref{sec:AGT}. In the next section~\ref{sec:Higgsing}, the discussion temporarily deviates from the CFT matters focusing rather on the interplay between the moduli spaces of the corresponding gauge theories and 5-brane web physics. We argue that the semi-degeneration of a primary field on the ($q$-deformed) CFT side mirrors a Higgsing of the $T_N$ theory on the 4D (5D) side.  A more CFT-oriented reader can skip this section, with the exception of \ref{subsec:domainofparameters}.  
The AGT genesis of Fateev-Litvinov formula for \textbf{W}$_3$ Toda 3-point function, via pinching an integration contour by a particular residue of the corresponding integrand and applying non-trivial summation theorems, is what section~\ref{sec:pinching} focuses upon. With the details of \textbf{W}$_4$ computation deferred to the appendix~\ref{app:subappT_4}, we then proceed to a discussion of the general \textbf{W}$_N$ case in section~\ref{sec:generalN}.
The conclusion and the outlook follow, whereas the remaining appendices are devoted to overview of notations and special functions, most importantly to describing and elaborating on the properties of the Kaneko-Macdonald-Warnaar $\bsl(N)$ hypergeometric functions which play a major role in our calculations.

\section{Toda CFT: a recap and a proposal}
\label{sec:FL}

In this section we briefly summarize some relevant facts about the Toda CFT, closely following \cite{Fateev:2005gs,Fateev:2007ab,Fateev:2008bm}. Furthermore, we spell out the Fateev-Litvinov formula for a special subset of Toda structure constants and present our proposal for the Toda 3-point functions of generic primary fields.

The Lagrangian of the $A_{N-1}$ Toda CFT is given by
\beq
\label{eq:Lagrangian}
L=\frac{1}{8\pi}\form{\partial_{\nu}\varphi}{\partial^{\nu}\varphi}+\mu\sum_{k=1}^{N-1}e^{b\form{e_k}{\varphi}},
\eeq
where $\varphi\colonequals \sum_{i=1}^{N-1}\varphi_i\omega_i$, with $e_k$, $\omega_k$ being the simple roots and the fundamental weights of sl$(N)$ respectively. 
The definition of the inner product $\form{\cdot}{\cdot}$ along with other
useful Lie-algebraic definitions and notations are collected in appendix \ref{subapp:sln} for the convenience of the reader.
The parameter $\mu$ is called the \textit{cosmological constant}, in analogy to the Liouville case ($N=2$) where it determines the constant curvature of a surface described by the classical equation of motion. The normalization of the Lagrangian is chosen in such a way that
\begin{align}
\varphi_i(z,\bar{z})\,\varphi_j(0,0)=-\delta_{ij}\,\text{log}|z|^2+\cdots & \hspace{4mm} \text{ at } z\rightarrow 0.
\end{align}
Following \cite{Fateev:2007ab,Fateev:2008bm}, we consider the correlators on a two-sphere, which prescribes putting a background charge at the north pole in order to render the Toda action finite:
\begin{align}
\varphi(z,\bar{z})=-\fQ\, \text{log}|z|+\cdots & \hspace{4mm} \text{ at } z\rightarrow \infty,
\end{align}
where $\fQ\colonequals Q\rho=(b+b^{-1})\rho$ with the Weyl vector $\rho$ defined in \eqref{eq:defWeylvector}.

Analyzing the path integral of the theory \eqref{eq:Lagrangian}, one can argue that the Toda CFT must have an exchange symmetry $b\leftrightarrow b^{-1}$ on a quantum level which simultaneously sends the cosmological constant to its dual $\tilde{\mu}$, defined as
\beq
\left(\pi\tilde{\mu}\gamma(b^{-2})\right)^b\stackrel{!}{=}\left(\pi\mu\gamma(b^{2})\right)^{\frac{1}{b}}\Longrightarrow 
\tilde{\mu}=\frac{\left(\pi\mu\gamma(b^2)\right)^{\nicefrac{1}{b^2}}}{\pi\gamma(\nicefrac{1}{b^2})},
\eeq
where $\gamma(x)\colonequals \frac{\Gamma(x)}{\Gamma(1-x)}$.
As we mentioned in the introduction, the Toda CFT also has a \textbf{W}$_N$ higher spin chiral symmetry generated by the fields $W_2\equiv T$, $W_3, \ldots, W_N$ of spins $2,\dots,N$. The primaries under the full symmetry algebra $\textbf{W}_N\times \overline{\textbf{W}}_N$ are the exponential fields of spin zero labeled by a weight of sl$(N)$:
\beq
\label{eq:defprimaryfield}
V_{\balpha}\colonequals e^{\form{\balpha}{\varphi}}.
\eeq
In what follows, we will parametrize the fundamental weight decomposition of a weight $\balpha_i$ as
\beq
\label{eq:paramalphaN}
\balpha_i=N\sum_{j=1}^{N-1}\alpha_i^j\omega_j.
\eeq
By looking at the corresponding OPEs, one reads off the central charge $c$ of the Toda CFT and the conformal dimensions $\Delta(\alpha)$ of its primary fields:
\beq 
c=N-1+12\form{\fQ}{\fQ}=(N-1)\left(1+N(N+1)Q^2\right),\qquad  \Delta(\alpha)=\frac{\form{2\fQ-\balpha}{\balpha}}{2},
\eeq
with the anti-holomorphic conformal dimensions of the primary fields being equal to the holomorphic ones.

The conformal dimension, as well as the eigenvalues of all the other higher spin currents $W_k$ are invariant under the affine\footnote{One should not confuse the affine Weyl transformation, \textit{i.e.} Weyl reflections accompanied by two translations, with Weyl reflections belonging to the Weyl group of the affine Lie algebra.} Weyl transformations \eqref{eq:defaffineWeyltransformations} of the weights $\balpha_i$, which roughly means that several exponential fields correspond to the same 'physical' field.
The primary fields of Toda CFT  transform under an affine Weyl transformations $\balpha\rightarrow \fw\circ \balpha$ given in \eqref{eq:defaffineWeyltransformations} as
\beq
\label{eq:reflectionprimaryfield}
V_{\fw\circ \balpha} = \fR^{\fw}(\balpha) V_{\balpha}
\eeq
with the reflection amplitude $\fR$ given by the expression
\beq
\label{eq:defreflectionamplitude}
\fR^{\fw}(\balpha)\colonequals \frac{\fA(\balpha)}{\fA(\fw\circ\balpha)} \, 
\eeq
in terms of the function
\beq
\fA(\balpha)\colonequals \left(\pi \mu\gamma(b^2)\right)^{\frac{\form{\balpha-\fQ}{\rho}}{b}}\prod_{e>0}\Gamma\left(1-b\form{\balpha-\fQ}{e}\right)\Gamma\left(-b^{-1}\form{\balpha-\fQ}{e}\right).
\eeq

The two-point correlation functions of primary fields are fixed by conformal invariance and by the normalization \eqref{eq:defprimaryfield}. They read
\beq
\label{eq:twopointfunctions}
\left\langle V_{\balpha_1}(z_1,\bar{z}_1)V_{\balpha_2}(z_2,\bar{z}_2)\right\rangle=\frac{(2\pi)^{N-1}\delta(\balpha_1+\balpha_2-2\fQ)+\text{Weyl-reflections}}{|z_1-z_2|^{4\Delta(\balpha_1)}},
\eeq
where ``Weyl-reflections'' stands for additional $\delta$-contributions that come from the field identifications \eqref{eq:reflectionprimaryfield}. 

The coordinate dependence of 3-point functions of primary fields \eqref{eq:defprimaryfield} is fixed by conformal symmetry up to an overall coefficient $C(\balpha_1,\balpha_2,\balpha_3)$ called the 3-point structure constant:

\beq
\label{eq:3pointcorrelator}
\left\langle V_{\balpha_1}(z_1,\bar{z}_1)V_{\balpha_2}(z_2,\bar{z}_2)V_{\balpha_3}(z_3,\bar{z}_3)\right\rangle=\frac{C(\balpha_1,\balpha_2,\balpha_3)}{|z_{12}|^{2(\Delta_1+\Delta_2-\Delta_3)}|z_{13}|^{2(\Delta_1+\Delta_3-\Delta_2)}|z_{23}|^{2(\Delta_2+\Delta_3-\Delta_1)}}\,,
\eeq
where $z_{ij}\colonequals z_i-z_j$ and $\Delta_i$ is the conformal dimension of the primary $V_{\balpha_i}$.

Up to now, the CFT machinery has produced expressions only for a restricted subset of 3-point functions, as well as for some interesting physical limits of those, see \cite{Fateev:2005gs,Fateev:2007ab,Fateev:2008bm} for the state of the art. The formula of Fateev and Litvinov \cite{Fateev:2005gs} which we will quote in a moment gives the Toda structure constants for the particular \textit{semi-degenerate case} when one of the fields contains a null-vector at level one, implying that the corresponding weight becomes proportional to the first $\omega_1$ or to the last $\omega_{N-1}$ fundamental weight of sl$(N)$. Specifically, if one sets\footnote{We use a slightly different convention than \cite{Fateev:2005gs}. One has to rescale $\varkappa\rightarrow \frac{\varkappa}{N}$ to match the expressions.} $\balpha_1=N\varkappa \omega_{N-1}$, the structure constants read
 \beq
\begin{split}
\label{eq:FLTodacorr}
C(N \varkappa \omega_{N-1},\balpha_2,\balpha_3)=&\left(\pi\mu\gamma(b^2)b^{2-2b^2}\right)^{\frac{\form{2\fQ-\sum_{i=1}^3\balpha_i}{\rho}}{b}}\times \\&\times\frac{\Upsilon'(0)^{N-1}\Upsilon(N\varkappa)\prod_{e>0}\Upsilon(\form{\fQ-\balpha_2}{e})\Upsilon(\form{\fQ-\balpha_3}{e})}{\prod_{i,j=1}^N\Upsilon(\varkappa+\form{\balpha_2-\fQ}{h_i}+(\balpha_3-\fQ, h_j))} \, ,
\end{split}
\eeq
where the function $\Upsilon$ is an entire function defined in appendix~\ref{subapp:special}.

Before presenting our formula for the 3-point functions, we need to introduce the $q${\it -deformed Toda theory}.
 Albeit no Lagrangian description of the $q$-deformed version of Toda field theory has been found yet, many quantities of this conjectural deformation are algebraically well-defined, in full analogy to the Toda CFT
(see \cite{Awata:1996dx} and references therein). 
While the $q$-deformed Toda CFTs are vastly unexplored, for the $q$-deformed Liouville case a bit more is known   \cite{Awata:2009ur,Awata:2010yy,Schiappa:2009cc,Mironov:2011dk,Itoyama:2013mca,Bao:2011rc,Nieri:2013yra,Bao:2013pwa,Nieri:2013vba,Aganagic:2013tta,Aganagic:2014oia,
 Taki:2014fva,Dijkgraaf:2009pc,Cheng:2010yw,Tan:2013xba}.
The details of our working definition for the $q$-deformed Toda are presented in section 3.4 of  \cite{Mitev:2014isa}.
The building blocks of our proposal are $q$-deformed functions who reproduce  the known limit as $q\colonequals e^{-\beta}\rightarrow 1$, keep the same symmetries and transformation properties as well as the
 poles and zeros\footnote{To be more precise, the $q$-deformed functions have a whole tower of zeroes/poles for each zero/pole of the undeformed function.
The tower is generated by beginning with the undeformed  zero/pole and translating it by
$r\frac{2\pi i }{\log q} = - r \frac{2\pi i }{\beta}$, where $r$ is a positive integer.} of  the undeformed ones. 
In the Toda CFT, the dependence on the cosmological constant $\mu$
 is fully fixed by a Ward identity coming from the path integral formulation. The absence of a path integral formulation  for the $q$-deformed Toda implies that such quantities as structure constants of the theory are ambiguous up to a function of $\mu$, $b$ and $q$.  Due to this, we define the $q$-deformed structure constants here up to the $\pi \mu \gamma(b^2)$ term, having $q$-deformed only the part respecting the symmetry $b\leftrightarrow b^{-1}$:
 \beq
\begin{split}
\label{eq:qdefFLTodacorr}
C_q(N\varkappa \omega_{N-1},\balpha_2,\balpha_3)\cong &\left(\frac{\big(1-q^b\big)^2\big(1-q^{b^{-1}}\big)^{2b^2}}{(1-q)^{2(1+b^2)}}\right)^{\frac{\form{2\fQ-\sum_{i=1}^3\balpha_i}{\rho}}{b}} \\&\times\frac{\Up'(0)^{N-1}\Up(N\varkappa)\prod_{e>0}\Up(\form{\fQ-\balpha_2}{e})\Up(\form{\fQ-\balpha_3}{e})}{\prod_{i,j=1}^N\Up(\varkappa+\form{\balpha_2-\fQ}{h_i}+\form{\balpha_3-\fQ}{h_j})}\, ,
\end{split}
\eeq
where the function $\Up$ is a $q$-deformation of $\Upsilon$, also defined in appendix~\ref{subapp:special}.
To match with the undeformed Toda structure constants in the limit $q\rightarrow 1$, one has to set, respectively:
\beq
\label{Cq2C}
C_q(\balpha_1,\balpha_2,\balpha_3)\stackrel{q\rightarrow 1}{\longrightarrow}\left(\pi \mu\gamma(b^2)\right)^{-\frac{\form{2\fQ-\sum_{i=1}^3\balpha_i}{\rho}}{b}}C(\balpha_1,\balpha_2,\balpha_3) \,.
\eeq
In our calculations, we will reproduce the $q$-deformed Fateev-Litvinov formula \eqref{eq:qdefFLTodacorr} which then gives the undeformed one \eqref{eq:FLTodacorr} upon taking the limit $q\rightarrow 1$ and reintroducing the $\mu$-dependence as in \eqref{Cq2C}.

We finish this section with our proposal for the 3-point function of  of generic primary fields of the Toda theory
\begin{multline}
 \label{eq:3pointfunctionsastopologicalstrings}
 C(\balpha_1,\balpha_2,\balpha_3)=\,\text{const}\times \left(\pi \mu\gamma(b^2)b^{2-2b^2}\right)^{\frac{\form{2\fQ-\sum_{i=1}^3\balpha_i}{\rho}}{b}} \\\times \lim_{\beta\rightarrow 0}\beta^{-2Q\sum_{i=1}^3\form{\balpha_i}{\rho}} \oint\prod_{i=1}^{N-2}\prod_{j=1}^{N-1-i}\left[\frac{d\tilde{A}_i^{(j)}}{2\pi i \tilde{A}_i^{(j)}}\left|M(\ft,\fq)\right|^{2}\right] \left|\calZ_N^{\text{top}}\right|^2
 \end{multline}
where by ``const'' we mean an overall  function of only $b$ that is independent of the weights of the CFT primaries.
To spell out the details of the right-hand side, in particular the topological string amplitude $\calZ_N^{\text{top}}$,
we require some notions and notations which will come in the next section. The impatient reader may skip the explanations and proceed straight to the formulae \eqref{eq:identificationparameters}, \eqref{eq:normsquared}, \eqref{eq:finalexpressionforZTN}, \eqref{eq:ZTNperturbative}, \eqref{eq:ZTNsum}, \eqref{eq:borderAdefMNL} consulting also appendices \ref{subapp:special}, \ref{app:finiteproduct} for definitions of the encountered special functions.

\section{AGT dictionary}
\label{sec:AGT}

According to the AGT-W correspondence \cite{Alday:2009aq,Wyllard:2009hg}, the correlation functions of the 2D Toda CFT are obtained from the partition functions of the corresponding 4D $\mathcal{N}=2$ gauge theories as
\begin{align}
\label{eq:AGTWrelation}
\calZ^{S^4}=\int [da] \Big|\calZ_{\textrm{Nek}}^{\textrm{4D}}(a,m,\tau,\epsilon_{1,2})\Big|^2 
\propto 
\langle V_{\boldsymbol{\alpha}_1}(z_1)\cdots V_{\boldsymbol{\alpha}_n}(z_n)
\rangle_{\textrm{Toda}}
\, ,
\end{align}
where the Omega deformation parameters are related to the Toda coupling constant\footnote{We also use the notation $\epsilon_+=\epsilon_1+\epsilon_2$.  When we specialize $\epsilon_1=b$ and $\epsilon_2=b^{-1}$ in order to connect the topological string expressions to the Toda expressions, we have $\epsilon_+=b+b^{-1}=Q$.} via $\epsilon_1=b$ and $\epsilon_2=b^{-1}$. Moreover, $a$ stands for the set of Coulomb moduli of the theory, $m$ for the masses of the hypermultiplets and $\tau$ for the coupling constants. The correspondence relates the masses $m$ to the weights $\boldsymbol{\alpha}_i$ and the couplings constants $\tau$ to the insertion points $z_i$ of the primary fields.
In particular, the conformal blocks of the 2D CFTs are given by the appropriate Nekrasov instanton partition functions \cite{Alday:2009aq,Wyllard:2009hg} and the 3-point structure constants  by the partition functions of the $T_N$ superconformal theories on $S^4$ \cite{Kozcaz:2010af,Bao:2013pwa}.

\begin{figure}[ht]
 \centering
  \includegraphics[height=3.5cm]{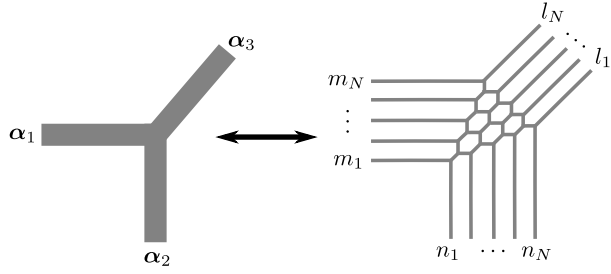}
  \caption{\it This figure depicts the identification of the $\balpha$ weights appearing on the Toda CFT side with the position of the flavor branes on the $T_N$ side, here drawn for the case $N=5$.}
  \label{fig:3point}
\end{figure}

A similar relation between 5D gauge theories and  2D $q$-CFT exists
 \cite{Awata:2009ur,Awata:2010yy,Schiappa:2009cc,Mironov:2011dk,Itoyama:2013mca,Bao:2011rc,Nieri:2013yra,Bao:2013pwa,Nieri:2013vba,Aganagic:2013tta,Aganagic:2014oia,
 Taki:2014fva,Dijkgraaf:2009pc,Cheng:2010yw,Tan:2013xba}, which relates the 5D Nekrasov partition functions on $S^4 \times S^1$ to correlation functions the of $q$-deformed Liouville/Toda field theory:
\begin{align}
\calZ^{S^4\times S^1}=\int [da] \Big|
\calZ_{\textrm{Nek}}^{\textrm{5D}}(a,m,\tau, \beta,\epsilon_{1,2})\Big|^2
\propto  \langle
V_{\boldsymbol{\alpha}_1}(z_1)\cdots V_{\boldsymbol{\alpha}_n}(z_n)
\rangle_{q\textrm{-Toda}} \, ,
\end{align}
where $\beta= - \log q$ is the circumference  of the $ S^1$. The exponentiated Omega background parameters
\beq
\fq=e^{-\beta \epsilon_1}\, ,\qquad \ft=e^{\beta \epsilon_2}\, ,
\eeq
are used in this case.
The partition function on  $S^4\times S^1$ is
the 5D superconformal index, 
which as discussed  in \cite{Iqbal:2012xm} 
 can also be computed using topological string theory techniques
\beq
\calZ^{S^4\times S^1}= \int [d a] \, |\calZ_{\textrm{Nek}}^{\textrm{5D}} (a)|^2\propto \int [d a] \, |\calZ_{\textrm{top}} (a)|^2 \, .
\eeq
In \cite{Bao:2013pwa} we computed the partition functions of the 5D $T_N$ theories on $S^4\times S^1$ (see also \cite{Hayashi:2013qwa})   and suggested that they should be interpreted as the 3-point structure constants of $q$-deformed Toda. We read them off from the toric-web diagrams of the $T_N$ junctions of \cite{Benini:2009gi} by  employing the refined topological vertex formalism of \cite{Awata:2005fa,Iqbal:2007ii}.  In a subsequent paper \cite{Mitev:2014isa}, part one of the present series of papers, we showed how the 4D limit, corresponding to $\beta \rightarrow 0$ or $q\rightarrow 1$, is to be taken. We thus obtained the partition function of the 4D $T_N$ theories on $S^4$
 \beq
\label{eq:def4DlimitcalZ}
\calZ_{N}^{S^4}=\text{const}\times \lim_{\beta\rightarrow 0} \beta^{-\frac{\chi_N}{\epsilon_1\epsilon_2}}\calZ_{N}^{S^4\times S^1} \, ,
\eeq
where by ``const'' we mean a function of $\epsilon_1$, $\epsilon_2$ that is independent of the mass parameters of the theory.
The degree of divergence was determined as proportional to the quadratic Casimir of $\SU(N)^3$ 
\beq
\label{eq:chiNgeneral}
\chi_N=-\sum_{1\leq i<j\leq N}\left[(m_i-m_j)^2+(n_j-n_i)^2+(l_i-l_j)^2\right]=-N\sum_{i=1}^3 \form{\balpha_i-\fQ}{\balpha_i-\fQ} \,,
\eeq
where $\fQ\colonequals Q\rho=(b+b^{-1})\rho$ with the $\SU(N)$ Weyl vector $\rho$ defined in \eqref{eq:defWeylvector}. After the first equality of  \eqref{eq:chiNgeneral}, we have introduced the mass parameters $m_i$, $n_i$ and $l_i$ of the $T_N$ theory, which, as shown in figure~\ref{fig:3point}, are connected to the Toda theory parameters \cite{Mitev:2014isa}
 \beq
\label{eq:identificationparameters}
\begin{split}
m_i&=\form{\balpha_1-\fQ}{h_i}=N\sum_{j=i}^{N-1}\alpha_1^j-\sum_{j=1}^{N-1}j\alpha_1^j-\frac{N+1-2i}{2}Q\,,\\
n_i&=-\form{\balpha_2-\fQ}{h_i}=-N\sum_{j=i}^{N-1}\alpha_2^j+\sum_{j=1}^{N-1}j\alpha_2^j+\frac{N+1-2i}{2}Q\,,\\
l_i&=-\form{\balpha_3-\fQ}{h_{N+1-i}}=-N\sum_{j=N+1-i}^{N-1}\alpha_3^j+\sum_{j=1}^{N-1}j\alpha_3^j-\frac{N+1-2i}{2}Q \, .
\end{split}
\eeq
It is important to note, that the mass parameters are not all independent, but obey
\beq
\sum_{i=1}^Nm_i=\sum_{i=1}^Nn_i=\sum_{i=1}^Nl_i=0\, ,
\eeq
which is reflected in the fact that the sum of the weights $h_i$ of the fundamental $\SU(N)$ representation is zero.
Then the structure constants of three primary operators in the $q$-Toda theory are given by the $T_N$ partition functions on $S^4\times S^1$ as 
\begin{align}
\label{eq:main5Dequality}
C_q(\balpha_1,\balpha_2,\balpha_3)=\text{const}\times \left[\prod_{j=1}^3Y_q(\balpha_j)\right] (1-q)^{-\chi_N} \calZ_{N}^{S^4\times S^1}\,,
\end{align}
where by ``const'' we mean a function of $\epsilon_1$, $\epsilon_2$ and $\beta$ that is independent of the mass parameters of the theory. We stress that the superconformal index $\calZ_{N}^{S^4\times S^1}$ is invariant under the affine Weyl transformations \eqref{eq:defWeyltransformations} and that  all the non-trivial Weyl transformation properties of the structure constants are captured by  the following special functions:
\beq
\label{eq:deffunctionYqdef}
Y_q(\balpha)\colonequals \left[\frac{\big(1-q^b\big)^{2b^{-1}}\big(1-q^{b^{-1}}\big)^{2b}}{(1-q)^{2Q}}\right]^{-\form{\balpha}{\rho}}\prod_{e>0}\Up\left(\form{\fQ-\balpha}{e}\right)\,,
\eeq
with the functions $\Up$ defined in \eqref{eq:defUp} and the product taken over all positive roots $e$ of $\SU(N)$. The partition function on  $S^4\times S^1$, or the superconformal index,  for the $T_N$ theory is given by an integral over the refined topological string amplitude with an integration measure containing the refined MacMahon function\footnote{See \eqref{eq:McMahonUp} for the definition of the refined MacMahon function $M(\ft,\fq)$.} $M(\ft,\fq)$ \cite{Iqbal:2012xm} 
\beq
\label{eq:defindex}
\calZ_N^{S^4\times S^1}\colonequals \oint\prod_{i=1}^{N-2}\prod_{j=1}^{N-1-i}\left[\frac{d\tilde{A}_i^{(j)}}{2\pi i \tilde{A}_i^{(j)}}\left|M(\ft,\fq)\right|^{2}\right] \left|\frac{\calZ_N^{\text{top}}}{\calZ_{N}^{\text{dec}}}\right|^2\,.
\eeq
Here, we have removed the decoupled degrees of freedom, referred to as ``non-full spin content'' in \cite{Bao:2013pwa}, 
\beq
\label{NonFullSpin}
\begin{split}
\left|\calZ_{N}^{\text{dec}}\right|^2&\colonequals \prod_{1\leq i<j\leq N}\Big|\calM(\tilde{M}_i\tilde{M}_j^{-1})
\calM(\nicefrac{\ft}{\fq}\tilde{N}_i\tilde{N}_j^{-1})\calM(\tilde{L}_i\tilde{L}_j^{-1})\Big|^2\\
&=\text{const}\times \prod_{k=1}^3\left(1-q\right)^{N\form{\balpha_k}{\balpha_k-2\fQ}}\left(\big(1-q^b\big)^{2b^{-1}}\big(1-q^{b^{-1}}\big)^{2b}\right)^{\form{\balpha_k}{\rho}}Y_q(\balpha_k)\,,
\end{split}
\eeq
where the function $\calM$ is defined in \eqref{eq:defcalMeverywhere}.
\begin{figure}[t]
 \centering
  \includegraphics[height=3.5cm]{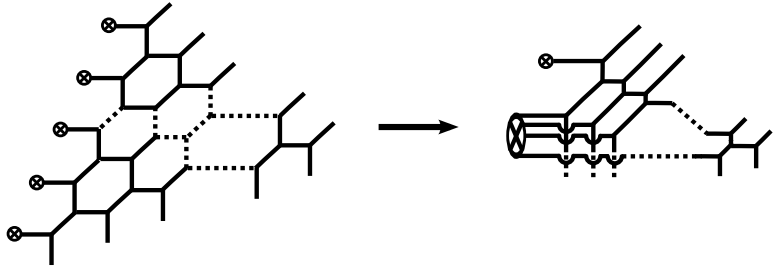}
  \caption{\it The figure illustrates the desired Higgsing procedure for the general $T_N$ diagram. We denote $7$-branes by crossed circles. The left part of the figure shows the original $T_N$ 5-brane web diagram, while the right one depicts the web diagram obtained by letting $N-1$ of the  left 5-branes terminate on the same 7-brane.}
  \label{fig:TNHiggsing}
\end{figure}
Interestingly enough, as noted in \cite{Mitev:2014isa}, these degrees of freedom are responsible for the Weyl covariance of the Toda structure constants. Here and elsewhere, we shall use the shorthand notation
\beq
\label{eq:normsquared}
|f(U_1,\ldots, U_r;\ft,\fq)|^2\colonequals f(U_1,\ldots, U_r;\ft,\fq)f(U_1^{-1},\ldots, U_r^{-1};\ft^{-1},\fq^{-1})\,.
\eeq
Inserting \eqref{eq:defindex} into \eqref{eq:main5Dequality}, we find the nice expression
\beq
\label{eq:finalCqformula}
C_q(\balpha_1,\balpha_2,\balpha_3)=\text{const}\times \oint\prod_{i=1}^{N-2}\prod_{j=1}^{N-1-i}\left[\frac{d\tilde{A}_i^{(j)}}{2\pi i \tilde{A}_i^{(j)}}\left|M(\ft,\fq)\right|^{2}\right] \left|\calZ_N^{\text{top}}\right|^2\, .
\eeq
The topological string amplitude is $\calZ_N^{\text{top}}$ obtained from the $T_N$ web-diagram by using the refined topological vertex formalism and reads
\beq
\label{eq:finalexpressionforZTN}
\calZ_N^{\text{top}}=\calZ_N^{\text{pert}}\calZ_N^{\text{inst}}\, ,
\eeq
where the ``perturbative'' partition function\footnote{We put the words  ``perturbative'' and ``instanton'' inside quotation marks because for the $T_N$ there is no notion of  instanton expansion, since there is no coupling constant.} is
\begin{align}
\label{eq:ZTNperturbative}
\calZ_N^{\text{pert}}&\colonequals \prod_{r=1}^{N-1}\prod_{1\leq i<j\leq N-r}\frac{\calM\Big(\frac{\tilde{A}_i^{(r-1)}\tilde{A}_j^{(r-1)}}{\tilde{A}_{i-1}^{(r-1)}\tilde{A}_{j+1}^{(r-1)}}\Big)}{\calM\Big(\sqrt{\frac{\ft}{\fq}}\frac{\tilde{A}_i^{(r-1)}\tilde{A}_{j-1}^{(r)}}{\tilde{A}_{i-1}^{(r-1)}\tilde{A}_{j}^{(r)}}\Big)\calM\Big(\sqrt{\frac{\ft}{\fq}}\frac{\tilde{A}_i^{(r)}\tilde{A}_j^{(r-1)}}{\tilde{A}_{i-1}^{(r)}\tilde{A}_{j+1}^{(r-1)}}\Big)}\prod_{1\leq i<j\leq N-r-1}\calM\Big(\frac{\ft}{\fq}\frac{\tilde{A}_i^{(r)}\tilde{A}_{j}^{(r)}}{\tilde{A}_{i-1}^{(r)}\tilde{A}_{j+1}^{(r)}}\Big),
\end{align}
and the ``instanton'' one is
\begin{align}
\label{eq:ZTNsum}
\calZ_N^{\text{inst}}&\colonequals\nonumber\\& \sum_{\boldsymbol{\nu}}\prod_{r=1}^{N-1}\prod_{i=1}^{N-r}\left(\frac{\tilde{N}_r \tilde{L}_{N-r}}{\tilde{N}_{r+1}\tilde{L}_{N-r+1}}\right)^{\frac{|\nu_i^{(r)}|}{2}} 
\prod_{r=1}^{N-1}\prod_{1\leq i\leq j\leq N-r}\left[\frac{\tN_{\nu_i^{(r-1)}\nu_j^{(r)}}\Big(a_i^{(r-1)}+a_{j-1}^{(r)}-a_{i-1}^{(r-1)}-a_j^{(r)}-\nicefrac{\epsilon_+}{2}\Big)}{\tN_{\nu_i^{(r-1)}\nu_{j+1}^{(r-1)}}\Big(a_i^{(r-1)}+a_j^{(r-1)}-a_{i-1}^{(r-1)}-a_{j+1}^{(r-1)}\Big)}\right.\nonumber\\&\times\left. \frac{\tN_{\nu_i^{(r)}\nu_{j+1}^{(r-1)}}\Big(a_i^{(r)}+a_j^{(r-1)}-a_{i-1}^{(r)}-a_{j+1}^{(r-1)}-\nicefrac{\epsilon_+}{2}\Big)}{\tN_{\nu_i^{(r)}\nu_j^{(r)}}\Big(a_i^{(r)}+a_{j-1}^{(r)}-a_{i-1}^{(r)}-a_{j}^{(r)}-\epsilon_+\Big)}\right],
\end{align}
where the $a_{i}^{(j)}$ are defined via $\tilde{A}_i^{(j)}=e^{-\beta a_{i}^{(j)}}$, while the $\tN_{\lambda\mu}$ are given in \eqref{eq:deftN}. The summation goes over  $\frac{N(N-1)}{2}$ partitions $\nu_i^{(r)}$, $r=1,\dots ,N-1$, $i=1,\dots ,N-r$. The ``interior'' Coulomb moduli $\tilde{A}_j^{(i)}=e^{-\beta a_i^{(j)}}$ are independent, while the  ``border'' ones are given by
\beq
\label{eq:borderAdefMNL}
\tilde{A}_{i}^{(0)} = \prod_{k=1}^i \tilde{M}_k\,,
\qquad
\tilde{A}_{0}^{(i)} = \prod_{k=1}^i \tilde{N}_k\, ,
\qquad
\tilde{A}_{i}^{(N-i)}  = \prod_{k=1}^i \tilde{L}_k\,  ,
\eeq
where $\tilde{M}_k\colonequals e^{-\beta m_k}$ and similarly for $\tilde{N}_k$  and $\tilde{L}_k$. See appendix~\ref{appA} for more details on the parametrization of the $T_N$ junction.

\begin{figure}[t]
   \centering
   \includegraphics[height=3.2cm]{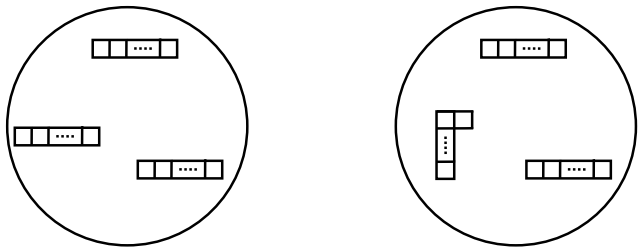}
   \caption{On the left we depict the sphere with three full punctures that corresponds to the un-Higgsed $T_N$ with $\SU(N)^3$ global symmetry. On the right we show the sphere with two full punctures and one L-shaped $\{N-1,1\}$ puncture. This particular Higgsing of $T_N$ leads to a theory with with $\SU(N)\times \SU(N)\times \text{U}(1)$ global symmetry. 
   The partition function of this theory will lead to the Toda 3-point function with one semi-degenerate primary insertion.}
   \label{fig:SemiDeg}
\end{figure}

The formula \eqref{eq:3pointfunctionsastopologicalstrings} (correspondingly, \eqref{eq:main5Dequality}) for the structure constants of three primary fields of ($q$-deformed) Toda CFT, has the correct symmetry properties, the zeros that it should and, for $N=2$, gives the known answer for the Liouville CFT \cite{Mitev:2014isa}. However, it is very implicit, requiring to perform $\frac{N(N-1)}{2}$ sums over the partitions $\nu_i^{(j)}$, followed by a $ \frac{(N-1)(N-2)}{2}$-dimensional\footnote{It is the number of faces of the left diagram in figure~\ref{fig:TNHiggsing}.} integral over the Coulomb moduli $\tilde{A}_i^{(j)}$ and finally to take the 4D ($q\rightarrow 1$) limit \eqref{eq:def4DlimitcalZ}. In the subsequent parts of the paper we will show how to derive the special case \eqref{eq:FLTodacorr}, known due to Fateev and Litvinov \cite{Fateev:2005gs,Fateev:2007ab,Fateev:2008bm}, from our formula \eqref{eq:3pointfunctionsastopologicalstrings}. This provides a strong check of our general proposal.

\section{Semi-degeneration from Higgsing the \texorpdfstring{$T_N$}{TN} theories}
\label{sec:Higgsing}

\begin{figure}[t]
   \centering
   \includegraphics[width=14cm]{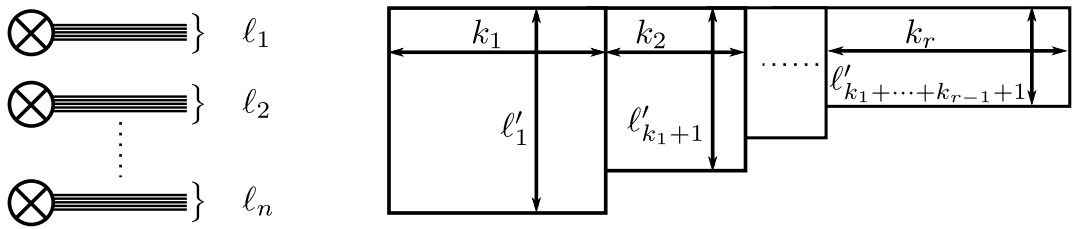}
   \caption{On the left part of this figure, we see $N$ 5-branes ending on $n$ 7-branes in bunches of $\ell_1,\ldots ,\ell_n$ 5-branes each. On the right side of the figure, we depict the Young diagram $\{\ell'_1,\ell'_2,\dots,\ell'_n\}$ that gives the flavor symmetry of the corresponding puncture. Having $n$  bunches of 5-branes, each ending of a 7-brane leads to a puncture in the Gaiotto curve with flavor symmetry $S(\text{U}(k_1)\times \cdots \times \text{U}(k_r))$, where the widths $k_i$ of the boxes are equal to the numbers of stacks with the same number of branes per stack.}
   \label{fig:Stacks}
\end{figure}

In this section we argue that a particular way of Higgsing  the $T_N$ theories, as depicted in figure \ref{fig:TNHiggsing}, corresponds to the degeneration with one simple and two full punctures. On the Toda side, this is equivalent to the semi-degeneration of Fateev and Litvinov. On the gauge theory side, the partition function of the theory  with one simple and two full punctures is the partition function of $N^2$ free hypermultiplets. Our discussion is based on the physics of $(p,q)$ 5-brane webs and their symmetries.
 In particular, we identify which  Higgsing mechanism corresponds to the Fateev and Litvinov semi-degeneration by introducing 7-branes on the  5-brane web. Finally, in this section, we discuss the domain in which the mass parameters, or Toda weights,  take value, which will dictate the contour for the integral \eqref{eq:3pointfunctionsastopologicalstrings}.

In the next sections we will use the intuition acquired here to explicitly substitute the values dictated by the web diagram, \eqref{Higgsed_A's}  and \eqref{Higgsed_Q's},  in \eqref{eq:main5Dequality} so as to obtain the  formula \eqref{eq:FLTodacorr} by Fateev and Litvinov.

\subsection{Higgsing the $T_N$ -  Review}

The physics of  the  $(p,q)$ 5-brane webs that we will need in the context of this section is studied in  \cite{Benini:2009gi,Hayashi:2013qwa,Hayashi:2014wfa,Hayashi:2014hfa}. We give a short review of their relevant results. A very useful way of realizing 4D $\mathcal{N}=2$ quiver gauge theories in  string theory is by using type IIA string theory and the Hanany-Witten construction \cite{Hanany:1996ie} of D4 branes suspended between NS5 branes \cite{Witten:1997sc}. This configuration can be lifted to M-theory, where both the D4 and the NS5 branes become a single M5 brane with non-trivial topology, physically realizing the Seiberg-Witten curve in which all the low energy data are encoded \cite{Witten:1997sc}. Similarly,  5D  $\mathcal{N}=1$ gauge theories can be realized using type IIB string theory with D5 branes suspended between NS5 branes forming $(p,q)$ 5-brane webs \cite{Aharony:1997ju,Aharony:1997bh}.
A large class of $\mathcal{N}=2$ SCFTs, called class $\mathcal{S}$, can be reformulated (from the realization in \cite{Witten:1997sc} with a single M5 brane with non-trivial topology) as a compactification of $N$ M5 branes on a sphere \cite{Gaiotto:2009we}.  This point of view  is very useful since intersections of these  $N$ M5 branes with other M5 branes can be thought of as insertions of defect operators on the world volume of the M5 branes and thus punctures on the sphere. The name  {\it simple puncture} is used for defects that are obtained from the intersection of the original $N$ M5 branes with a single M5 brane (originating from D4's ending on an NS5 in the Hanany-Witten construction), while  {\it full or maximal punctures} stem from defects corresponding to intersections with $N$ semi-infinite M5 branes  (external flavor semi-infinite D4's  in \cite{Witten:1997sc}).

{\it More general punctures}, naturally labeled by Young diagrams consisting of $N$ boxes, are also possible
\cite{Gaiotto:2009we,Gaiotto:2009gz}. In the $(p,q)$ 5-brane web language, they  can be described when additional 7-branes are introduced  \cite{Benini:2009gi}.  Semi-infinite $(p,q)$ 5-branes  are equivalent to    $(p,q)$ 5-branes ending on  $(p,q)$ 7-branes \cite{DeWolfe:1999hj}.
Consider $N$ 5-branes and let them end on $n$ 7-branes, as shown on the left of figure~\ref{fig:Stacks}. The $j^{\text{th}}$ 7-brane carries $\ell_j$ 5-branes. We define the numbers $\ell_j'$ as a permutation of the $\ell_j$ such that they are ordered
\beq
\ell'_1\geq \ell'_2\geq\cdots \geq \ell_n'\, ,
\eeq
and arrange them as the columns of a Young diagram\footnote{In this article, we draw the Young diagrams in the English notation. By $\{c_1,\ldots, c_r\}$ we mean a Young diagram with $r$ columns for which the  $j$-th column has $c_j$ boxes, $j=1,\ldots, r$. Furthermore, we use the notation $\{a^b\}$ for the partition $\{a,\ldots, a\}$ with $b$ columns.}  $\{\ell'_1,\ell'_2,\dots,\ell'_n\}$, see the right hand side of figure~\ref{fig:Stacks}.
As we started with  $N$ 5-branes, the $\ell_j'$s must obey the condition $\sum_{j=1}^n\ell_j'=N$. The integers $k_a$ are defined recursively
\beq
k_a=\{\#\,\ell_j': \ell_j'=\ell'_{k_1+\cdots k_{a-1}+1}\}\, ,
\eeq
and are equal to the number of columns of equal height. Since the diagonal $\text{U}(1)$ of the whole set of the $N$  5-branes is not realized on the low energy theory \cite{DeWolfe:1999hj},  the flavor symmetry of the corresponding puncture in the Gaiotto curve is $\text{S}(\text{U}(k_1)\times \cdots\times \text{U}(k_r))$  \cite{Gaiotto:2009we}. 

The Coulomb branch of the $T_N$ theories, corresponding to normalizable deformations of the web which do not change its shape at infinity, has dimension equal to the number of faces in the $T_N$ web diagram, see the left part of figure \ref{fig:TNHiggsing}, and  has dimension
 $\frac{(N-1)(N-2)}{2}$, as it should \cite{Gaiotto:2009gz}.
Moreover,  the dimension of
 the Higgs branch of the $T_N$ theories, known to be $\frac{3N^2-N-2}{2}$ \cite{Gaiotto:2009gz}, was obtained by terminating all the external semi-infinite 5-branes on 7-branes and  counting the independent degrees of freedom for moving them around on the web-plane \cite{Benini:2009gi}.
 Finally,  the global symmetry $\SU(N)^3$ of the $T_N$ theories is realized on the 7-branes.
\begin{figure}[t]
 \centering
  \includegraphics[height=3cm]{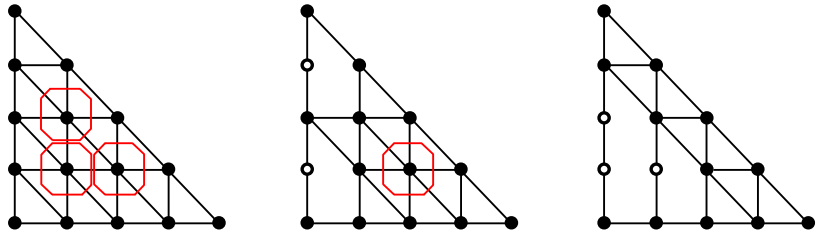}
  \caption{\it In this figure we present the dot diagrams of $T_4$ with three different Higgsings. On the left we have the un-Higgsed dot diagram with three full punctures, $\SU(4)^3$ global symmetry and three Coulomb moduli. In the middle, the four D5 branes end on two D7 branes with two D5 branes on each, which corresponds to the Young diagram $\{2,2\}$. This theory has apparent global symmetry $\SU(4)^2\times \SU(2)$ and one closed polygon corresponding to one leftover Coulomb modulus.
  Finally, on the right we have the fully-Higgsed theory with three D5 branes on the first D7 brane and one D5 brane on the second D7. This theory has no  Coulomb moduli left. 
  }
  \label{fig:T4dot}
\end{figure}

Higgsed $T_N$ theories can also be understood in this way \cite{Benini:2009gi}.
Beginning with the $T_N$ 5-brane webs which correspond to the sphere with three full punctures (labeled by the Young diagrams $\{1^N\}$) and grouping
the $N$ parallel 5 branes of the punctures
 into smaller bunches (labeled by the Young diagrams 
 $\{\ell'_1,\ell'_2,\dots,\ell'_n\}$),  5-brane configurations which realize 5D  theories with $E_{6,7,8}$ flavor symmetry were obtained.
These theories have Coulomb and Higgs branches of smaller dimension than the original $T_N$ which can be counted using a generalization of the s-rule \cite{Mikhailov:1998bx,DeWolfe:1998bi,Bergman:1998ej} from the so-called dot diagrams\footnote{The dot diagrams are the dual graphs of the web diagrams with the additional information about the 7-branes encoded in white and black dots.}, 
see also \cite{Hayashi:2013qwa,Hayashi:2014wfa,Hayashi:2014hfa}.
For us, the important result from  \cite{Benini:2009gi}   is that
the dimension of the Higgs moduli space of a puncture corresponding to the Young diagram depicted in figure \ref{fig:Stacks} is 
\beq
\mbox{dim}_{\mathbb{H}} \mathcal{M}^p_H = \sum^n_{j=1} \left( j-1 \right) \ell_j  \, ,
\eeq
and that the Coulomb branch is the number of closed dual polygons in the dot diagram.

\subsection{The Fateev-Litvinov degeneration from Higgsing}

We need to decide which puncture (Young diagram $\{\ell'_1,\ell'_2,\dots,\ell'_n\}$) corresponds to the Fateev-Litvinov semi-degenerate primary operator.
This puncture should have only $\text{U}(1)$ symmetry (for $N>2$).
Thus, it can be obtained by grouping the $N\,$5-branes in two  bunches of unequal number of 5-branes,  $N-1$ and $1$ respectively, forming the L-shaped Young diagram $\{N-1,1\}$ shown in figure~\ref{fig:SemiDeg}.  For $N=2$, the puncture has  an $\SU(2)$ flavor symmetry, while for $N\geq 3$ the flavor symmetry gets reduced to $\text{U}(1)$, as required for the semi-degenerate field. This Young diagram $\{N-1,1\}$ corresponds to the simple punctures discussed before. 
The Higgs moduli space of this configuration has
$\mbox{dim}_{\mathbb{H}} \mathcal{M}^{\mbox{\footnotesize{semi-deg}}}_H =1$ 
which is  consistent with the fact that we have only one parameter $\varkappa$ in the CFT side. 
Finally, the dot diagrams tell us that the dimension of the Coulomb branch in this case is zero, which, as we will see later, is consistent with what one gets  by just substituting \eqref{eq:alphadeg} in \eqref{eq:main5Dequality}.

\begin{figure}[t]
   \centering
   \includegraphics[height=3.4cm]{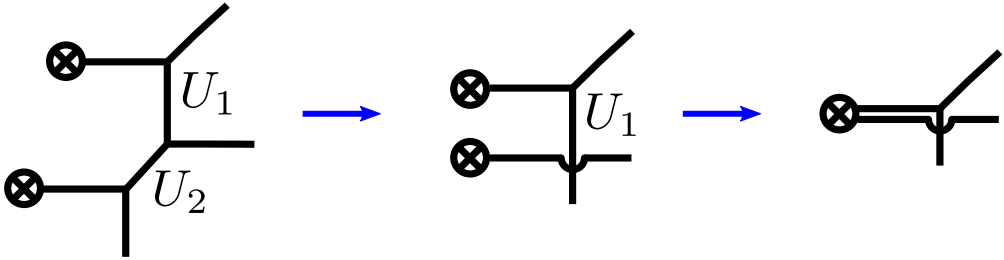}
   \caption{This figure shows the way two 5-branes are brought on the same 7-brane \cite{Hayashi:2013qwa}.}
   \label{fig:Higgsing}
\end{figure}

Now, let us discuss what happens with the K\"ahler moduli  that parametrize the $T_N$ partition functions as we bring together $N-1$ parallel horizontal external D5 branes on a single D7 brane. 
These we will then translate in the language of mass parameters $m_i, n_i, l_i$ ($i=1,\dots,N$) and Coulomb moduli $a_r$ ($r=1,\dots,\nicefrac{(N-1)(N-2)}{2}$)  using the dictionary of appendix \ref{app:notation} and in particular equation \eqref{eq:PQR} and, finally, to the Toda weights $\balpha_{1,2,3}$ using \eqref{eq:identificationparameters}.
We follow closely the discussion  in \cite{Hayashi:2014wfa}.  
For simplicity, we begin with two parallel D5 branes that originally end on different D7 branes. This process is depicted in figure  \ref{fig:Higgsing}. First we need to shrink $u_2$ of $U_2=e^{-\beta u_2}$ to zero while still having two 7-branes.  In the process of sending the $u_1$ of $U_1=e^{-\beta u_1}$ to zero, one of the two D7 branes will meet a D5 brane and the two parallel D5 branes will fractionate on the D7 branes.
After moving the cut piece to infinity it effectively decouple from the
rest of the web.

For the unrefined topological strings, {\it i.e.} for $\epsilon_2=-\epsilon_1$, shrinking the length of a 5-brane that is parametrized\footnote{The parameter $u$ in the exponent is the length of the 5-brane segment.} by $U=e^{-\beta u}$
corresponds to setting $U = 1$. 
This is not true any more in the case of the refined topological string where zero size will correspond either to  $U =\sqrt{\nicefrac{\ft}{\fq}}$ or $U =\sqrt{\nicefrac{\fq}{\ft}}$  \cite{Dimofte:2010tz,Taki:2010bj,Aganagic:2011sg,Aganagic:2012hs}. It turns out that both choices are equivalent as is extensively discussed in  \cite{Hayashi:2014wfa}. 
In this paper we wish to consider only the parameter space that corresponds to Toda CFT with $Q=\epsilon_1+\epsilon_2>0$, {\it i.e.} $\nicefrac{\ft}{\fq}>1$, and thus we have to pick $U =\sqrt{\nicefrac{\ft}{\fq}}$.

For the $T_3$ case the situation is exactly the same as the simple example depicted in figure \ref{fig:Higgsing}. The following two K\"ahler parameters 
\beq
\label{zeroforceT3}
Q_{m;1}^{(1)}=\bA^{-1}\tilde{M}_1\tilde{N}_1  \quad \mbox{and}  \quad Q_{l;1}^{(1)}=\bA\tilde{M}_2^{-1}\tilde{N}_1^{-1}
\eeq
are the ones we have to shrink, where $\bA\equiv \tilde{A}_1^{(1)}$ is the Coulomb modulus of $T_3$. See appendix \ref{app:notation}  for notations and figure \ref{fig:T3Higgsing} for the web diagram of $T_3$. Thus, we have to set 
\beq
\label{eq:shrinkingQmQlforT3}
Q_{m;1}^{(1)} = Q_{l;1}^{(1)}  =\sqrt{\frac{\ft}{\fq}} \, .
\, 
\eeq
In general for $T_N$ as depicted in figure \ref{fig:TNparam} we must tune
\beq
 \label{Higgsed_Q's}
Q_{m;i}^{(j)}=Q_{l;i}^{(j)}=\sqrt{\frac{\ft}{\fq}} \qquad \mbox{with}  \qquad i=1,\ldots, N-2, \quad j=1,\ldots, N-1-i\, .
\eeq
Going back to the Toda side, we wish to semi-degenerate the weight $\balpha_1$, {\it i.e.}  set it to
\beq
\label{eq:alphadeg}
\balpha_1=N\varkappa\omega_{N-1}
\qquad  \Longleftrightarrow \qquad
  m_i \, = \,  \left\{\begin{array}{ll}\varkappa-\frac{N+1-2i}{2}Q\,  &\,  i=1,\ldots, N-1\,, \\-(N-1)\varkappa+\frac{N-1}{2}Q\,  & \,  i=N\,, \end{array}\right.
\eeq
where the implications from \eqref{eq:identificationparameters} of the semi-degeneration on the mass  parameters are written on the right.
For the  $T_3$ case that implies for the exponentiated mass parameters that
 \beq
 \tilde{M}_1=\frac{\ft}{\fq} \tilde{K}  =e^{-\beta (\varkappa - Q)} \qquad \mbox{and}  \qquad  \tilde{M}_2= \tilde{K}
 \eeq
which is consistent with
 \eqref{zeroforceT3} and \eqref{eq:shrinkingQmQlforT3}
when  the Coulomb moduli is tuned to the value
\beq
\tA = \sqrt{ \frac{\ft}{\fq}   } \tilde{K} \tilde{N}_1 \, .
\eeq
This is compatible with the statement that after Higgsing, the $T_3$ the dimension of the Coulomb branch is zero, and also with the fact that we will discuss in next section,  the contour integral gets pinched once one substitutes \eqref{eq:alphadeg} in \eqref{eq:main5Dequality}.
In the general $T_N$ case, Higgsing forces the Coulomb parameters to become
\beq
 \label{Higgsed_A's}
\tilde{A}_i^{(j)}=\left(\frac{\ft}{\fq}\right)^{\frac{i(N-i-j)}{2}}\tilde{K}^i\prod_
{k=1}^{j}\tilde{N}_k\, , 
\eeq
where $i,j=1,\ldots, N-2 $, $i+j\leq N-1$ and $\tilde{K}=e^{-\beta\varkappa}$. This implies that the K\"ahler parameters obey \eqref{Higgsed_Q's}.

At the level of partition functions, the Fateev-Litvinov formula for the special 3-point functions can be identified with the partition function of $N^2$ free hypermultiplets, after removal of the decoupled degrees of freedom \eqref{NonFullSpin}. We know from \cite{Gomis:2011pf,Bao:2013pwa}, that the partition function of a single free hypermultiplet is given by 
\begin{align}
\label{OneFreeHyper}
\calZ_{\text{free hyper}}^{S^4}
& =\frac{1}{\Upsilon(m-\frac{\epsilon_+}{2})}\,,
\nonumber\\
\calZ_{\text{free hyper}}^{S^4\times S^1} 
&=\frac{1}{|\calM(e^{-\beta m} \sqrt{\frac{\ft}{\fq}} )|^2}=\frac{(1-q)^{-\frac{m^2}{\epsilon_1\epsilon_2}}}{\left|\calM\left(\sqrt{\frac{\ft}{\fq}};\ft,\fq\right)\right|^{2}}\frac{1}{\Upsilon_q(m-\frac{\epsilon_+}{2})}\,.
\end{align}
Thus, the 5D superconformal index of $N^2$ free hypermultiplets is  the  product  of $N^2$  such partition functions
 \beq
\label{NFreeHyper}
\calZ_{N^2 \text{ free hypers}}^{S^4\times S^1}  =\frac{1}{\prod_{i,j=1}^{N}\left|\calM\left(\sqrt{\frac{\ft}{\fq}}e^{-\beta m_{ij}}\right)\right|^2}\,.
\eeq  
Up to factors that for now we drop and using \eqref{NonFullSpin}, we can identify
 \begin{multline}
\frac{C_q(N \varkappa \omega_{N-1},\balpha_2,\balpha_3) }{\left|\calZ_{N}^{\text{dec}}\right|^2}
\sim
\frac{1}{\prod_{i,j=1}^N\Upsilon_q(\varkappa+\form{\balpha_2-\fQ}{h_i}+(\balpha_3-\fQ, h_j))}  \sim 
 \calZ_{N^2 \text{ free hypers}}^{S^4\times S^1}\, .
\end{multline}
From this knowledge, one could go ahead and guess some of the complicated summation formulas like \eqref{eq:summationidentityT3}, as was done by
\cite{Hayashi:2014wfa} for the $T_3$ case.

\subsection{The domain of the parameters restricts the contour}
\label{subsec:domainofparameters}

An important step we will have to take is to perform the contour integral in \eqref{eq:3pointfunctionsastopologicalstrings}. For that we need to carefully discuss the domain in which our parameters take values. On the Toda side, this type of conditions is obtained by considering the physicality of the \textbf{W}$_N$ Toda weights $\balpha$ is in order. Denoting by $\Delta(\balpha)$ the conformal dimension of the primary field $V_{\balpha}$, the formula for the 2-point functions 
\beq
\label{eq:twopointfunctions}
\left\langle V_{\balpha'}(z',\bar{z}')V_{\balpha}(z,\bar{z})\right\rangle=\frac{(2\pi)^{N-1}\delta(\balpha+\balpha'-2\fQ)+\text{Weyl-reflections}}{|z-z'|^{4\Delta(\balpha)}}\,,
\eeq
tells us that requiring that $V_{\balpha'}$ be the conjugate field to $V_{\balpha}$ leads to the following reality condition\footnote{See section 4 and 11 of \cite{Teschner:2001rv} for a detailed discussion of the physicality condition in the Liouville case. }
\beq
\label{eq:Todaphysicalitycondition}
\Re(\balpha)=\fQ\qquad \Longleftrightarrow\qquad  m_i, n_i, l_i\in i\mathbb{R}\,.
\eeq
The physicality condition for the Toda weights \eqref{eq:Todaphysicalitycondition} implies through the dictionary \eqref{eq:identificationparameters} that the mass parameters are purely imaginary. 
On the $(p,q)$ 5-brane web diagram side,
 distances are measured by the real part of the mass parameters, see equations (2.7-2.12) of \cite{Bao:2011rc} for a review of our conventions. 
 When the 5-branes are on top of each other, {\it i.e.} when their distance is 
 zero\footnote{In the refined topological vertex, the Seiberg-Witten curve is replaced by its quantum version in which zero distance is understood as integer multiples of $\epsilon_+$.}, $T_N$ has $\SU(N)^3$ symmetry \cite{Benini:2009gi}
  and we  can have physical Toda theory states. Since $\fQ=Q\sum_{i=1}^{N-1}\omega_i$ and since semi-degeneration requires that $\balpha=N\varkappa \omega_{N-1}$, we see that semi-degeneration/Higgsing is incompatible with the physicality condition \eqref{eq:Todaphysicalitycondition}. This agrees with a CFT intuition \cite{Teschner:2001rv}.

\begin{figure}[ht]
 \centering
  \includegraphics[height=2.6cm]{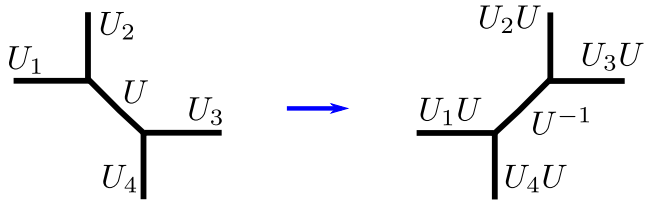}
\caption{The figure illustrates the change of the K\"aher parameters upon flopping. }
  \label{fig:flopping}
\end{figure}

We wish to conclude this section by stressing that the formulas we are dealing with have different domains with different convergent expansions depending on the values of the masses, just like in \eqref{eq:defcalMeverywhere}. In the topological string language they correspond to different geometries that are related to each other by flopping. For each K\"ahler parameter $U$, we distinguish between the region $|U|>1$ and the one with $|U|<1$; to each we associate a different $(p,q)$ 5-brane web diagram. Going from one region to the other involves ``flopping'' which transforms the K\"ahler parameters as depicted in figure~\ref{fig:flopping}. See  \cite{Mitev:2014jza} for a recent discussion of the topic.
In the next section, we explain how the contour in \eqref{eq:main5Dequality} is to be chosen and we argue that the contour is dictated by the choice of the flopping frame.

\section{The semi-degenerate \texorpdfstring{$\textbf{W}_3$}{W3} 3-point functions}
\label{sec:pinching}

In this section we explicitly derive the Fateev-Litvinov result for the semi-degenerate 3-point functions of the sl$(3)$ Toda theory from our general formula.  To succeed in this calculation, we need to do two things: to evaluate the contour integral in \eqref{eq:finalCqformula} and to perform the sum in \eqref{eq:ZTNsum}.
For general values of the parameters, infinitely many poles contribute to the contour integral, but luckily in the semi-degeneration limit only two of them do for the sl$(3)$ case. This is due to a phenomenon known as ``pinching'', which we illustrate in the beginning of the section with a very simple example. Then, we show that in the sl$(3)$ case, there are two possible poles where the contour can be pinched,  each of them corresponding to a different flopping frame of the $T_3$ geometry. From this observation, we infer three different possible choices for the contour in \eqref{eq:finalCqformula}. We compute the integral for each of them and find the same result. Finally, we show that for the particular residues that contribute it is possible to compute the sum in the ``instanton'' factor.

\medskip

\begin{figure}[ht]
 \centering
  \includegraphics[height=3cm]{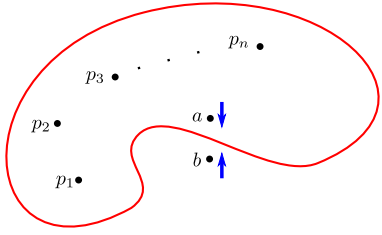}
  \caption{\it The figure shows an example of contour pinching. As the poles at $a$ and at $b$ collide, the contour integral diverges, which is why we regulate it by multiplying with $a-b$. In the limit $a\rightarrow b$, the integral is given by a single residue.}
  \label{fig:pinchingexample}
\end{figure}

Let us first make a simple example to illustrate pinching. 
Let $g$ be a meromorphic function in a domain $D\subset\mathbb{C}$ that has only simple poles at the points $a$, $b$ and $p_i$, meaning that it can be written as
\beq
g(z)=\frac{f(z)}{(z-a)(z-b)\prod_i (z-p_i)},
\eeq
where $f$ is a holomorphic  function in $D$. Let $\mathcal{C}$ be a closed contour in $D$ that encircles $a$ as well as the $p_i$ but not $b$. We write $a=p+\delta$ and $b=p-\delta$ and take the limit $\delta\rightarrow 0$, thus letting the two points $a$ and $b$ collide on the contour $\mathcal{C}$ on both sides, as depicted in figure~\ref{fig:pinchingexample}.
If we now compute the contour integral of $g$ around $\mathcal{C}$ and multiply it by $a-b$, we obtain
\beqa
\label{eq:trivialcontourintegral}
(a-b)\ointctrclockwise_{\mathcal{C}}\frac{dz}{2\pi i }g(z)&=&\frac{f(a)}{\prod_{i}(a-p_i)}+\sum_i\frac{(a-b)f(p_i)}{(p_i-a)(p_i-b)\prod_{j\neq i}(p_i-p_j)}\nonumber\\
&\stackrel{\delta\rightarrow0}\longrightarrow & \frac{f(p)}{\prod_{i}(p-p_i)}=\lim_{a\rightarrow b}\left[(a-b)\text{Res}(g(z),a)\right]\,.
\eeqa
Thus, in the limit $a\rightarrow b$, the contour gets pinched at the point $a=b=p$ and the integral is given by a single residue. This is essentially the contour integral version of the identity $\lim_{\varepsilon\rightarrow 0}\frac{\varepsilon}{(x+i \varepsilon)(x-i \varepsilon)}=\pi \delta(x)$. This example can also be easily generalized to the case in which $g$ has not only simple poles, but we will not need it.

We now want to explain how this simple example applies to our integral formulas for the correlation functions of sl$(3)$.
\begin{figure}[h]
 \centering
  \includegraphics[height=3.8cm]{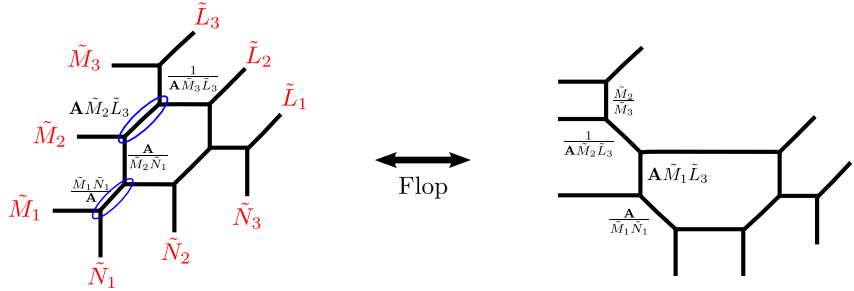}
\caption{This figure shows the two flopping frames for $T_3$. One can obtain the right geometry from the left one by applying two flopping moves, see figure~\ref{fig:flopping}, to the encircled segments.}
  \label{fig:T3Higgsing}
\end{figure}
In the sl$(3)$ case, our contour integral formula \eqref{eq:finalCqformula} for the structure constants reads
\beq
\label{eq:contourforT3}
C_q(\balpha_1,\balpha_2,\balpha_3)=\text{const}\times \ointctrclockwise\frac{d\bA}{2\pi i \bA}\left|M(\ft,\fq)\right|^{2}|\calZ_{3}^{\text{top}}|^2\, ,
\eeq
where $A_1^{(1)}\equiv \bA=e^{-\beta \ba}$ while the integrand is
\begin{align}
\label{eq:horribleT3amplitude}
&|\calZ_{3}^{\text{top}}|^2=|\calZ_{3}^{\text{pert}}|^2|\calZ_{3}^{\text{inst}}|^2=
\left|\frac{ \prod_{1\leq i<j\leq3}\calM\left(\frac{\tilde{M}_i}{\tilde{M}_j}\right) }{
 \prod_{k=1}^3\biggl[\calM\left(\sqrt{\frac{\ft}{\fq}}\bA\tilde{M}_k\tilde{L}_3\right)  \calM\left(\sqrt{\frac{\ft}{\fq}} \frac{\bA}{\tilde{M}_k\tilde{N}_1}\right)\biggl]}
 \frac{\calM\left(\frac{\bA^2\tilde{L}_3}{\tilde{N}_1}\right) \calM\left(\frac{\tilde{N}_1}{\bA^2\tilde{L}_3}\right)}{\calM\left(\sqrt{\frac{\ft}{\fq}}\frac{\bA\tilde{N}_2}{\tilde{L}_1}\right)\calM\left(\sqrt{\frac{\ft}{\fq}}\frac{\bA\tilde{N}_3}{\tilde{L}_2}\right) }\right|^2\nonumber\\
&
\times\, \left|\sum_{\boldsymbol{\nu}}\, \left(\frac{\tilde{N}_1\tilde{L}_2}{\tilde{N}_2\tilde{L}_3}\right)^{\frac{|\nu_1^{(1)}|+|\nu_2^{(1)}|}{2}} \left(\frac{\tilde{N}_2\tilde{L}_1}{\tilde{N}_3\tilde{L}_2}\right)^{\frac{|\nu_1^{(2)}|}{2}} 
\frac{\prod_{k=1}^3\left[\tN_{\nu_1^{(1)}\emptyset}\left(\ba-m_k-n_1-\nicefrac{Q}{2}\right)\tN_{\emptyset\nu_2^{(1)}}\left(\ba+m_k+l_3-\nicefrac{Q}{2}\right)\right]}{\tN_{\nu_1^{(1)}\nu_1^{(1)}}\left(0\right)\tN_{\nu_2^{(1)}\nu_2^{(1)}}\left(0\right)\tN_{\nu_1^{(2)}\nu_1^{(2)}}\left(0\right)}\right.\nonumber\\
&
\times\,\left. 
\frac{\tN_{\nu_1^{(1)}\nu_1^{(2)}}\left(\ba+n_2-l_1-\nicefrac{Q}{2}\right)\tN_{\nu_1^{(2)}\nu_2^{(1)}}\left(\ba+n_3-l_2-\nicefrac{Q}{2}\right)  }{\tN_{\nu_1^{(1)}\nu_2^{(1)}}\left(2\ba-n_1+l_3\right)\tN_{\nu_2^{(1)}\nu_1^{(1)}}\left(-2\ba+n_1-{l_3}\right)}\right|^2,
\end{align}
with the sum going over all partitions $\boldsymbol{\nu}=\{\nu_1^{(1)},\nu_2^{(1)},\nu_1^{(2)}\}$.
Since we wish to evaluate the contour integral \eqref{eq:contourforT3} in the semi-degenerate limit $\balpha_1=3\varkappa \omega_2$,  we introduce a regulator $\delta$ and  parametrize  the three masses labeling the positions of the branes on the left as
\beq
\label{eq:paramdegeneration}
m_1=\varkappa+\delta-Q\,,\qquad m_2=\varkappa-\delta\,,\qquad m_3=-2\varkappa+Q\,,
\eeq
which implies that the exponentiated masses $\tilde{M}_i=e^{-\beta m_i}$ are
\beq
\tilde{M}_1=\frac{\ft}{\fq}\tilde{K} e^{-\beta \delta}\,,\qquad \tilde{M}_2=\tilde{K} e^{\beta \delta}\,,\qquad \tilde{M}_3=\frac{\fq}{\ft}\tilde{K}^{-2}\,,
\eeq
with $\tilde{K}=e^{-\beta \varkappa}$.  The semi-degenerate limit then corresponds to $\delta\rightarrow 0$. For these values of the masses, the numerator of $|\calZ_{3}^{\text{top}}|^2$ in \eqref{eq:horribleT3amplitude} goes to zero, just like the term $a-b$ in the simple example \eqref{eq:trivialcontourintegral} above, since  
\beq
\label{eq:semidegenerationnumerator}
| \calM(\tilde{M}_1\tilde{M}_2^{-1})|^2=(1-e^{-2\beta\delta})\times \text{reg.}\approx \delta\times \text{reg.}\, ,
\eeq 
where ``reg'' are terms that don't vanish for $\delta\rightarrow 0$. 

The next step is to analyze the poles in the integrand of \eqref{eq:horribleT3amplitude} and determine which ones will contribute in the semi-degenerate limit. We make the assumption\footnote{This can be supported by a following simple observation. The integral in our formula \eqref{eq:3pointfunctionsastopologicalstrings} for the Toda three-point function should be regarded as a complicated deformation of a conventional Mellin-Barnes contour integral of ratio of gamma functions multiplying a hypergeometric function. The ``perturbative'' part of the integrand corresponds to the deformed gamma functions, whereas the ``instanton'' part is the analogue of the hypergeometric function. As the usual hypergeometric function is an entire function of its parameters, it cannot give residue contributions to the value of the Mellin-Barnes integral. It is natural to expect the same property for its deformation.} that only poles from the ``perturbative'' part, {\it i.e.} the first line of \eqref{eq:horribleT3amplitude}, are relevant for this computation, which will be justified by the final result. 

Due to the vanishing of the numerator \eqref{eq:semidegenerationnumerator}, we need to have pinching in order to get a non-zero answer. As we learned from the simple example at the beginning of the section, we need to find poles that lie on different sides of the contour and that collide when the regulator is removed. The poles in the integrand  come from the zeroes of the functions $|\calM(U)|^2$ in the first line of \eqref{eq:horribleT3amplitude}. Since, in order to obtain the Toda theory from topological strings we wish to have $b>0$, so that $|\fq|<1$ and $|\ft|>1$,  we get from \eqref{eq:defcalMeverywhere} the expression
\beq
\label{eq:T3zeronumerator}
|\calM(U;\ft,\fq)|^2=\calM(U;\ft,\fq)\calM(U^{-1};\ft^{-1},\fq^{-1})=\prod_{i,j=1}^{\infty}(1-U\ft^{-i}\fq^j) (1-U^{-1}\ft^{1-i}\fq^{j-1})\,.
\eeq
Thus, the zeroes of $|\calM(U)|^2$ are to be found on the points 
\beq
\label{eq:zeroesofcalMsquared}
U=\ft^{-m}\fq^n\,,\qquad U=\ft^{m+1}\fq^{-n-1}\,,
\eeq
for $m,n\in \mathbb{N}_0=\{0,1,2,\ldots\}$. We see that there are two classes of poles of $|\calZ^{\text{top}}|^2$, namely those  that condense around zero in the $\bA$ complex plane and those that condense around infinity. 

\begin{figure}[ht]
 \centering
  \includegraphics[height=5.1cm]{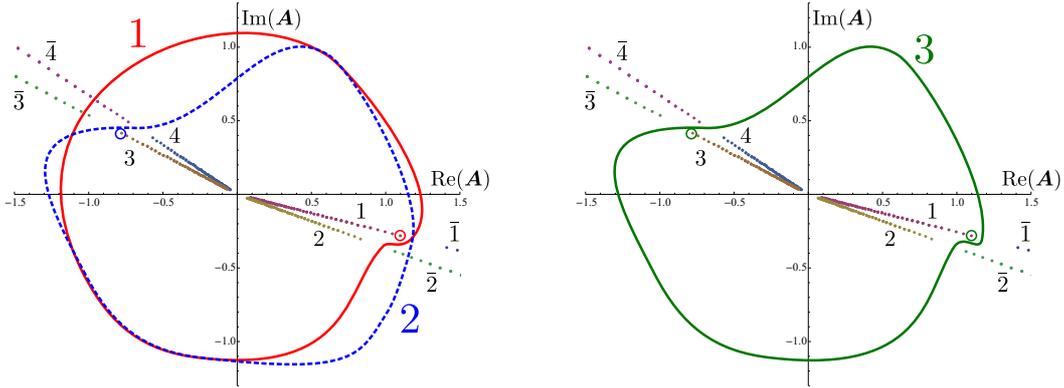}
  \caption{\it  As the variable $\delta$ is sent to zero, the contour gets pinched between two zeroes and the contributions are given by finite number of residues, indicated by circles. The set of poles are labeled according to \eqref{eq:polesofintegrandandpinching}. The first contour picks up a single residue from the line of poles $1$, the second one a single residue from the line of poles $3$, while the third one, shown on the right, picks both of these residues.}
  \label{fig:Contour}
\end{figure}

When we then take the limit $\delta\rightarrow 0$, some poles from the exterior of the contour integral with coincide with some from the interior, leading to a divergence that will cancel the zero of \eqref{eq:T3zeronumerator}, just like in the simple example of equation \eqref{eq:trivialcontourintegral}. We easily see that the relevant terms in the denominator of the first line of \eqref{eq:horribleT3amplitude}
are
 \beq
 \label{eq:calMpinching}
 \left|\calM\left(\sqrt{\frac{\ft}{\fq}} \bA\tilde{M}_1^{-1}\tilde{N}_1^{-1}\right)
\calM\left(\sqrt{\frac{\ft}{\fq}} \bA\tilde{M}_2^{-1}\tilde{N}_1^{-1}\right) \calM\left(\sqrt{\frac{\ft}{\fq}}\bA\tilde{M}_1\tilde{L}_3\right)\calM\left(\sqrt{\frac{\ft}{\fq}}\bA\tilde{M}_2 \tilde{L}_3\right)\right|^2\, .
\eeq
The other zeroes in the denominator will not pinch the integral once the regulator $\delta$ is set to zero and can be ignored, just like the $p_i$ terms in \eqref{eq:trivialcontourintegral}. Numbering the functions $\calM$ as $1$ to $4$ in \eqref{eq:calMpinching} from left to right, using \eqref{eq:zeroesofcalMsquared} and the parametrization \eqref{eq:paramdegeneration}, we know that we have first order poles in the integrand if
\begin{align}
\label{eq:polesofintegrandandpinching}
 &(1)& &\bA=\tilde{K}\tilde{N}_1e^{-\beta \delta}\ft^{-m+\frac{1}{2}}\fq^{n-\frac{1}{2}}\,,& &(\bar{1})& &\bA=\tilde{K}\tilde{N}_1e^{-\beta \delta}\ft^{m+\frac{3}{2}}\fq^{-n-\frac{3}{2}}\,,&\nonumber\\
&(2)& &\bA=\tilde{K}\tilde{N}_1e^{\beta \delta}\ft^{-m-\frac{1}{2}}\fq^{n+\frac{1}{2}}\,,& &(\bar{2}) &&\bA=\tilde{K}\tilde{N}_1e^{\beta \delta}\ft^{m+\frac{1}{2}}\fq^{-n-\frac{1}{2}}\,, & \nonumber\\ &(3)&
&\bA=\tilde{K}^{-1}\tilde{L}_3^{-1}e^{\beta \delta}\ft^{-m-\frac{3}{2}}\fq^{n+\frac{3}{2}}\,,& &(\bar{3})& &\bA=\tilde{K}^{-1}\tilde{L}_3^{-1}e^{\beta \delta}\ft^{m-\frac{1}{2}}\fq^{-n+\frac{1}{2}}\,, & \nonumber\\
&(4)& &\bA=\tilde{K}^{-1}\tilde{L}_3^{-1}e^{-\beta \delta}\ft^{-m-\frac{1}{2}}\fq^{n+\frac{1}{2}}\,,&  &(\bar{4})& &\bA=\tilde{K}^{-1}\tilde{L}_3^{-1}e^{-\beta \delta}\ft^{m+\frac{1}{2}}\fq^{-n-\frac{1}{2}}\,, &
\end{align}
for $m,n\in \mathbb{N}_0$. We have labeled with a $\bar{\cdot}$ those sets of poles that coalesce around $\bA=\infty$. We see in figure~\ref{fig:Contour} that there are two places where the towers of poles collide, namely where the first pole of the tower $1$ hits the first pole of the tower $\bar 2$ and where the first pole of the tower $3$ hits the first of the tower $\bar 4$.
Now the time has come for us to choose the form of the contour. Given the fact that we need to pinch the contour, we have three possible options, depicted in ~\ref{fig:Contour}. We will compute the integral for each of the three choices.

\begin{figure}[ht]
 \centering
  \includegraphics[height=2.5cm]{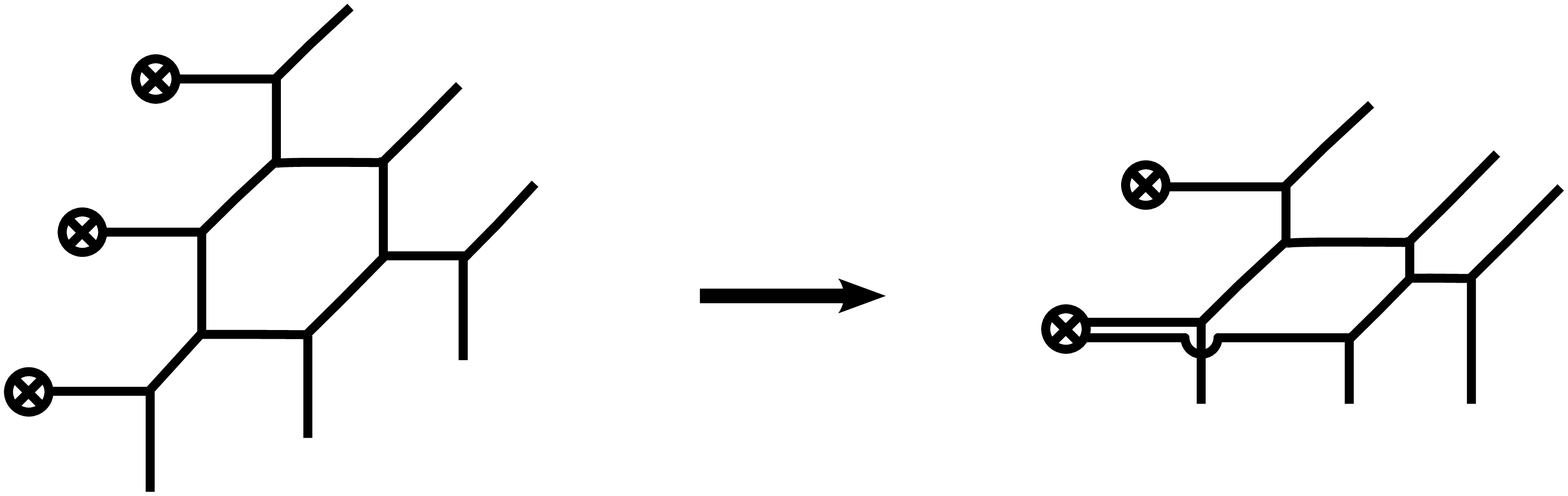}
  \caption{\it The figure shows the Higgsed geometry corresponding to the residue \eqref{eq:residuesforT3}. For this residue, the K\"ahler parameters take the values \eqref{eq:shrinkingQmQlforT3}.}
  \label{fig:T3HiggsingResidue1}
\end{figure}

We begin with the first contour, see figure~\ref{fig:Contour}, passing between the towers $1$ and $\bar 2$ but avoiding the pinching of the towers $3$ and $\bar 4$. 
We see that, due to set of poles $1$ colliding with the set of poles $\bar{2}$ for $m=n=0$, the integral gets pinched as $\delta\rightarrow 0$ and that the result is given by the residue at 
\beq
\label{eq:residuesforT3}
\bA=\sqrt{\frac{\ft}{\fq}} \tilde{K}\tilde{N}_1e^{-\beta \delta}\, .
\eeq
It is very important as a guiding principle to note that this first contour corresponds to Higgsing in the flopping frame of figure~\ref{fig:T3HiggsingResidue1}.
We furthermore see that for the choice of contour in figure~\ref{fig:Contour}, the fact that for $\delta\rightarrow 0$ we get an overlap between a pole from $3$ and a pole from $\bar{4}$ is of no consequence since they both lie of the same side of the contour.

Let us now compute the residue of $|\calZ^{\text{top}}_3|^2$ at \eqref{eq:residuesforT3} directly. We first need a couple of technical results. We can use the fact that for a function $f$ that has no pole at $B\ft^k\fq^l$, we have
\beqa
\label{eq:residueformula}
\text{Res}\left(\frac{f(\bA)}{\bA|\calM(\bA B^{-1})|^2},\bA=B\ft^k\fq^l\right)=\frac{\tg_{-k,l}}{\left|M(\ft,\fq)\right|^{2}}f(B\ft^k\fq^l)\, .
\eeqa
Here $\left|M(\ft,\fq)\right|^{2}$ is the norm squared of the refined MacMahon function defined in \eqref{eq:McMahonUp} and the function $\tg_{kl}$ is defined as
\beq
\label{eq:deftextsfg}
\tg_{kl}(\ft,\fq)\colonequals \lim_{U\rightarrow 1}\frac{\left|\calM(U)\right|^2}{\left|\calM(U\ft^{-k}\fq^l)\right|^2}=\prod_{i=1}^k\frac{(\ft^{-i}\fq^{l+1};\fq)_{\infty}}{(\ft^{i}\fq^{-l};\fq)_{\infty}}\prod_{j=1}^{l}\frac{(\ft^{-1}\fq^{j};\ft^{-1})_{\infty}}{(\fq^{-j};\ft^{-1})_{\infty}}\, ,
\eeq
where we have used the shift properties \eqref{eq:calMshift} of the $\calM$ functions and the last equality is only valid for  $k,l\in \mathbb{N}_0$. 
The above expression can be continued for negative $k$ and $l$ with $\tg_{kl}=-\tg_{-k-1,-l-1}$. In particular $\tg_{-n,0}=\tg_{0,-n}=0$ for $n\geq 1$. 
Thus, we can now finally write down the residue for the pole \eqref{eq:residuesforT3} using \eqref{eq:residueformula}
\begin{align}
\label{eq:calZ3topdegenerate1}
 \lim_{\delta\rightarrow 0}\ointctrclockwise\frac{d\bA}{2\pi i \bA}\left|M(\ft,\fq)\right|^{2}|\calZ_{3}^{\text{top}}|^2&=\left|M(\ft,\fq)\right|^{2}\text{Res}\left(|\calZ_{3}^{\text{top}}|^2,\bA=\sqrt{\frac{\ft}{\fq}} \tilde{K}\tilde{N}_1e^{-\beta \delta}\right)\\
&
=\frac{\left|\calM(\tilde{K}^{-3})\right|^2}{\left|\prod_{k=1}^3\calM\big(\frac{\tilde{N}_k\tilde{L}_{4-k}}{\tilde{K}}\big)\right|^2}\left|Z^{\text{inst}}_3\right|^2_{\big|\bA=\sqrt{\frac{\ft}{\fq}} \tilde{K}\tilde{N}_1}\, .\nonumber
\end{align}
One can observe that due to \eqref{eq:tNdelta}, the sum over $\nu_1^{(1)}$ in $\left|\calZ^{\text{inst}}_3\right|^2_{\big|\bA=\sqrt{\frac{\ft}{\fq}} \tilde{K}\tilde{N}_1}$ drops out and we obtain the following result for the ``instanton'' partition function 
\begin{multline}
\label{eq:Z3instdeg1}
\left(\calZ^{\text{inst}}_3\right)_{\big|\bA=\sqrt{\frac{\ft}{\fq}} \tilde{K}\tilde{N}_1}=\sum_{\nu_1,\nu_2} \left(\frac{ \tilde{N}_2\tilde{L}_1}{\tilde{N}_3\tilde{L}_2 }\right)^{\frac{|\nu_1|}{2}}
\left(\frac{ \tilde{N}_1\tilde{L}_2}{ \tilde{N}_2\tilde{L}_3}\right)^{\frac{|\nu_2|}{2} }\\\times \frac{ \tN_{\nu_1\emptyset}(n_3+l_1-\varkappa) \tN_{\nu_2\nu_1}(n_2+l_2-\varkappa ) \tN_{\emptyset\nu_2}(n_1+l_3-\varkappa) }{\tN_{\nu_1\nu_1}(0) \tN_{\nu_2\nu_2}(0) }\,,
\end{multline}
where we denoted $\nu_1^{(2)}\equiv \nu_1$, $\nu_2^{(1)}\equiv\nu_2$. 

\begin{figure}[ht]
 \centering
  \includegraphics[height=2.5cm]{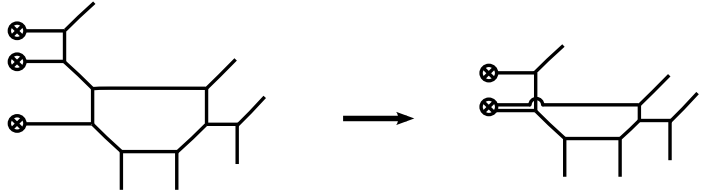}
  \caption{\it The figure shows the Higgsed geometry corresponding to the residue \eqref{eq:residuesforT3part2}.}
  \label{fig:T3HiggsingResidue2}
\end{figure}

Next, we also need to compute the the contour integral for the second contour, depicted in blue in figure~\ref{fig:Contour}. We find that the result is given by the residue of $\left|\calZ_3^{\text{top}}\right|^2$ at 
\beq
\label{eq:residuesforT3part2}
\bA=\sqrt{\frac{\fq}{\ft}} \tilde{K}^{-1}\tilde{L}_3^{-1}e^{-\beta \delta}\,,
\eeq
which, together with \eqref{eq:paramdegeneration} implies for $\delta\rightarrow 0$ the Higgsed geometry shown in figure~\ref{fig:T3HiggsingResidue2}.
Computing the residue, we find that the ``perturbative'' contribution, {\it i.e.} the prefactor of $\left|\calZ_3^{\text{inst}}\right|^2$ in \eqref{eq:Z3instdeg1}, is the same as before. Furthermore, we find after relabeling $\nu_2^{(1)}\leftrightarrow \nu_1^{(1)}$ and using \eqref{eq:propertiestN} that the ``instanton'' contribution in \eqref{eq:Z3instdeg1} is unchanged, {\it i.e.}
\beq
\left(\calZ^{\text{inst}}_3\right)_{\big|\bA=\sqrt{\frac{\ft}{\fq}} \tilde{K}\tilde{N}_1}=\left(\calZ^{\text{inst}}_3\right)_{\big|\bA=\sqrt{\frac{\fq}{\ft}} \tilde{K}^{-1}\tilde{L}_3^{-1}}\, .
\eeq
Finally, for the third contour, shown on the right hand side of figure~\ref{fig:Contour}, we simply find the sum of the results of contour one and two.

In order to complete the computation, we need to calculate the sum in \eqref{eq:Z3instdeg1} over the two remaining partitions. For this purpose, we shall use the following identity that we shall state in full generality in section~\ref{sec:generalN} and prove in appendix~\ref{appB}:
\begin{align}
\label{eq:summationidentityT3}
&\sum_{\nu_1,\nu_2} \left(V_1 \sqrt{U_1 U_2}\right)^{|\nu_1|} \left(V_2 \sqrt{U_2 U_3}\right)^{|\nu_2|}
\frac{\tN_{\nu_1\emptyset}\left(u_1-\nicefrac{Q}{2}\right)
\tN_{\nu_2\nu_1}\left(u_2-\nicefrac{Q}{2}\right)
\tN_{\emptyset \nu_2}\left(u_3-\nicefrac{Q}{2}\right) 
 }{\tN_{\nu_1\nu_1}\left(0\right)\tN_{\nu_2\nu_2}\left(0\right)}  \\
& = 
\frac{
\calM\big(U_1 V_1\big)\calM\big(\frac{\ft}{\fq} V_1 U_2 \big)\calM\big(U_2 V_2\big)\calM\big(\frac{\ft}{\fq} V_2 U_3 \big)
\calM\big(U_1 V_1 U_2 V_2\big)
\calM\big(\frac{\ft}{\fq}  V_1  U_2 V_2 U_3\big)}
{\calM\big(\sqrt{\frac{\ft}{\fq}} V_1\big)\calM\big(\sqrt{\frac{\ft}{\fq}} V_2\big)
\calM\big(\sqrt{\frac{\ft}{\fq}} U_1 V_1 U_2 \big)\calM\big(\sqrt{\frac{\ft}{\fq}} V_1 U_2 V_2  \big)\calM\big(\sqrt{\frac{\ft}{\fq}} U_2 V_2 U_3 \big)
\calM\big(\sqrt{\frac{\ft}{\fq}} U_1 V_1 U_2 V_2 U_3 \big)}, \nonumber
\end{align}
where $U_i:=e^{-\beta u_i}$.
Upon making the following substitutions in \eqref{eq:summationidentityT3}
\beq
U_k=\sqrt{\frac{\fq}{\ft}}\frac{\tilde{N}_{4-k} \tilde{L}_k}{\tilde K},\qquad  
V_1=\sqrt{\frac{\ft}{\fq}}\frac{\tilde K}{\tilde{N}_3 \tilde{L}_2}, \qquad V_2=\sqrt{\frac{\ft}{\fq}}\frac{\tilde K}{\tilde{N}_2 \tilde{L}_3}, 
\eeq
where $k=1,2,3$,  we arrive at
\beq
\left(\calZ^{\text{inst}}_3\right)_{\big|\bA=\sqrt{\frac{\ft}{\fq}} \tilde{K}\tilde{N}_1} = 
\frac{
\calM\big(\frac{\tilde{L}_1}{\tilde{L}_2}\big)\calM\big(\frac{\tilde{L}_2}{\tilde{L}_3}\big)
\calM\big(\frac{\tilde{L}_1}{\tilde{L}_3}\big)\calM\big(\frac{\ft}{\fq} \frac{\tilde{N}_1}{\tilde{N}_2}\big)
\calM\big(\frac{\ft}{\fq} \frac{\tilde{N}_2}{\tilde{N}_3}\big)\calM\big(\frac{\ft}{\fq} \frac{\tilde{N}_1}{\tilde{N}_3}\big)}
{\calM\big(\frac{\tilde{N}_1 \tilde{L}_1}{\tilde K}\big)\calM\big(\frac{\tilde{N}_1 \tilde{L}_2}{\tilde K}\big)
\calM\big(\frac{\tilde{N}_2 \tilde{L}_1}{\tilde K}\big)\calM\big( \frac{\ft}{\fq} \frac{\tilde K}{\tilde{N}_2 \tilde{L}_3}\big)
\calM\big(\frac{\ft}{\fq} \frac{\tilde K}{\tilde{N}_3 \tilde{L}_2}\big)\calM\big(\frac{\ft}{\fq} \frac{\tilde K}{\tilde{N}_3 \tilde{L}_3}\big)}.
\eeq
Inserting the above into \eqref{eq:calZ3topdegenerate1}, we arrive at
\beqa
\label{eq:calZ3topdegenerate2}
\lim_{\delta\rightarrow 0}\ointctrclockwise\frac{d\bA}{2\pi i \bA}\left|M(\ft,\fq)\right|^{2}|\calZ_{3}^{\text{top}}|^2&=&\,\frac{\left|\calM(\tilde{K}^{-3})\prod_{1\leq i<j\leq 3}\calM\left(\nicefrac{\tilde{N}_j}{\tilde{N}_i}\right)\calM\left(\nicefrac{\tilde{L}_i}{\tilde{L}_j}\right)\right|^2}{\left|\prod_{i,j=1}^3\calM(\tilde{N}_i\tilde{L}_j\tilde{K}^{-1})\right|^2}\nonumber\\
&=&\,\frac{(1-q)^{\varphi_3}}{\Lambda^2}\frac{\Up(3\varkappa)\prod_{1\leq i<j\leq 3}\Up(n_i-n_j)\Up(l_{4-i}-l_{4-j})}{\prod_{i,j=1}^3\Up(\varkappa-n_i-l_{4-j})},
\eeqa
where we have used \eqref{eq:defUp}, \eqref{eq:Lambda} and defined the exponent
\begin{align}
\varphi_3&=\left(\frac{Q}{2}-3\varkappa\right)^2+\sum_{1\leq i<j\leq 3}\left[\left(\frac{Q}{2}+n_j-n_i\right)^2+\left(\frac{Q}{2}+l_{4-j}-l_{4-i}\right)^2\right]-\sum_{i,j=1}^3\left(\frac{Q}{2}+n_i+l_{4-j}-\varkappa\right)^2\nonumber\\
&=2Q\left(3\varkappa+\sum_{i=1}^3 i(n_i+l_{4-i})\right)-\frac{Q^2}{2}=-2Q\form{2\fQ-\sum_{i=1}^3\balpha_i}{\rho}-\frac{Q^2}{2}\, ,
\end{align}
where in the last line we have used our sl$(3)$ conventions, see appendix~\ref{subapp:sln} and equation \eqref{eq:identificationparameters}. Now we employ \eqref{eq:MactoLambda} and rearrange the prefactors of \eqref{eq:calZ3topdegenerate2} to obtain the $q$-deformed \textbf{W}$_3$ Fateev-Litvinov structure constants  \eqref{eq:qdefFLTodacorr} in the form conjectured by \cite{Mitev:2014isa}:
\begin{align}
\label{eq:qdefFLTodacorr-two}
&C_q(3\varkappa\omega_2,\balpha_2,\balpha_3)=\nonumber\\&=\left(\beta\left|M(\ft,\fq)\right|^{2}\right)^{2}\left(\big(1-q^b\big)^{2b^{-1}}\big(1-q^{b^{-1}}\big)^{2b}\right)^{\form{2\fQ-\sum_{i=1}^3\balpha_i}{\rho}}\lim_{\delta\rightarrow 0}\ointctrclockwise\frac{d\bA}{2\pi i \bA}\left|M(\ft,\fq)\right|^{2}|\calZ_{3}^{\text{top}}|^2\\
&=\left(\frac{\big(1-q^b\big)^{2b^{-1}}\big(1-q^{b^{-1}}\big)^{2b}}{(1-q)^{2Q}}\right)^{\form{2\fQ-\sum_{i=1}^3\balpha_i}{\rho}}\frac{\Up'(0)^{2}\Up(3\varkappa)\prod_{e>0}\Up(\form{\fQ-\balpha_2}{e})\Up(\form{\fQ-\balpha_3}{e})}{\prod_{i,j=1}^{3}\Up(\varkappa+(\balpha_2-\fQ,\,h_i)+(\balpha_3-\fQ,\, h_j))}\, .\nonumber
\end{align}
Taking here the 4D limit $q \rightarrow 1$ and reintroducing the cosmological constant dependence according to \eqref{Cq2C} leads to the Fateev-Litvinov formula \eqref{eq:FLTodacorr} for $N=3$. To conclude, we see that any of the three contours that we presented leads to the desired formula, up to a factor of two for the third one that should be absorbed in the proportionality constant.

\section{The general \texorpdfstring{$\textbf{W}_N$}{WN} case}
\label{sec:generalN}

Having computed the structure constants for the \textbf{W}$_3$ case in the previous section, we now want to turn our attention to the general case. Starting with \textbf{W}$_4$, corresponding to the $T_4$ gauge theory, one has to deal with multiple integrals and their residues. We relegate the investigation of the subtleties associated to those to appendix~\ref{app:subappT_4}. 

We begin with the following conjecture concerning the choice of the contour. As we saw in the last section, there are multiple choices that we believe all lead to the same final result, up to a multiplicity factor. Hence, here  we make the simplest possible choice of the contour, corresponding to the flopping frame of figure~\ref{fig:TNHiggsing}, which pinches at just one pole.

We parametrize the masses as follows
\beq
\label{eq:regularizationTN}
\tilde{M}_i=\left(\frac{\ft}{\fq}\right)^{\frac{N+1-2i}{2}}\tilde{K}d_i \qquad \text{for} \,\,\, i=1,\dots, N-1,\qquad \tilde{M}_N=\left(\frac{\fq}{\ft}\right)^{\frac{N-1}{2}}\frac{1}{\tilde{K}^{N-1}},
\eeq
where the $d_i=e^{-\beta \delta_i}$ are regulators satisfying $\prod_{i=1}^{N-1}d_i=1$ and $\tilde{K}=e^{-\beta\varkappa}$. The numerator of $\left|\calZ^{\text{top}}\right|^2$ has a zero of order $\frac{(N-2)(N-1)}{2}$ in the limit $\delta_i\rightarrow 0$ since
\beq
\prod_{1\leq i<j\leq N}\Big|\calM\Big(\frac{\tilde{M}_i}{\tilde{M}_j}\Big)\Big|^2=\text{reg}\times\prod_{1\leq i<j\leq N-1}\Big|\calM\left(\Big(\frac{\ft}{\fq}\Big)^{j-i}\frac{d_i}{d_j}\right)\Big|^2\, ,
\eeq
and $\left|\calM\left(\big(\nicefrac{\ft}{\fq}\big)^n\right)\right|^2=0$ for $n\geq 0$. These zeroes can all be canceled by divergences coming from the  pinching of the $\frac{(N-2)(N-1)}{2}$ integrals if we choose the contour carefully, see for instance figure~\ref{fig:ContourT4A1} for an example in the $T_4$ case. Thus the final answer is obtained by taking the residues in the integration variables ${\tilde{A}_i^{(j)}}$ at
\beq
\tilde{A}_i^{(j)}=\left(\frac{\ft}{\fq}\right)^{\frac{i(N-i-j)}{2}}\tilde{K}^i\prod_{k=1}^{j}\tilde{N}_k\, .
\eeq
Computing the residues, we obtain the result:
\begin{align}
\label{eq:calZNtopdegeneratefinalintegral}
& \lim_{\delta_a\to 0}\ointctrclockwise\prod_{i=1}^{N-2}\prod_{j=1}^{N-1-i}\biggl[\frac{d\tilde{A}_i^{(j)}}{2\pi i \tilde{A}_i^{(j)}}|M(\ft,\fq)|^2\biggr]|\calZ_{N}^{\text{top}}|^2=\nonumber\\
&=\frac{\left|\calM(\tilde{K}^{-N})\right|^2}{\left|\prod_{k=1}^N\calM\big(\frac{\tilde{N}_k\tilde{L}_{N+1-k}}{\tilde{K}}\big)\right|^2}\,\times\Bigg|\sum_{\nu_1,\dots , \nu_{N-1}} \biggl[\prod_{i=1}^{N-1} \left(\frac{\tilde{N}_{N-i}\tilde{L}_i}{\tilde{N}_{N-i+1}\tilde{L}_{i+1}}\right)^{\frac{|\nu_i|}{2}}\biggr ] \\  
& \times \frac{\tN_{\nu_1\emptyset}\left(n_N+l_1-\varkappa\right) \biggl [\prod_{i=1}^{N-2}\tN_{\nu_{i+1}\nu_i}\left(n_{N-i}+l_{i+1}-\varkappa\right)\biggr ]
\tN_{\emptyset \nu_{N-1}}\left(n_1+l_{N}-\varkappa\right) }{\prod_{i=1}^{N-1}\tN_{\nu_i\nu_i}\left(0\right)}
\Bigg|^2\, .\nonumber
\end{align}
Here $\nu_i$ for $i=1,\dots, N-1$ denote the partitions corresponding to the $N-1$ brane junctions not affected by Higgsing at the given pole. For our choice of flopping frame, see figure~\ref{fig:TNHiggsing}, these partitions are readily identified as $\nu_i:=\nu_i^{(N-i)}$, $i=1,\dots , N-1$, see figure 4 of \cite{Mitev:2014isa} for the notation.

The remaining sums in \eqref{eq:calZNtopdegeneratefinalintegral} will be now performed by using the summation identity \eqref{T_N-identity} proven in appendix~\ref{subapp:summation}, which we reproduce here for convenience.\vspace{0.1cm}\\
\textbf{Theorem}
\begin{multline}
\label{T_N-identity-main}
\sum_{\nu_1,\dots , \nu_{N-1}}\left[ \prod_{i=1}^{N-1} \frac{\left( V_i \sqrt{U_i U_{i+1}}\right)^{|\nu_i|}}{\tN_{\nu_i\nu_i}\left(0\right)}\right] \tN_{\nu_1\emptyset}\left(u_1-\nicefrac{\epsilon_+}{2}\right) \left[\prod_{i=1}^{N-2}\tN_{\nu_{i+1}\nu_i}\left(u_{i+1}-\nicefrac{\epsilon_+}{2}\right)\right] \tN_{\emptyset\nu_{N-1}}\left(u_N-\nicefrac{\epsilon_+}{2}\right)=\\
=\prod_{i=1}^{N-1}\prod_{j=1}^{N-i}\frac{\calM\big( \prod_{s=j}^{i+j-1}U_sV_s \big)\calM\big(\frac{\ft}{\fq}\frac{U_{i+j}}{U_j}\cdot \prod_{s=j}^{i+j-1}U_sV_s \big)}{\calM\big( \sqrt{\frac{\ft}{\fq}} U_{i+j}  \prod_{s=j}^{i+j-1}U_sV_s \big)\calM\big(\sqrt{\frac{\ft}{\fq}} \frac{1}{U_{j}}  \prod_{s=j}^{i+j-1}U_sV_s \big)}.
\end{multline}
Setting the parameters here to be equal to
\begin{align}
U_i=\sqrt{\frac{\fq}{\ft}}\frac{\tilde N_{N-i+1}\tilde L_i}{\tilde K}\,,  \qquad 
V_j=\sqrt{\frac{\ft}{\fq}}\frac{\tilde K}{\tilde N_{N-j+1}\tilde L_{j+1}}\,, 
\end{align}
for $i=1,\cdots N$ and $j=1,\cdots N-1$, one straightforwardly obtains:
\begin{align}
\label{eq:summationformulafinalform}
&\sum_{\nu_1,\dots , \nu_{N-1}} \biggl[\prod_{i=1}^{N-1} \left(\frac{\tilde{N}_{N-i}\tilde{L}_i}{\tilde{N}_{N-i+1}\tilde{L}_{i+1}}\right)^{\frac{|\nu_i|}{2}}\biggr ] \nonumber\\  
& \times \frac{\tN_{\nu_1\emptyset}\left(n_N+l_1-\varkappa\right) \biggl [\prod_{i=1}^{N-2}\tN_{\nu_{i+1}\nu_i}\left(n_{N-i}+l_{i+1}-\varkappa\right)\biggr ]
\tN_{\emptyset \nu_{N-1}}\left(n_1+l_{N}-\varkappa\right) }{\prod_{i=1}^{N-1}\tN_{\nu_i\nu_i}\left(0\right)} \nonumber \\
& = \prod_{1 \leq i < j \leq N}\frac{\calM\big( \frac{\tilde L_i}{\tilde L_j}\big)\calM\big(\frac{\ft}{\fq}\frac{\tilde N_{N-j+1}}{\tilde N_{N-i+1}}\big)}{\calM\big( \frac{\tilde N_{N-j+1}\tilde L_i}{\tilde K}\big)\calM\big(\frac{\ft}{\fq}\frac{\tilde K}{\tilde N_{N-i+1}\tilde L_{j}}\big)}\, .
\end{align}

Substituting \eqref{eq:summationformulafinalform} in \eqref{eq:calZNtopdegeneratefinalintegral} and expressing everything in term of the $\Up$ functions through formula \eqref{eq:defUp} one obtains
\begin{align}
& \lim_{\delta_a\to 0}\ointctrclockwise\prod_{i=1}^{N-2}\prod_{j=1}^{N-1-i}\biggl[\frac{d\tilde{A}_i^{(j)}}{2\pi i \tilde{A}_i^{(j)}}|M(\ft,\fq)|^2\biggr]|\calZ_{N}^{\text{top}}|^2=\nonumber\\
&=\frac{(1-q)^{\varphi_N}}{\Lambda^{N-1}}\frac{\Up(N\varkappa)\prod_{1\leq i<j\leq N} [\Up(n_i-n_j)\Up(l_{N+1-i}-l_{N+1-j})]}{\prod_{i,j=1}^{N}\Up(\varkappa-n_i-l_{N+1-j})}\
\end{align}
where the exponent
\beq
\begin{split}
\varphi_N=&\left(\frac{Q}{2}-N\varkappa\right)^2+\sum_{1\leq i<j\leq N}\left[\left(\frac{Q}{2}+n_j-n_i\right)^2+\left(\frac{Q}{2}+l_{N+1-j}-l_{N+1-i}\right)^2\right]\\&-\sum_{i,j=1}^N\left(\frac{Q}{2}+n_i+l_{N+1-j}-\varkappa\right)^2
\end{split}
\eeq
after a little algebra simplifies into
\beq
\varphi_N=2Q\left(\frac{N(N-1)}{2}\varkappa+\sum_{i=1}^N i(n_i+l_{N+1-i})\right)-\frac{N-1}{4}Q^2=-2Q\form{2\fQ-\sum_{i=1}^3\balpha_i}{\rho}-\frac{N-1}{4}Q^2\, .
\eeq
Now we will employ our sl$(N)$ conventions, see appendix~\ref{subapp:sln}, equation  \eqref{eq:identificationparameters} as well as equations \eqref{eq:Lambda}, \eqref{eq:MactoLambda}
and rearrange the prefactors to obtain the  the $q$-deformed Fateev-Litvinov 3-point function in the form conjectured by \cite{Mitev:2014isa}:
\begin{align}
\label{eq:qdefFLTodacorr-N}
&C_q(N\varkappa\omega_{N-1},\balpha_2,\balpha_3)=\left(\beta\left|M(\ft,\fq)\right|^{2}\right)^{N-1}\left(\big(1-q^b\big)^{2b^{-1}}\big(1-q^{b^{-1}}\big)^{2b}\right)^{\form{2\fQ-\sum_{i=1}^3\balpha_i}{\rho}}\nonumber\\&\qquad\qquad \qquad \qquad \qquad \times\lim_{\delta_a\to 0}\ointctrclockwise\prod_{i=1}^{N-2}\prod_{j=1}^{N-1-i}\biggl[\frac{d\tilde{A}_i^{(j)}}{2\pi i \tilde{A}_i^{(j)}}|M(\ft,\fq)|^2\biggr]|\calZ_{N}^{\text{top}}|^2\\
&=\left(\frac{\big(1-q^b\big)^{2b^{-1}}\big(1-q^{b^{-1}}\big)^{2b}}{(1-q)^{2Q}}\right)^{\form{2\fQ-\sum_{i=1}^3\balpha_i}{\rho}}\frac{\Up'(0)^{N-1}\Up(N\varkappa)\prod_{e>0}\Up(\form{\fQ-\balpha_2}{e})\Up(\form{\fQ-\balpha_3}{e})}{\prod_{i,j=1}^{N}\Up(\varkappa+(\balpha_2-\fQ,\,h_i)+(\balpha_3-\fQ,\, h_j))}\,.\nonumber
\end{align}
Taking here the 4D limit $q \rightarrow 1$ and reintroducing the cosmological constant dependence according to \eqref{Cq2C} leads to the Fateev-Litvinov formula \eqref{eq:FLTodacorr} for general $N$.

\section{Conclusions and outlook}

This paper is a second in the series of papers proposing a general formula \eqref{eq:3pointfunctionsastopologicalstrings} for primary 3-point functions of Toda CFT. Here we provided a very convincing check  of \eqref{eq:3pointfunctionsastopologicalstrings} by reproducing an important known special case when one of the primaries has a null-vector at level one, a result due to Fateev and Litvinov \cite{Fateev:2005gs}. Before giving an outlook of interesting problems we would like to be addressed next, let us briefly summarize the main points of this note.

After introducing a required background material, we discussed
in section~\ref{sec:Higgsing} how the degeneration of the primary fields on the Toda side corresponds to Higgsing on the $(p,q)$ 5-brane web diagram side. Committing to the choice of the flopping frame which then dictates the form of the contour, we demonstrated that, in the semi-degenerate limit, the contour integral expressing Toda structure constants is given by a single residue. This considerably simplified the flow of the subsequent calculation. Using a summation formula derived from $q$-binomial identities \eqref{T_N-identity-main} for Kaneko-Macdonald-Warnaar sl$(n)$ hypergeometric functions, we proved that the sums over partitions still present in the residues can be computed exactly. Eventually, our result \eqref{eq:qdefFLTodacorr-N} indeed gives the expression of Fateev and Litvinov \eqref{eq:FLTodacorr} after one takes the $q\rightarrow 1$ limit and reintroduces \eqref{Cq2C} the dependence on the cosmological constant $\mu$ that is fixed from a corresponding Ward identity.

Reproducing the Fateev-Litvinov formula is a powerful test in support of our proposal for 3-point functions of generic Toda exponential fields.
We would, of course, like to obtain further checks of \eqref{eq:main5Dequality} which is currently the work in progress. There are two natural steps to take here. The first one involves placing a more general semi-degenerate field to the 3-point function. Specifically for  $\textbf{W}_3$, if a semi-degenerate condition reads $\balpha_1=N\varkappa\omega_2-mb\omega_1$, where $m$ is a positive integer, it corresponds to a primary field having a null-vector on a level $m+1>1$. The Toda 3-point functions containing such a field are also known from \cite{Fateev:2008bm}. In fact, these are the best of the CFT knowledge for the 3-point functions of generic primaries. The corresponding formula (see (3.11) and appendix B of \cite{Fateev:2008bm}) involves two very different pieces: a straightforward generalization of \eqref{eq:FLTodacorr} and a $4m$-dimensional Coulomb integral. This intriguing factorization indeed looks like to be reproducible from our general perspective.

The second natural step is matching the known semi-classical asymptotics \cite{Fateev:2007ab}. We observe that in such a limit the combinatorial functions $\tN_{\lambda\mu}$ factorize as
\beq
\tN_{\lambda\mu}(m;b,b^{-1})\stackrel{b\rightarrow \infty}{\longrightarrow}\tN_{\lambda\emptyset}(m;b,b^{-1})\tN_{\emptyset\mu}(m;b,b^{-1})\,.
\eeq
The sums over partitions thus disentangle, and proper generalizations of hypergeometric identities for the case of sl$(2)$ KM hypergeometric functions can be found to perform them. 
In fact, this step could then serve as a launch pad for a more ambitious goal of guessing a still unknown 'Lagrangian' for the $q$-deformed Toda theory. One would have to begin here by looking for the
Lagrangian description of the $q$-deformed Liouville theory, returning to the work of \cite{Nieri:2013yra,Nieri:2013vba}.
It could well be that the 2D space has to be made non-commutative  \cite{Olshanetsky:1993sw,Chaichian,Lavagno}.

Having checked the known cases, it is very interesting to go beyond them, the ultimate goal being to compute the contour integral in \eqref{eq:finalCqformula} exactly for generic values of the parameters. This will mean a considerable simplification of our general formula for the 3-point functions of Toda primaries. Doing so requires finding a closed form expression for the ``instanton'' sum of \eqref{eq:ZTNsum}, meaning that a suitable generalization of the KMW sl$(n)$ hypergeometric functions, as well as corresponding summation identities for them, have to be found. As an exercise to do before going for this serious problem, one could like to compute the corresponding sums for the cases with
 $E_{6,7,8}$ flavor symmetry studied in \cite{Benini:2009gi,Hayashi:2013qwa,Hayashi:2014wfa,Hayashi:2014hfa} 
 which are obtained from the general $T_N$ by a less severe Higgsing than the one we perform here.

Putting the above into the perspective of a full solution of the Toda theory, let us mention the remaining ingredients of it. First, a well-known fact is that, unlike the Virasoro case, the $\textbf{W}_N$ symmetry is not restrictive enough to constrain the 3-point functions of descendent fields from those of primaries \cite{Bowcock:1993wq}. The number of corresponding Ward identities is simply too small to find from them the descendent structure constants. This means that in order to find all the 3-point correlators, one needs to calculate independently the 3-point structure constants of two primaries and one descendent. It is however rather straightforward from the topological strings point of view.

The second remaining ingredient of a complete solution of Toda CFT are the conformal blocks. The paper \cite{Fateev:2011hq} describes the particular family of blocks which can be obtained by gluing the Fateev-Litvinov 3-point functions \eqref{eq:FLTodacorr}. Gluing the general ($q$-deformed) Toda 3-point functions in the same way would give the general conformal blocks of the ($q$-deformed) Toda CFT. Addressing this problem for $q$-Liouville, that is a starting point in such an investigation, is the work in progress \cite{FutoshiMasato}. Due to many uncertainties in properly defining a $q$-deformed Liouville (Toda) theory, such a finding would then as well work in opposite direction, allowing to know more about the $q$-deformed AGT-W correspondence and its relation to topological strings (see \cite{AwataDI}). The novel identities for Kaneko-Macdonald-Warnaar sl$(n)$ hypergeometric functions could probably be as helpful here as they were in the present note, to sum up known and new expressions for conformal blocks.

We finish with two remarks on the gauge theory side.
The degeneration we study in this paper, and in general Higgsing, should also be understood on the 4D/5D gauge theory side using  a generalization of the AGT correspondence with additional co-dimension two half-BPS surface defects \cite{Gukov:2006jk} as in \cite{Gomis:2007fi,Alday:2009fs,Gaiotto:2009fs,Dimofte:2010tz}.  See also \cite{Gaiotto:2012xa,Gadde:2013dda}.
The partition functions with half-BPS surface operators can be obtained form certain 2D partition functions \cite{Gomis:2014eya}. This 2D/4D relation  has its $q$-deformation to a
 3D/5D relation that was initiated by \cite{Nieri:2013yra} and further studied by \cite{Aganagic:2013tta,Nieri:2013vba,Aganagic:2014oia}.
See \cite{Gaiotto:2014ina} for the latest advancements on the subject.

 Lastly, by observing that the Higgsed geometry corresponding to the degeneration, see the right side of figure~\ref{fig:TNHiggsing}, is related to the strip geometry $\tilde{T}_N$,  see figure 7  in \cite{Kozcaz:2010af}, by the Hanany-Witten effect. We refer the interested reader to \cite{Kozcaz:2010af,Bonelli:2011fq} for a nice discussion on the subject. The invariance of the topological string amplitude
under the Hanany-Witten transition is non-trivial. It would be important to see how one can relate formula \eqref{eq:qdefFLTodacorr} for the $q$-deformed structure constants to the topological string amplitude for the strip, see equation (4.66) of \cite{Bao:2013pwa}.

\section*{Acknowledgments}

We would  like to thank first our collaborators on closely related projects Masato Taki and Futoshi Yagi. We are indebted to Volker Schomerus and Futoshi Yagi for reading the draft of this paper and making helpful comments. In addition, we are thankful to Can Koz\c{c}az, Fabrizio Nieri, J\"org Teschner and Dan Xie for insightful comments and discussions. We furthermore gratefully acknowledge support from the Simons Center for Geometry and Physics, Stony Brook University, as well as of the C.N.\ Yang Institute for Theoretical Physics,  where some of the research for this paper was performed. V.M.\ acknowledges the support of the Marie Curie International Research Staff Exchange Network UNIFY of the European Union's Seventh Framework Programme [FP7-People-2010-IRSES] under grant agreement n°269217, which allowed him to visit Stony Brook University. E.P.\ is partially supported by  the Marie Curie action FP7-PEOPLE-2010-RG. M.I.\ thanks the Research Training Association RTG1670 for partial support. The research leading to these results has received funding from the People Programme (Marie Curie Actions) of the European Union's Seventh Framework Programme FP7/2007-2013/ under REA Grant Agreement No 317089 (GATIS).

\appendix

\setcounter{equation}{0}
\section{Notations, conventions and special functions} \label{appA}
\numberwithin{equation}{section}

In this appendix, we summarize our conventions and the main properties of the special functions that we use the most.

\subsection{Parametrization of the \texorpdfstring{$T_N$}{TN} junction}
\label{app:notation}

\begin{figure}[t]
  \centering
\includegraphics[height=10cm]{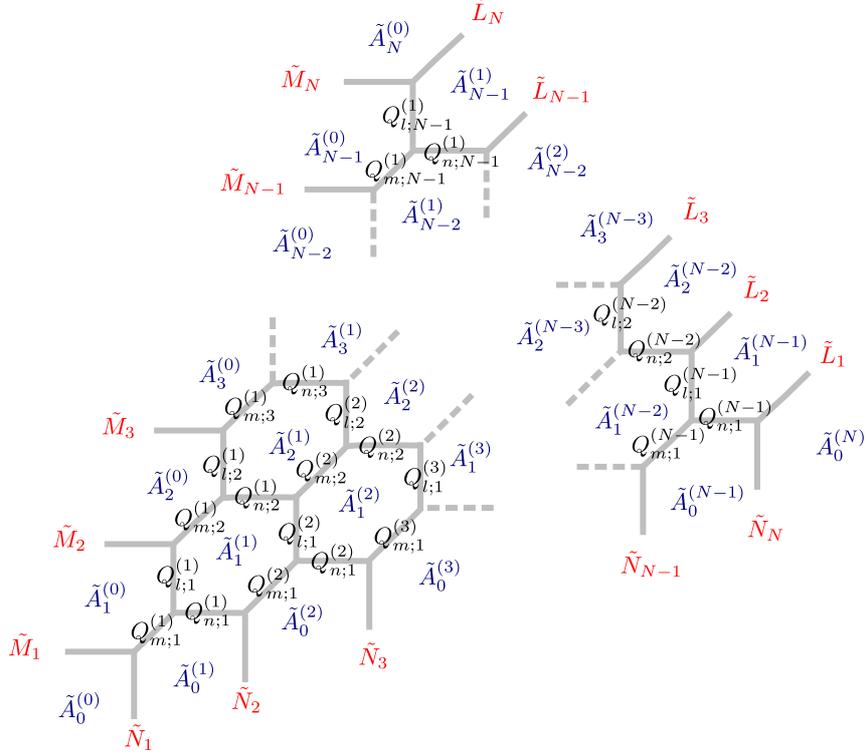}
  \caption{Parametrization for $T_N$. We denote the K\"ahler moduli parameters corresponding to the horizontal lines as $Q_{n;i}^{(j)}$, to the vertical lines as $Q_{l;i}^{(j)}$, and to tilted lines as $Q_{m;i}^{(j)}$. We denote the breathing modes as $\tilde{A}^{(j)}_i$. The index $j$ labels the strips in which the diagram can be decomposed. }
  \label{fig:TNparam}
\end{figure}
We gather in this appendix all necessary formulas for the parametrizations of the K\"ahler moduli of the $T_N$. 
First, the ``interior'' Coulomb moduli $\tilde{A}_j^{(i)}=e^{-\beta a_i^{(j)}}$ are independent, while the  ``border'' ones are given by
\beq
\label{eq:borderAdefMNL}
\tilde{A}_{i}^{(0)} = \prod_{k=1}^i \tilde{M}_k,
\qquad
\tilde{A}_{0}^{(j)} = \prod_{k=1}^j \tilde{N}_k,
\qquad
\tilde{A}_{i}^{(N-i)}  = \prod_{k=1}^i \tilde{L}_k.
\eeq
The parameters labeling the positions of the flavors branes obey the relations
\beq
\label{eq:externalparametersconstraints}
\prod_{k=1}^N\tilde{M}_k=\prod_{k=1}^N\tilde{N}_k=\prod_{k=1}^N\tilde{L}_k=1\Longleftrightarrow \sum_{k=1}^Nm_k=\sum_{k=1}^Nn_k=\sum_{k=1}^Nl_k=0.
\eeq
Therefore, $\tilde{A}_0^{(0)}=\tilde{A}_N^{(0)}=\tilde{A}_0^{(N)}=1$ and we can invert  relation \eqref{eq:borderAdefMNL} as
\beq
\label{eq:AfunctionMNL}
\tilde{M}_i=\frac{\tilde{A}_i^{(0)}}{\tilde{A}_{i-1}^{(0)}},\qquad \tilde{N}_i=\frac{\tilde{A}_0^{(i)}}{\tilde{A}_{0}^{(i-1)}},\qquad \tilde{L}_i=\frac{\tilde{A}_i^{(N-i)}}{\tilde{A}_{i-1}^{(N-i+1)}}.
\eeq
All placements are illustrated in figure~\ref{fig:TNparam}.
The   K\"ahler parameters associated to the edges of the $T_N$ junction are related to the $\tilde{A}_i^{(j)}$ as follows
\begin{align}
\label{eq:PQR}
Q_{n;i}^{(j)}
 = \frac{\tilde{A}_{i}^{(j)} \tilde{A}_{i-1}^{(j)}}
    {\tilde{A}_{i}^{(j-1)} \tilde{A}_{i-1}^{(j+1)}},
\qquad
Q_{l;i}^{(j)}
 = \frac{\tilde{A}_{i}^{(j)} \tilde{A}_{i}^{(j-1)}}
    {\tilde{A}_{i-1}^{(j)} \tilde{A}_{i+1}^{(j-1)}},
\qquad
Q_{m;i}^{(j)}
 = \frac{\tilde{A}_{i}^{(j-1)} \tilde{A}_{i-1}^{(j)}}
    {\tilde{A}_{i}^{(j)} \tilde{A}_{i-1}^{(j-1)}}.
\end{align}
For each inner hexagon of \eqref{fig:TNparam}, the following two constraints are satisfied
\begin{equation}
Q_{l;i}^{(j)} Q_{m;i+1}^{(j)} = Q_{m;i}^{(j+1)} Q_{l;i}^{(j+1)},
\qquad
Q_{n;i}^{(j)} Q_{m;i}^{(j+1)} = Q_{m;i+1}^{(j)} Q_{n;i+1}^{(j)} .
\end{equation}

\subsection{Conventions and notations for  \texorpdfstring{$\SU(N)$}{SU(N)}}
\label{subapp:sln}

For the convenience of the reader we summarize here our $\SU(N)$ conventions. The weights of the fundamental representation of $\SU(N)$ are $h_i$ with $\sum_{i=1}^N h_i=0$. We remind that the scalar product is defined via $\form{h_i}{h_j}=\delta_{ij}-\frac{1}{N}$.
The simple roots are 
\beq
e_k\colonequals h_k-h_{k+1 }\,, \qquad k=1,\ldots, N-1\,,
\eeq
and the positive roots $e>0$ are contained in the set
\beq
\Delta^+\colonequals \{h_i-h_j\}_{i<j=1}^N=\{e_i\}_{i=1}^{N-1}\cup\{e_i+e_{i+1}\}_{i=1}^{N-2}\cup\cdots \cup \{e_1+\cdots+e_{N-1}\}\,.
\eeq
The Weyl vector $\rho$ for $\SU(N)$ is given by 
\beq
\label{eq:defWeylvector}
\rho\colonequals \frac{1}{2}\sum_{e>0}e=\frac{1}{2}\sum_{i<j=1}^N(h_i-h_j)=\sum_{i=1}^N\frac{N+1-2i}{2}h_i=\omega_1+\cdots+\omega_{N-1},
\eeq
and it obeys $\form{\rho}{e_i}=1$ for all $i$. 
The $N-1$ fundamental weights $\omega_i$ of $\SU(N)$ are given by
\beq
\omega_i=\sum_{k=1}^i h_k\,, \qquad i=1,\dots, N-1
\eeq
and the corresponding finite dimensional representations are the $i$-fold antisymmetric tensor product of the fundamental representation. They obey the scalar products $\form{e_i}{\omega_j}=\delta_{ij}$, {\it i.e.} they are a dual basis. Furthermore, we find the following scalar products useful
\beq 
\label{eq:SUNscalarproducts1}
 \form{\rho}{h_j}=\frac{N+1}{2}-j,\qquad \form{\rho}{\omega_i}=\frac{i(N-i)}{2}, \qquad (h_j,\omega_i)=\left\{\begin{array}{ll}1-\frac{i}{N}& j\leq i\\-\frac{i}{N} & j>i \end{array}\right. \,,
\eeq
as well as
\beq
\label{eq:SUNscalarproducts2}
\form{\omega_i}{\omega_j}=\frac{\text{min}(i,j)\left(N-\text{max}(i,j)\right)}{N},\qquad \form{\rho}{\rho}=\frac{N(N^2-1)}{12}\,.
\eeq

The Weyl group of $\SU(N)$ is isomorphic to $S_N$ and is generated by the $N-1$ Weyl reflections associated to the simple roots. If $\balpha$ is a weight, we define the  Weyl reflections with respect to the simple root $e_i$
\beq
\label{eq:defWeyltransformations}
\fw_i\cdot \balpha \colonequals  \balpha-2\frac{\form{e_i}{\balpha}}{\form{e_i}{e_i}}e_i=\balpha-\form{e_i}{\balpha}e_i\,.
\eeq
Furthermore, we define the affine Weyl reflections with respect to $e_i$ as follows
\beq
\label{eq:defaffineWeyltransformations}
\fw_i\circ \balpha\colonequals \fQ+\fw_i\cdot(\balpha-\fQ)=\fw_i\cdot \balpha+Qe_i=\balpha-\form{\balpha-\fQ}{e_i}e_i\,,
\eeq
where $\fQ\colonequals Q\rho=(b+b^{-1})\rho$.

\subsection{Special functions}
\label{subapp:special}

In this section we gather the definitions and properties of the special functions used in the main text.

We begin with the function $\Upsilon(x)$ which is defined for $0<\Re(x)<Q=b+b^{-1}$ as the integral
\beq
\log \Upsilon(x)\colonequals  \int_{0}^{\infty}\frac{dt}{t}\left[\left(\frac{Q}{2}-x\right)^2e^{-t}-\frac{\sinh^2\left[\left(\frac{Q}{2}-x\right)\frac{t}{2}\right]}{\sinh \frac{b t}{2}\sinh \frac{t}{2b}}\right].
 \eeq 
It is clear from the definition that 
\beq
\Upsilon(x)=\Upsilon(Q-x), \qquad  \Upsilon\left(\frac{Q}{2}\right)=1. 
\eeq
One can show from the alternative definition below that the following shift identities are obeyed
\beq
\label{eq:shiftUpsilon4D}
\Upsilon(x+b)=\gamma(x b)\,b^{1-2bx}\,\Upsilon(x),\qquad \Upsilon(x+b^{-1})=\gamma(x b^{-1})\,b^{2xb^{-1}-1}\,\Upsilon(x).
\eeq
where $\gamma(x)\colonequals \frac{\Gamma(x)}{\Gamma(1-x)}$.  An useful implication is 
\beq
\label{eq:extrashiftsUpsilon4D}
\Upsilon(x+Q)=b^{2(b^{-1}-b)x}\frac{\Gamma\big(1+bx\big)\Gamma\big(b^{-1}x\big)}{\Gamma\big(1-bx\big)\Gamma\big(-b^{-1}x\big)}\Upsilon(x),
\eeq
which is used in the derivation of the reflection amplitude \eqref{eq:defreflectionamplitude}.
It follows from \eqref{eq:shiftUpsilon4D} that $\Upsilon$ is an entire function with zeroes at
\beq
\label{eq:zeroesofUpsilon}
x=-n_1 b -n_2 b^{-1},\quad \text { or } \quad x=(n_1+1) b +(n_2+1) b^{-1},
\eeq
where $n_i\in \mathbb{N}_0$. 

The function $\Upsilon$ can be connected to the Barnes Double Gamma function $\Gamma_2(x|\omega,\omega_2)$.
First, we define $\Gamma_2(x|\omega_1,\omega_2)$ via the \textit{analytic continuation} (the sum is only well-defined if $\Re(t)>2$) of 
\beq
\log \Gamma_2(s|\omega_1,\omega_2)=\left[\frac{\partial}{\partial t}\sum_{n_1,n_2=0}^{\infty}(s+n_1\omega_1+n_2\omega_2)^{-t}\right]_{t=0}.
\eeq
From this definition, one can prove (see A.54 of \cite{Nakayama:2004vk}) the \textit{difference property}
\beq
\frac{\Gamma_2(s+\omega_1|\omega_1,\omega_2)}{\Gamma_2(s|\omega_1,\omega_2)}=\frac{\sqrt{2\pi}}{\omega_2^{\frac{s}{\omega_2}-\frac{1}{2}}\Gamma\Big(\frac{s}{\omega_2}\Big)},\qquad \frac{\Gamma_2(s+\omega_2|\omega_1,\omega_2)}{\Gamma_2(s|\omega_1,\omega_2)}=\frac{\sqrt{2\pi}}{\omega_1^{\frac{s}{\omega_1}-\frac{1}{2}}\Gamma\Big(\frac{s}{\omega_1}\Big)}.
\eeq
In order to express the $\Upsilon$ function using the Barnes double Gamma function, we have to first define  the \textit{normalized} function
\beq
\label{eq:defGammabviaBarnes}
 \Gamma_b(x)\colonequals \frac{ \Gamma_2(x|b,b^{-1})}{ \Gamma_2(\frac{Q}{2}|b,b^{-1})}.
\eeq
The log of the function $\Gamma_b(x)$ has an integral representation as
\beq
\log\Gamma_b(x)=\int_{0}^{\infty}\frac{dt}{t}\left(\frac{e^{-x t}-e^{-\frac{Q t}{2}}}{(1-e^{-t b })(1-e^{-t b^{-1}})}-\frac{\left(\frac{Q}{2}-x\right)^2}{2}e^{-t}-\frac{\frac{Q}{2}-x}{t}\right).
\eeq
Then, using \eqref{eq:defGammabviaBarnes} we can express the $\Upsilon(x)$ as 
\beq
\label{eq:defUpsilonproduct}
\Upsilon(x)= \frac{1}{\Gamma_b(x)\Gamma_b(Q-x)}.
\eeq
This, together with the difference properties of $\Gamma_2$ proves the shift identities \eqref{eq:shiftUpsilon4D}.

We proceed by defining some $q$-deformed special functions we need in the main text, such as shifted factorials\footnote{A good source for the properties of the shifted factorials is \cite{Nishizawa:2001}.} 
\beq
\label{eq:defshiftedfactorial1}
(U;q)_p:=\prod_{i=1}^p (1-Uq^{i-1})
\eeq
for positive $p$, which is continued to negative $p$ according to 
\beq
\label{eq:defshiftedfactorial2}
(U;q)_p=\frac{1}{(Uq^p;q)_{-p}}\,.
\eeq
In particular for $p\rightarrow \infty$, and for arbitrary number of $q$'s, we have (we require for convergence that $|q_i|<1$ for all $i$)
\beq
\label{eq:defshiftedfactorial3}
(U;q_1,\ldots, q_r)_{\infty}\colonequals \prod_{i_1=0,\ldots, i_r=0}^{\infty}(1-U q_1^{i_1}\cdots q_r^{i_r})\,. 
\eeq
We can extend the definition of the shifted factorial for all values of $q_i$ by imposing the relations
\beq
\label{eq:shiftedfactorialinversion}
(U;q_1,\ldots,q_i^{-1},\ldots, q_r)_{\infty}=\frac{1}{(Uq_i;q_1,\ldots, q_r)_{\infty}}\,.
\eeq
Furthermore, they obey the following shifting properties
\beq
\label{eq:shiftingfactorials}
(q_j U;q_1,\ldots, q_r)_{\infty}=\frac{(U;q_1,\ldots, q_r)_{\infty}}{(U;q_1,\ldots,q_{j-1},q_{j+1},\ldots, q_r)_{\infty}}\,.
\eeq
We then define the function $\calM(U;\ft,\fq)$ as
\beq
\label{eq:defcalMeverywhere}
\calM(U;\ft,\fq)\colonequals (U \fq ;\ft,\fq)_{\infty}^{-1}=\left\{\begin{array}{ll}
\prod_{i,j=1}^{\infty}(1-U\ft^{i-1}\fq^j)^{-1} & \text{ for } |\ft|<1, |\fq|<1\\
\prod_{i,j=1}^{\infty}(1-U\ft^{i-1}\fq^{1-j}) & \text{ for } |\ft|<1, |\fq|>1\\
\prod_{i,j=1}^{\infty}(1-U\ft^{-i}\fq^j) & \text{ for } |\ft|>1, |\fq|<1\\
\prod_{i,j=1}^{\infty}(1-U\ft^{-i}\fq^{1-j})^{-1} & \text{ for } |\ft|>1, |\fq|>1\\ \end{array}\right.\,,
\eeq
converging for all $U$. This function can be written as a plethystic exponential 
\beq
\label{eq:defMpexp}
\calM(U;\ft,\fq)=\exp\left[\sum_{m=1}^{\infty} \frac{U^m}{m}\frac{\fq^m}{(1-\ft^m)(1-\fq^m)}\right]\,,
\eeq
which converges for all $\ft$ and all $\fq$ provided that $|U|<\fq^{-1+\theta(|\fq|-1)}\ft^{\theta(|\ft|-1)}$. Here and elsewhere $\theta(x)=1$ if $x>0$ and is zero otherwise. The following identity is obvious from the definition
\beq
\label{eq:exchangerelationMN}
\calM(U;\fq,\ft)=\calM(U\nicefrac{\ft}{\fq};\ft,\fq)\,.
\eeq
From the analytic properties of the shifted factorials \eqref{eq:shiftedfactorialinversion}, we read the identities
\beq
\label{eq:inversionidentities}
\calM(U;\ft^{-1},\fq)=\frac{1}{\calM(U\ft;\ft,\fq)}, \qquad \calM(U;\ft,\fq^{-1})=\frac{1}{\calM(U\fq^{-1};\ft,\fq)}\,,
\eeq
while from \eqref{eq:shiftingfactorials} we take the following shifting identities
\beq
\label{eq:calMshift}
\calM(U\ft;\ft,\fq)=(U \fq; \fq)_{\infty} \calM(U;\ft,\fq),\qquad \calM(U\fq;\ft,\fq)=(U \fq; \ft)_{\infty} \calM(U;\ft,\fq)\,.
\eeq
We define the $q$-deformed $\Upsilon$ function as
\beq
\label{eq:defUp}
\begin{split}
\Up(x|\epsilon_1,\epsilon_2)=&(1-q)^{-\frac{1}{\epsilon_1\epsilon_2}\left(x-\frac{\epsilon_+}{2}\right)^2}\prod_{n_1,n_2=0}^{\infty}\frac{(1-q^{x+n_1\epsilon_1+n_2\epsilon_2})(1-q^{\epsilon_+-x+n_1\epsilon_1+n_2\epsilon_2})}{(1-q^{\nicefrac{\epsilon_+}{2}+n_1\epsilon_1+n_2\epsilon_2})^2}\\
=&(1-q)^{-\frac{1}{\epsilon_1\epsilon_2}\left(x-\frac{\epsilon_+}{2}\right)^2}\left|\frac{\calM(q^{-x};\ft,\fq)}{\calM(\sqrt{\frac{\ft}{\fq}};\ft,\fq)}\right|^2\,,
\end{split}
\eeq
where we have used the definition \eqref{eq:normsquared} for the norm squared. From time to time we will use the short-hand notation
\beq
\label{eq:Lambda}
\Lambda\colonequals \left|\calM\left(\sqrt{\frac{\ft}{\fq}};\ft,\fq\right)\right|^2\,.
\eeq
If follows from the definition \eqref{eq:defUp} that $\Up(\nicefrac{\epsilon_+}{2}|\epsilon_1,\epsilon_2)=1$, that $\Up(x|\epsilon_1,\epsilon_2)=\Up(\epsilon_+-x|\epsilon_1,\epsilon_2)$ and  that $\Up(x|\epsilon_1,\epsilon_2)=\Up(x|\epsilon_2,\epsilon_1)$. Furthermore, from the shifting identities for $\calM$, we can easily prove that 
\beq
\label{eq:shiftidentitiesqUpsilon}
\Up(x+\epsilon_1|\epsilon_1,\epsilon_2)=\left(\frac{1-q}{1-q^{\epsilon_2}}\right)^{1-2\epsilon_2^{-1}x}\gamma_{q^{\epsilon_2}}(x\epsilon_2^{-1})\Up(x|\epsilon_1,\epsilon_2)\,,
\eeq
together with a similar equation for the shift with $\epsilon_2$. Here, we have used the definition of the $q$-deformed $\Gamma$ and $\gamma$ functions
\beq
\label{eq:defGammaq}
\Gamma_q(x)\colonequals (1-q)^{1-x}\frac{(q;q)_{\infty}}{(q^x;q)_{\infty}},\qquad \gamma_q(x)\colonequals\frac{\Gamma_q(x)}{\Gamma_q(1-x)}=(1-q)^{1-2x}\frac{(q^{1-x};q)_{\infty}}{(q^{x};q)_{\infty}}\,,
\eeq
valid for $|q|<1$. They obey  $\Gamma_q(x+1)=\frac{1-q^x}{1-q}\Gamma_q(x)$, implying  $\gamma_q(x+1)=\frac{(1-q^x)(1-q^{-x})}{(1-q)^2}\gamma_q(x)$.  Because of the normalization of $\Up(x|\epsilon_1,\epsilon_2)$ and since the factors of the right hand side of \eqref{eq:shiftidentitiesqUpsilon} have a well defined limit for $q\rightarrow 1$, we find by comparing functional identities that
\beq
\label{Y4Dlimit}
\Up({x}|\epsilon_1,\epsilon_2)\stackrel{q\rightarrow 1}{\longrightarrow } \Upsilon(x|\epsilon_1,\epsilon_2)\colonequals \frac{\Gamma_2\big(\frac{\epsilon_+}{2}|\epsilon_1,\epsilon_2\big)^2}{\Gamma_2\big(x|\epsilon_1,\epsilon_2\big)\Gamma_2\big(\epsilon_+-x|\epsilon_1,\epsilon_2\big)}\,.
\eeq
where $\Gamma_2$ is the Barnes Double Gamma function. 
In particular, the usual  function $\Upsilon(x)$ introduced in \cite{Zamolodchikov:1995aa} is equal to $\Upsilon(x|b,b^{-1})$. We shall often just write $\Up(x)$ instead of $\Up(x|\epsilon_1,\epsilon_2)$ and indicate in the text whether the $\epsilon_i$ parameters are arbitrary or whether $b=\epsilon_1=\epsilon_2^{-1}$. 

We will also need to evaluate the derivative of $\Up(x)$ at $x=0$. Since the zero of $\Up(x)$ at $x=0$ is due to the factor $(1-q^x)$ in the numerator of \eqref{eq:defUp}, we find that the only piece of the derivative that survives is 
\beq
\label{eq:derivativeofUp}
\Up'(0)=\frac{\beta}{1-q}\Up(b)\,. 
\eeq
From this formula we can then obtain an identity useful for the calculations of the main text. Let us define the norm squared of the refined McMahon function following \cite{Iqbal:2012xm}:
\beq
\label{eq:McMahonUp}
\begin{split}
|M(\ft,\fq)|^2&\colonequals \lim_{U\rightarrow 1}\frac{|\calM(U;\ft,\fq)|^2}{1-U^{-1}}=|\calM(\fq^{-1};\ft,\fq)|^2=(1-q)^{\frac{\left(\epsilon_1-\epsilon_2\right)^2}{4\epsilon_1\epsilon_2}}\Lambda\Up(\epsilon_1)\,.
\end{split}
\eeq
Then, from \eqref{eq:Lambda} and \eqref{eq:derivativeofUp} we get for $\epsilon_1=b$ and $\epsilon_2=b^{-1}$
\beq
\label{eq:MactoLambda}
\left|M(\ft,\fq)\right|^{2}=\frac{1}{\beta}(1-q)^{\left(\frac{Q}{2}\right)^2}\Lambda \Up'(0)\,.
\eeq

\subsection{Combinatorial special functions}
\label{app:finiteproduct}

We shall use in the following 
\beq
\label{eq:part-statistics}
|\lambda|\colonequals \sum_{i=1}^{\ell(\lambda)}\lambda_i,\qquad ||\lambda||^2\colonequals \sum_{i=1}^{\ell(\lambda)}\lambda_i^2,\qquad n(\lambda)\colonequals\sum_{i=1}^{\ell(\lambda)}(i-1)\lambda_i=\frac{||\lambda^t||^2-|\lambda|}{2}\,, 
\eeq
where $\ell(\lambda)$ is the number of rows of the partition $\lambda$. We also define the relative arm-length $a_{\mu}(s)$, arm-colength $a'_{\mu}(s)$, leg-length $l_{\mu}(s)$ and leg-colength $l'_{\mu}(s)$ of a given box $s$ of the partition $\lambda$ with respect to another partition $\mu$ as:
\beq
\label{eq:arms-legs}
a_{\mu}(s)\colonequals\mu_i-j\,, \qquad a'_{\mu}(s)\colonequals j-1\, ,
\qquad l_{\mu}(s)\colonequals\mu^{t}_j-i\,,\quad   l'_{\mu}(s)\colonequals i-1\,.
\eeq
It is of course also possible to have $\lambda=\mu$. The $(\fq,\ft)$-deformed factorial of $U$ depending on a partition $\lambda$ is then given as a following product over its boxes:
\beq
\label{eq:qtfactorial}
(U;\fq, \ft)_{\lambda}\colonequals\prod_{i=1}^{\ell(\lambda)}(U \ft^{1-i};\fq)_{\lambda_i}=\prod_{s\in\lambda}(1-U\fq^{a'(s)}\ft^{-l'(s)})\,.
\eeq

The next piece of notation that we need are the $(\fq,\ft)$-deformations of the hook product of a Young diagram $\lambda$. There are two inequivalent ways for this number to be deformed to a two-variable polynomial, namely:
\beq
\label{eq:defhooks}
h_{\lambda}(\fq,\ft)\colonequals\prod_{s\in\lambda}(1-\fq^{a(s)}\ft^{l(s)+1})\,,\qquad h'_{\lambda}(\fq,\ft)\colonequals\prod_{s\in\lambda}(1-\fq^{a(s)+1}\ft^{l(s)})\,.
\eeq
Our last definition is that of the 5D uplift of Nekrasov functions, which we write as
\begin{align}
\label{eq:deftN}
\tN_{\lambda\mu}(u;\epsilon_1,\epsilon_2)\colonequals&\prod_{(i,j)\in \lambda}2\sinh\frac{\beta}{2}\left[u+\epsilon_1(\lambda_i-j+1)+\epsilon_2(i-\mu^t_j)\right]\nonumber
\\&
\times \prod_{(i,j)\in \mu}2\sinh\frac{\beta}{2}\left[u+\epsilon_1(j-\mu_i)+\epsilon_2(\lambda^t_j-i+1)\right]\\
& =\prod_{s\in \lambda}2\sinh\frac{\beta}{2}\left[u+\epsilon_1\left(a_{\lambda}(s)+1\right)-\epsilon_2 l_{\mu}(s)\right]
\prod_{s\in \mu}2\sinh\frac{\beta}{2}\left[u-\epsilon_1 a_{\mu}(s)+\epsilon_2\left(l_{\lambda}(s)+1\right)\right]\nonumber
\end{align}
where the products are taken over boxes of partitions $\lambda$ and $\mu$, respectively. By pulling some factors out of the products, the definition can also be rewritten as
\begin{align}
\label{eq:deftN-1}
\tN_{\lambda\mu}(u;\epsilon_1,\epsilon_2)\colonequals&\left(\sqrt{\frac{\ft}{\fq}}\frac{1}{U}\right)^{\frac{|\lambda|+|\mu|}{2}}\ft^{\frac{||\lambda^t||^2-||\mu^t||^2}{4}}\fq^{\frac{||\mu||^2-||\lambda||^2}{4}}\prod_{(i,j)\in \lambda}\biggl(1-U\ft^{\mu_j^t-i}\fq^{\lambda_i-j+1}\biggr)\nonumber
\\&
\times \prod_{(i,j)\in \mu}\biggl(1-U\ft^{-\lambda_j^t+i-1}\fq^{-\mu_i+j}\biggr),
\end{align}
where $U=e^{-\beta u}$. For particular values of the parameter $u$, the introduced functions behave like Kronecker$-\delta$ functions, namely
\beq
\label{eq:tNdelta}
\tN_{\lambda\emptyset}(-\epsilon_+)=\tN_{\emptyset\lambda}(0)=\delta_{\lambda\emptyset},
\eeq
where $\epsilon_+=\epsilon_1+\epsilon_2$. 
Furthermore, they obey the exchange identities
\begin{align}
\label{eq:propertiestN}
\tN_{\lambda\mu}(u;-\epsilon_2,-\epsilon_1)&=\tN_{\mu^t\lambda^t}(u-\epsilon_+;\epsilon_1,\epsilon_2),\nonumber\\
\tN_{\lambda\mu}(-u;\epsilon_1,\epsilon_2)&=(-1)^{|\lambda|+|\mu|}\tN_{\mu\lambda}(u-\epsilon_+;\epsilon_1,\epsilon_2),\\
\tN_{\lambda\mu}(u;\epsilon_2,\epsilon_1)&=\tN_{\lambda^t\mu^t}(u;\epsilon_1,\epsilon_2).\nonumber
\end{align}
Finally, there are two relations involving the functions we just defined, namely
\begin{align}
\label{warnaartotop-1}
\frac{1}{h_{\lambda}(\fq,\ft)h'_{\lambda}(\fq,\ft)}=\frac{(-1)^{|\lambda|}\ft^{-\frac{||\lambda^t||^2}{2}}\fq^{-\frac{||\lambda||^2}{2}}}{\tN_{\lambda\lambda}\left(0\right)}
\end{align}
as well as
\begin{align}
\label{warnaartotop-2}
&(U)_{\lambda}\equiv (U;\fq,\ft)_{\lambda}=\left(\sqrt{\frac{\ft}{\fq}}U\right)^{\frac{|\lambda|}{2}}\ft^{-\frac{||\lambda^t||^2}{4}}\fq^{\frac{||\lambda||^2}{4}}\tN_{\lambda\emptyset}\left(u-\epsilon_+\right),
\end{align}
where $U=e^{-\beta u}$.

\section{The \texorpdfstring{$\bsl(N)$}{sl(N)} Kaneko-Macdonald-Warnaar hypergeometric functions} \label{appB}
\numberwithin{equation}{section}

This appendix contains the derivation of the summation formula \eqref{T_N-identity-main} used in the main text. It exploits a binomial identity for the Kaneko-Macdonald-Warnaar extension of basic hypergeometric functions \cite{Warnaar} which generalizes the Kaneko-Macdonald $\bsl(2)$  identity of \cite{Kaneko, Macdonald, Baker-Forrester}.

\subsection{The sl$(N)$ KMW hypergeometric functions and their \texorpdfstring{$\fq$}{q}-binomial identity}

The {\it Macdonald polynomials} $P_{\lambda}(\fx;\fq,\ft)$ (in the case of infinite alphabet $\fx$ referred as the Macdonald symmetric functions)
are labeled by a number partition $\lambda=(\lambda_1, \dots, \lambda_{\ell(\lambda)})$ and form an especially convenient basis in the ring of symmetric functions of $\fx=(\fx_1,\fx_2,\dots)$ over the field $\mathbb{F}=\mathbb{Q}(\fq,\ft)$ of rational functions in two variables $\fq$ and $\ft$ \cite{MacdonaldSymmetric}.

Having many nice properties, the Macdonald polynomials are applied in various areas of contemporary mathematics. One of them is the theory of $\bsl(N)$ {\it Kaneko-Macdonald-Warnaar analogues of basic hypergeometric functions}. These functions, of type $(r+1,r)$, are defined as
\begin{align}
\label{hyperdef}
&\pFq{r+1}{r}{A_1 \pFcomma\dots  \pFcomma A_{r+1}}{B_1\pFcomma \dots \pFcomma B_r}{\fq\pFcomma\ft;\fx^{(1)}\pFcomma \dots \pFcomma \fx^{(N-1)}}:=\nonumber\\
&\sum_{\lambda^{(1)},\dots , \lambda^{(N-1)}}'
\frac{(A_1 \pFcomma\dots  \pFcomma A_{r+1};\fq\pFcomma\ft)_{\lambda^{(N-1)}}}{(\fq \ft^{k_{N-1}-1}\pFcomma B_1\pFcomma \dots \pFcomma B_r;\fq\pFcomma\ft)_{\lambda^{(N-1)}}}
\prod_{s=1}^{N-1}\biggl[\ft^{n(\lambda^{(s)})}\frac{(\fq \ft^{k_{s}-1};\fq\pFcomma\ft)_{\lambda^{(s)}}}{h'_{\lambda^{(s)}}(\fq\pFcomma\ft)} P_{\lambda^{(s)}}(\fx^{(s)};\fq\pFcomma\ft)\biggr]\\
&\times \prod_{s=1}^{N-2} \prod_{i=1}^{k_s} \prod_{j=1}^{k_{s+1}}
\frac{(\fq \ft^{j-i-1+k_{s}-k_{s+1}};\fq)_{\lambda_i^{(s)}-\lambda_j^{(s+1)}}}{(\fq \ft^{j-i+k_{s}-k_{s+1}};\fq)_{\lambda_i^{(s)}-\lambda_j^{(s+1)}}},\nonumber 
\end{align}
where the integer parameters $k_s$ are such that $0\equiv k_0< k_1<  k_2<  \cdots<  k_{N-1}$ and the summations are performed over partitions $\lambda^{(s)}$, $1\leq s \leq N-1$ satisfying $k_s \geq\ell(\lambda^{(s)})$.  We have used here the definitions \eqref{eq:defshiftedfactorial1}, \eqref{eq:part-statistics},
\eqref{eq:qtfactorial},
\eqref{eq:defhooks}. The prime symbol above marks the fact that entries of the partitions giving a non-zero contribution to the sum all satisfy an additional condition $\lambda^{(s)}_i\geq \lambda^{(s+1)}_{i-k_s+k_{s+1}}$ for $1\leq i\leq k_s$. It provides a convenient visualization of the multiple sum as running over single skew plane partitions of shape $\eta-\nu$, where $\eta=(k_{N-1}^{N-1})$ is a rectangle and $\nu=(k_{N-1}-k_{1},\dots, k_{N-1}-k_{N-2})$.

In the following, it will be enough to restrict ourselves to a so-called {\it principal specialization} of a Macdonald polynomial, for which the string of arguments $\fx$ is set to $\tilde \fx:=z(1,t, \dots , t^{k-1})$:
\begin{align}
P_{\lambda}(\tilde \fx;\fq,\ft)=z^{|\lambda|}\ft^{n(\lambda)}\frac{(\ft^k;\fq,\ft)_{\lambda}}{h_{\lambda}(\fq,\ft)}.
\end{align}
The corresponding specialization of the $\bsl(N)$ multiple $q${\it-binomial theorem} is then written as:
\vspace{0.1cm}\\
\textbf{Theorem} [See \cite{Warnaar}, Cor.\ 3.1]
\begin{align}
\pFq{1}{0}{A}{-}{\fq\pFcomma\ft;\tilde \fx^{(1)}\pFcomma \dots \pFcomma \tilde \fx^{(N-1)}}=&\prod_{s=1}^{N-1}\prod_{i=1}^{k_s-k_{s-1}}\frac{(Az_s\cdots z_{N-1}\ft^{i+s+k_{s-1}+\cdots+k_{N-2}-N};\fq)_{\infty}}{(z_s\cdots z_{N-1}\ft^{i+s+k_{s-1}+\cdots+k_{N-2}-N};\fq)_{\infty}} \label{q-binomial}\\
& \times \prod_{1\leq s\leq r\leq N-2}\prod_{i=1}^{k_s-k_{s-1}}\frac{(\fq z_s\cdots z_{r}\ft^{i+s-r+k_{s-1}+\cdots+k_{r}-k_{r+1}-2};\fq)_{\infty}}{(z_s\cdots z_{r}\ft^{i+s-r+k_{s-1}+\cdots+k_{r-1}-1};\fq)_{\infty}}, \nonumber
\end{align}
where $\tilde \fx^{(s)}:=z_s(1,\ft,\dots ,\ft^{k_s-1})$ for $1\leq s\leq N-1$ and ``$-$'' indicates the absence of the parameters $B_i$ in the definition \eqref{hyperdef}.

\subsection{The summation formula}
\label{subapp:summation}

It will be convenient for the subsequent argument to rewrite the above formula \eqref{q-binomial} in the topological string conventions. This turns out to be possible due to the identities \eqref{eq:defcalMeverywhere}, \eqref{warnaartotop-1}, \eqref{warnaartotop-2} and the following lemma:
\vspace{0.1cm}\\
\textbf{Lemma}
\begin{align}
\label{warnaartotop-3}
\prod_{i=1}^{k_1} \prod_{j=1}^{k_2}\frac{(A \ft^{j-i})_{\lambda_{1,i}-\lambda_{2,j}}}{(A \ft^{j-i+1})_{\lambda_{1,i}-\lambda_{2,j}}}=\ft^{\frac{k_1 |\lambda_2|-k_2|\lambda_1|}{2}}\frac{\tN_{\lambda_2\lambda_1}\left(-a\right)}{\tN_{\lambda_2\emptyset}\left(-a-k_1\epsilon_2\right)\tN_{\emptyset\lambda_1}\left(-a+k_2\epsilon_2\right)},
\end{align}
where $\ell(\lambda_1)\leq k_1$, $\ell(\lambda_2)\leq k_2$ and $A\colonequals e^{-\beta a}$.

\begin{proof}
Let us first notice that by using definition \eqref{eq:deftN-1} as well as exchange identities \eqref{eq:propertiestN}, the right-hand side of the above formula can be written as a following product:
\begin{multline}
\ft^{\frac{k_1 |\lambda_2|-k_2|\lambda_1|}{2}}\frac{\tN_{\lambda_2\lambda_1}\left(-a\right)}{\tN_{\lambda_2\emptyset}\left(-a-k_1\epsilon_2\right)\tN_{\emptyset\lambda_1}\left(-a+k_2\epsilon_2\right)}\\=\prod_{(i,j)\in \lambda_1}\frac{1-A\frac{\ft}{\fq}\ft^{\lambda_{2,j}^t-i}\fq^{\lambda_{1,i}-j+1}}{1-A\frac{\ft}{\fq}\ft^{k_2-i}\fq^{\lambda_{1,i}-j+1}}\prod_{(i,j)\in \lambda_2}\frac{1-A\frac{\ft}{\fq}\ft^{-\lambda_{1,j}^t+i-1}\fq^{-\lambda_{2,i}+j}}{1-A\frac{\ft}{\fq}\ft^{-k_1+i-1}\fq^{-\lambda_{2,i}+j}}.
\end{multline}
In proving the lemma, we will deal with formal power series in variables $\ft$ and $\fq$, so that we will not be concerned with issues of convergence of the intermediate expressions, requiring only that $\ft,\fq\neq 1$. We also extend the entries of partitions $\lambda_1$ and $\lambda_2$, such that
\beq
\label{eq:part-extension}
\lambda_{1,i}\colonequals 0, \,\, i>\ell(\lambda_1), \qquad \lambda_{2,i}\colonequals 0, \,\, i>\ell(\lambda_2)\,
\eeq
and for now assume $\ell(\lambda_1)=k_1$, $\ell(\lambda_2)=k_2$.

So, let us start with the following obvious identity:
\begin{align}
\sum_{i,j=1}^{\infty}\ft^{j-i}\left(1-\fq^{\lambda_{1,i}-\lambda_{2,j}}\right)=\biggl(\sum_{i=1}^{k_1}\sum_{j=1}^{k_2}+\sum_{i=k_1+1}^{\infty}\sum_{j=1}^{k_2}+\sum_{i=1}^{k_1}\sum_{j=k_2+1}^{\infty}\biggr)\,\ft^{j-i}\left(1-\fq^{\lambda_{1,i}-\lambda_{2,j}}\right).
\end{align}
Taking the last two sums of the right-hand side, shifting their summation indices and using convention \eqref{eq:part-extension}, one gets:
\begin{align}
&\biggl(\sum_{i=k_1+1}^{\infty}\sum_{j=1}^{k_2}+\sum_{i=1}^{k_1}\sum_{j=k_2+1}^{\infty}\biggr)\,\ft^{j-i}\left(1-\fq^{\lambda_{1,i}-\lambda_{2,j}}\right)=\sum_{i=1}^{\infty}\sum_{j=1}^{k_2}\ft^{j-i-k_1}\left(1-\fq^{-\lambda_{2,j}}\right)+\sum_{i=1}^{k_1}\sum_{j=1}^{\infty}\ft^{j-i+k_2}\left(1-\fq^{\lambda_{1,i}}\right)\nonumber\\
&=\frac{1}{\ft^{-1}-1}\biggl(-\sum_{j=1}^{k_2}\ft^{j-1-k_1}\left(1-\fq^{-\lambda_{2,j}}\right)+\sum_{i=1}^{k_1}\ft^{-i+k_2}\left(1-\fq^{\lambda_{1,i}}\right)\biggr),
\end{align}
where in the last step we used the sum of an infinite geometric progression. Substituting this back and multiplying the whole expression by $\ft^{-1}-1$, we obtain:
\begin{align}\label{lemma-mark1}
(\ft^{-1}-1)\sum_{i,j=1}^{\infty}\ft^{j-i}\left(1-\fq^{\lambda_{1,i}-\lambda_{2,j}}\right)=&(\ft^{-1}-1)\sum_{i=1}^{k_1}\sum_{j=1}^{k_2}\ft^{j-i}\left(1-\fq^{\lambda_{1,i}-\lambda_{2,j}}\right)\nonumber\\
&-\sum_{j=1}^{k_2}\ft^{j-1-k_1}\left(1-\fq^{-\lambda_{2,j}}\right)+\sum_{i=1}^{k_1}\ft^{-i+k_2}\left(1-\fq^{\lambda_{1,i}}\right).
\end{align}
Now we will use the following identity which the reader can find for instance in \cite{Awata:2005fa}:
\begin{align}
-(\ft^{-1}-1)\sum_{i=1}^{\infty}\fq^{\lambda_{1,i}}\ft^{1-i}=(\fq^{-1}-1)\sum_{i=1}^{\infty}\ft^{-\lambda_{1,i}^t}\fq^{i}.
\end{align}
Multiplying it by $\sum_{j=1}^{\infty}\ft^{j-1}\fq^{-\lambda_{2,j}}$ and subtracting from the result the same with $\lambda_1$, $\lambda_2$ set to zero, we find:
\begin{align}
&(\ft^{-1}-1)\sum_{i,j=1}^{\infty}\ft^{j-i}\left(1-\fq^{\lambda_{1,i}-\lambda_{2,j}}\right)=(\fq^{-1}-1)\sum_{i,j=1}^{\infty}\ft^{j-1}\fq^{i}\left(\ft^{-\lambda_{1,i}^t}\fq^{-\lambda_{2,j}}-1\right).
\end{align}
Substituting this back as a left-hand side of \eqref{lemma-mark1} and dividing everything by $\fq^{-1}-1$, we obtain the following:
\begin{align}
\sum_{i,j=1}^{\infty}\ft^{j-1}\fq^{i}\left(\ft^{-\lambda_{1,i}^t}\fq^{-\lambda_{2,j}}-1\right)
=&\sum_{i=1}^{k_1}\sum_{j=1}^{k_2}\fq\left(\ft^{j-i-1}-\ft^{j-i}\right)\frac{1-\fq^{\lambda_{1,i}-\lambda_{2,j}}}{1-\fq}\nonumber\\
&+\sum_{j=1}^{k_2}\fq^{1-\lambda_{2,j}}\ft^{j-1-k_1}\frac{1-\fq^{\lambda_{2,j}}}{1-\fq}+\sum_{i=1}^{k_1}\fq\ft^{-i+k_2}\frac{1-\fq^{\lambda_{1,i}}}{1-\fq},
\end{align}
where one can now use the formula for finite geometric progression to get rid of the fractions in the right-hand side:
\begin{align}\label{lemma-mark2}
\sum_{i,j=1}^{\infty}\left(\ft^{j-1-\lambda_{1,i}^t}\fq^{i-\lambda_{2,j}}-\ft^{j-1}\fq^{i}\right)
=&\sum_{i=1}^{k_1}\sum_{j=1}^{k_2}\sum_{l=1}^{\lambda_{1,i}-\lambda_{2,j}}\left(\ft^{j-i-1}-\ft^{j-i}\right)\fq^l\nonumber\\
&+\sum_{j=1}^{k_2}\sum_{i=1}^{\lambda_{2,j}}\ft^{j-1-k_1}\fq^{i-\lambda_{2,j}}+\sum_{i=1}^{k_1}\sum_{j=1}^{\lambda_{1,i}}\ft^{-i+k_2}\fq^j.
\end{align}
For clarity, the upper bound of the first summation on the right is written schematically, implying that for terms having $\lambda_{1,i}-\lambda_{2,j}<0$ the sum should be replaced by an equivalent corresponding to a negative Pochhammer symbol.

For the left-hand side one now should employ an identity from \cite{Nakajima-Yoshioka} (our $\ft$ and $\fq$ are interchanged with respect to the formula there):
\begin{align}
\sum_{i,j=1}^{\infty}\left(\ft^{j-1-\lambda_{1,i}^t}\fq^{i-\lambda_{2,j}}-\ft^{j-1}\fq^{i}\right)&=\sum_{s\in\lambda_1}\ft^{l_{\lambda_2}(s)}\fq^{a_{\lambda_1}(s)+1}+\sum_{s\in\lambda_2}\ft^{-l_{\lambda_1}(s)-1}\fq^{-a_{\lambda_2}(s)}\nonumber\\
&\equiv\sum_{(i,j)\in\lambda_1}\ft^{\lambda_{2,j}^t-i}\fq^{\lambda_{1,i}-j+1}+\sum_{(i,j)\in\lambda_2}\ft^{i-\lambda_{1,j}^t-1}\fq^{j-\lambda_{2,i}}.
\end{align}
Interchanging the indices in the second summand of the right-hand side of \eqref{lemma-mark2}, changing the summation order in the third summand and moving them to the left, one finally obtains:
\begin{align}
&\sum_{(i,j)\in\lambda_1}\left(\ft^{\lambda_{2,j}^t-i}-\ft^{k_2-i}\right)\fq^{\lambda_{1,i}-j+1}+\sum_{(i,j)\in\lambda_2}\left(\ft^{-\lambda_{1,j}^t+i-1}-\ft^{-k_1+i-1}\right)\fq^{-\lambda_{2,i}+j}\nonumber\\
&=\sum_{i=1}^{k_1}\sum_{j=1}^{k_2}\sum_{l=1}^{\lambda_{1,i}-\lambda_{2,j}}\left(\ft^{j-i-1}-\ft^{j-i}\right)\fq^l.
\end{align}
Substituting here $\ft,\fq\longrightarrow\ft^r,\fq^r$, multiplying by $\nicefrac{\left(A\frac{\ft}{\fq}\right)^r}{r}$ and using a series expansion of the logarithm, we get
\begin{align}
&\sum_{(i,j)\in \lambda_1}\text{ln }\biggl(\frac{1-A\frac{\ft}{\fq}\ft^{\lambda_{2,j}^t-i}\fq^{\lambda_{1,i}-j+1}}{1-A\frac{\ft}{\fq}\ft^{k_2-i}\fq^{\lambda_{1,i}-j+1}}\biggr)+\sum_{(i,j)\in \lambda_2}\text{ln }\biggl(\frac{1-A\frac{\ft}{\fq}\ft^{-\lambda_{1,j}^t+i-1}\fq^{-\lambda_{2,i}+j}}{1-A\frac{\ft}{\fq}\ft^{-k_1+i-1}\fq^{-\lambda_{2,i}+j}}\biggr)\nonumber\\
&=\sum_{i=1}^{k_1}\sum_{j=1}^{k_2}\text{ln }\biggl(\prod_{l=1}^{\lambda_{1,i}-\lambda_{2,j}}\frac{1-A\ft^{j-i}\fq^{l-1}}{1-A\ft^{j-i+1}\fq^{l-1}}\biggr).
\end{align}
Exponentiation concludes the proof.

\medskip

Remark. Tracing the above argument, one can see that it can be literally extended to the case $\ell(\lambda_1)\leq k_1$, $\ell(\lambda_2) \leq k_2$. This will be crucial for what follows.
\end{proof}

Having the lemma, we now can show that \eqref{q-binomial} is equivalent to:
\begin{align}
&\sum_{\lambda^{(1)},\dots , \lambda^{(N-1)}}'\biggl[\prod_{i=1}^{N-2} \left(\frac{z_i}{\ft}\ft^{\frac{k_{i-1}}{2}+k_i-\frac{k_{i+1}}{2}}\right)^{|\lambda^{(i)}|}\biggr]\cdot \left(\sqrt{A\frac{\ft}{\fq}}\frac{z_{N-1}}{\ft}\ft^{\frac{k_{N-2}+k_{N-1}}{2}}\right)^{|\lambda^{(N-1)}|}\nonumber\\
&\times \biggl[\prod_{i=1}^{N-1}\frac{\tN_{\lambda^{(i)}\lambda^{(i-1)}}\left((k_{i-1}-k_i)\epsilon_2-\epsilon_+\right)}{\tN_{\lambda^{(i)}\lambda^{(i)}}\left(0\right)}\biggr]\cdot\tN_{\emptyset\lambda^{(N-1)}}\left(-a\right)\label{q-binomial-2} \\
&=\prod_{1\leq i \leq j\leq N-2}\frac{\calM\big(\ft^{i-(j+1)+k_i-k_{j+1}}\cdot \prod_{s=i}^j(z_s \ft^{k_s})\big)\calM\big(\frac{\ft}{\fq}\cdot\ft^{(i-1)-j+k_{i-1}-k_{j}}\cdot \prod_{s=i}^j(z_s \ft^{k_s})\big)}{\calM\big(\ft\cdot\ft^{(i-1)-(j+1)+k_{i-1}-k_{j+1}}\cdot \prod_{s=i}^j(z_s \ft^{k_s})\big)\calM\big(\frac{1}{\fq}\cdot\ft^{i-j+k_{i}-k_{j}}\cdot \prod_{s=i}^j(z_s \ft^{k_s})\big)}\nonumber\\
&\times \prod_{i=1}^{N-1}\frac{\calM\big(\frac{A}{\fq}\cdot\ft^{i-(N-1)+k_i-k_{N-1}}\cdot \prod_{s=i}^{N-1}(z_s \ft^{k_s})\big)\calM\big(\frac{\ft}{\fq}\cdot\ft^{(i-1)-(N-1)+k_{i-1}-k_{N-1}}\cdot \prod_{s=i}^{N-1}(z_s \ft^{k_s})\big)}{\calM\big(\frac{A\ft}{\fq}\cdot\ft^{(i-1)-(N-1)+k_{i-1}-k_{N-1}}\cdot \prod_{s=i}^{N-1}(z_s \ft^{k_s})\big)\calM\big(\frac{1}{\fq}\cdot\ft^{i-(N-1)+k_{i}-k_{N-1}}\cdot \prod_{s=i}^{N-1}(z_s \ft^{k_s})\big)}.\nonumber
\end{align}

Finally, we are in position to prove the required summation formula:
\vspace{0.1cm}\\
\textbf{Theorem}
\begin{align}
\label{T_N-identity}
\sum_{\lambda^{(1)},\dots , \lambda^{(N-1)}} &\biggl[\prod_{i=1}^{N-1} \frac{\left( V_i \sqrt{U_i U_{i+1}}\right)^{|\lambda^{(i)}|}}{\tN_{\lambda^{(i)}\lambda^{(i)}}\left(0\right)}\biggr] \tN_{\lambda^{(1)}\emptyset}\left(u_1-\nicefrac{\epsilon_+}{2}\right)\nonumber\\&\times  \biggl[\prod_{i=1}^{N-2}\tN_{\lambda^{(i+1)}\lambda^{(i)}}\left(u_{i+1}-\nicefrac{\epsilon_+}{2}\right)\biggr] \tN_{\emptyset\lambda^{(N-1)}}\left(u_N-\nicefrac{\epsilon_+}{2}\right)\\
&=\prod_{i=1}^{N-1}\prod_{j=1}^{N-i}\frac{\calM\big( \prod_{s=j}^{i+j-1}(V_s U_s)\big)\calM\big(\frac{\ft}{\fq}\frac{U_{i+j}}{U_j}\cdot \prod_{s=j}^{i+j-1}(V_s U_s)\big)}{\calM\big( \sqrt{\frac{\ft}{\fq}} U_{i+j} \cdot \prod_{s=j}^{i+j-1}(V_s U_s)\big)\calM\big(\sqrt{\frac{\ft}{\fq}} \frac{1}{U_{j}} \cdot \prod_{s=j}^{i+j-1}(V_s U_s)\big)},\nonumber
\end{align}
with $N$ site parameters $U_i=e^{-\beta u_i}$ and $N-1$ link  parameters $V_j$. One can visualize the right-hand side of this formula by noticing that the arguments of numerator are precisely all the simply-connected combinations of even number of site and link parameters (multiplied by $\frac{\ft}{\fq}$ when starting with a link parameter), whereas the arguments of denominator represent all the simply-connected combinations of odd number of site and link parameters (multiplied by $\sqrt{\frac{\ft}{\fq}}$, single site parameters are excluded).

\begin{proof}
We use a so-called {\it specialization technique} \cite{MacdonaldSymmetric}.
Let us group all terms on the left having the same powers of $V_i$, $i=1,\dots, N-1$, {\it i.e.}\ grade our infinite sum with respect to a number of boxes of partitions we sum over. The coefficient of each combination of $V_1^{i_1}\cdots V_{N-1}^{i_{N-1}}$ is a polynomial in variables $U_i$, $i=1,\dots, N$ of degree $2(i_1+\dots+i_{N-1})$, having its coefficients in $\mathbb{F}$. Similarly, expanding the right-hand side as a series in $V_i$ and re-summing geometric progressions in $\fq, \ft$ into rational functions, we learn that the corresponding coefficients are as well polynomial in variables $U_i$ with coefficients in $\mathbb{F}$.

Let us now take any ordered combination of positive integers $k_i$, $k_1<\cdots < k_{N-1}$,  such that 
\beq
k_{i+1}-k_i\geq \ell(\lambda^{(i+1)}).
\eeq
One can see that the condition $\lambda^{(i)}_s\geq \lambda^{(i+1)}_{s-k_i+k_{i+1}}$ is trivially satisfied in this way, turning the corresponding skew plane partition into a horizontal strip plane partition. Making the following specialization of $U_i$ (remember that $k_0\equiv 0$):
\begin{align}
U_i=\sqrt{\frac{\ft}{\fq}}\, \ft^{k_i-k_{i-1}}, \qquad\qquad i=1,\dots , N-1
\end{align}
and reparametrizing the remaining variables as 
\begin{align}
V_{j}=\sqrt{\frac{\fq}{\ft}}\, \frac{z_{j}}{\ft}\, \ft^{k_{j-1}+k_{j}-k_{j+1}}, \qquad\qquad j=1,\dots , N-2
\end{align}
as well as
\begin{align}
U_N=\sqrt{\frac{\fq}{\ft}}\, \frac{1}{A}, \qquad\qquad V_{N-1}=\sqrt{\frac{\ft}{\fq}}\, A\, \frac{z_{N-1}}{\ft}\, \ft^{k_{N-2}}
\end{align}
one can readily check that formula \eqref{T_N-identity} then degenerates to the established $\bsl(N)$ $\fq$-binomial identity \eqref{q-binomial-2}.
Correspondingly, the above statement on equality of two polynomial coefficients translates into a statement on equality of corresponding polynomial coefficients of $z_1^{i_1}\cdots z_{N-1}^{i_{N-1}}$, which holds true.

We see that two polynomials in $N-1$ variables\footnote{According to the above specialization, $U_N$ can be kept generic.} coincide on an $(N-1)$-dimensional semilattice, meaning they just coincide. Term by term, this proves the theorem.

\end{proof}

Finally, let us remark that the summation formula \eqref{T_N-identity} for $N=2$
\beq
\sum_{\lambda^{(1)}} \left(V_1 \sqrt{U_1 U_2}\right)^{|\lambda^{(1)}|}
\frac{\tN_{\lambda^{(1)}\emptyset}\left(u_1-\nicefrac{\epsilon_+}{2}\right)
\tN_{\emptyset \lambda^{(1)}}\left(u_2-\nicefrac{\epsilon_+}{2}\right) 
 }{\tN_{\lambda^{(1)}\lambda^{(1)}}\left(0\right)}
 = 
\frac{
\calM\big(U_1 V_1\big)\calM\big(\frac{\ft}{\fq} V_1 U_2 \big)}
{\calM\big(\sqrt{\frac{\ft}{\fq}} V_1\big)
\calM\big(\sqrt{\frac{\ft}{\fq}} U_1 V_1 U_2 \big)}
\eeq
reproduces the non-trivial part of (5.3) of \cite{Kozcaz:2010af},
whereas, taken for $N=3$
\begin{align}
&\sum_{\lambda^{(1)},\lambda^{(2)}} \left(V_1 \sqrt{U_1 U_2}\right)^{|\lambda^{(1)}|} \left(V_2 \sqrt{U_2 U_3}\right)^{|\lambda^{(2)}|}
\frac{\tN_{\lambda^{(1)}\emptyset}\left(u_1-\nicefrac{\epsilon_+}{2}\right)
\tN_{\lambda^{(2)}\lambda^{(1)}}\left(u_2-\nicefrac{\epsilon_+}{2}\right)
\tN_{\emptyset \lambda^{(2)}}\left(u_3-\nicefrac{\epsilon_+}{2}\right) 
 }{\tN_{\lambda^{(1)}\lambda^{(1)}}\left(0\right)\tN_{\lambda^{(2)}\lambda^{(2)}}\left(0\right)} \nonumber \\
& = 
\frac{
\calM\big(U_1 V_1\big)\calM\big(\frac{\ft}{\fq} V_1 U_2 \big)\calM\big(U_2 V_2\big)\calM\big(\frac{\ft}{\fq} V_2 U_3 \big)
\calM\big(U_1 V_1 U_2 V_2\big)
\calM\big(\frac{\ft}{\fq}  V_1  U_2 V_2 U_3\big)}
{\calM\big(\sqrt{\frac{\ft}{\fq}} V_1\big)\calM\big(\sqrt{\frac{\ft}{\fq}} V_2\big)
\calM\big(\sqrt{\frac{\ft}{\fq}} U_1 V_1 U_2 \big)\calM\big(\sqrt{\frac{\ft}{\fq}} V_1 U_2 V_2  \big)\calM\big(\sqrt{\frac{\ft}{\fq}} U_2 V_2 U_3 \big)
\calM\big(\sqrt{\frac{\ft}{\fq}} U_1 V_1 U_2 V_2 U_3 \big)}, 
\end{align}
it is equivalent to the formula (6.7) conjectured in \cite{Hayashi:2013qwa}.

\section{Higgsing and iterated integrals for the \texorpdfstring{$\textbf{W}_4$}{W4} case}
\label{app:subappT_4}

We saw in section \ref{sec:pinching} how for $T_3$ the semi-degeneration of the mass parameters $m_i$ pinches the integral contour, so that the $\textbf{W}_3$ structure constants are given by a finite number of residues -- one or two depending on the choice of contour in figure~\ref{fig:Contour}. The purpose of this section is to show a similar computation in the $T_4$ case, in order to illustrate some of the complexities that arise when we are confronted with iterated contour integrals. For simplicity of notation, we set $\bA_1\equiv A_1^{(1)}$, $\bA_2\equiv A_2^{(1)}$ and $\bA_3\equiv A_1^{(2)}$. From \eqref{eq:ZTNperturbative}, we read the ``perturbative'' part of the the topological string partition function
\beqa
\label{eq:perturbativepartT4}
\left|\calZ_{4}^{\text{pert}}\right|^2 = &&\left|\frac{\prod_{1\leq i<j\leq 4}\calM\big(\frac{\tilde{M}_i}{\tilde{M}_j}\big) }
{ \prod_{k=1}^4 \biggl [\calM\left(\sqrt{\frac{\ft}{\fq}}\frac{\bA_1 }{\tilde{M}_k \tilde{N}_1}\right)\calM\left(\sqrt{\frac{\ft}{\fq}}\frac{ \bA_1 \tilde{M}_k}{\bA_2}\right)\calM\left(\sqrt{\frac{\ft}{\fq}}\bA_2\tilde{M}_k\tilde{L}_4 \right) \biggl ] }\right|^2\nonumber\\
&&\times\left| \frac{ \calM\left(\frac{\bA_1^2}{ \bA_2\tilde{N}_1}\right) \calM\left(\frac{\bA_2\tilde{N}_1}{\bA_1^2}\right)
\calM\left(\frac{\bA_1 \bA_2\tilde{L}_4}{ \tilde{N}_1}\right) \calM\left(\frac{ \tilde{N}_1 }{\bA_1 \bA_2\tilde{L}_4}\right)}
{\calM\left(\sqrt{\frac{\ft}{\fq}}\frac{\bA_1 \bA_3 }{\bA_2\tilde{N}_1 \tilde{N}_2 }\right)\calM\left( \sqrt{\frac{\ft}{\fq}}\frac{\bA_2 \bA_3 }{\bA_1\tilde{L}_1  \tilde{L}_2 }\right) \calM\left(\sqrt{\frac{\ft}{\fq}}\frac{ \bA_3\tilde{N}_3 }{\tilde{L}_1}\right)\calM\left(\sqrt{\frac{\ft}{\fq}}\frac{\bA_3\tilde{N}_4  }{\tilde{L}_2}\right) }\right|^2\nonumber\\
&&\times \left| \frac{\calM\left(\frac{ \bA_2^2\tilde{L}_4}{\bA_1}\right) \calM\left(\frac{ \bA_1}{  \bA_2^2\tilde{L}_4}\right)\calM\left(\frac{\bA_3^2}{ \tilde{N}_1 \tilde{N}_2\tilde{L}_1 \tilde{L}_2}\right) \calM\left(\frac{\tilde{N}_1 \tilde{N}_2\tilde{L}_1 \tilde{L}_2  }{ \bA_3^2}\right)}{  \calM\left(\sqrt{\frac{\ft}{\fq}}\frac{\bA_1\tilde{N}_2 }{\bA_3}\right)\calM\left(\sqrt{\frac{\ft}{\fq}}\frac{\bA_2 }{ \bA_3\tilde{L}_3}\right)\calM\left(\sqrt{\frac{\ft}{\fq}}\frac{\bA_1 \bA_3 }{\tilde{N}_1\tilde{L}_1\tilde{L}_2 }\right)   \calM\left(\sqrt{\frac{\ft}{\fq}}\frac{ \bA_2 \bA_3 \tilde{L}_4}{\tilde{N}_1 \tilde{N}_2}\right)}\right|^2\,.
\eeqa
{In addition, the ``instanton'' part \eqref{eq:ZTNsum} takes for $N=4$ the  form}
\beqa
\label{eq:instantonpartT4}
&&\calZ_4^{\text{inst}}=\sum_{\boldsymbol{\nu}}\left(\frac{ \tilde{N}_1\tilde{L}_3}{\tilde{N}_2\tilde{L}_4 }\right)^{\frac{|\nu_{1}^{(1)}|+|\nu_{2}^{(1)}|+|\nu_{3}^{(1)}|}{2}} \left(\frac{\tilde{N}_2\tilde{L}_2 }{\tilde{N}_3\tilde{L}_3 }\right)^{\frac{|\nu_{1}^{(2)}|+|\nu_{2}^{(2)}|}{2}} \left(\frac{\tilde{N}_3\tilde{L}_1 }{\tilde{N}_4\tilde{L}_2 }\right)^{\frac{|\nu_{1}^{(3)}|}{2} }\nonumber\\
&&\times \frac{\tN_{\nu_{2}^{(1)}\nu_{2}^{(2)}}\left(-\ba_1+\ba_2+\ba_3-l_1-l_2-\frac{Q}{2}\right)
\tN_{\nu_{2}^{(2)}\nu_{3}^{(1)}}\left(\ba_2-\ba_3-l_3-\frac{Q}{2}\right)
\tN_{\emptyset\nu_{2}^{(1)}}\left(\ba_1-\ba_2+m_1-\frac{Q}{2}\right)}{\tN_{\nu_{1}^{(1)}\nu_{1}^{(1)}}(0) 
\tN_{\nu_{1}^{(2)}\nu_{1}^{(2)}}(0) 
\tN_{\nu_{1}^{(3)}\nu_{1}^{(3)}}(0) 
\tN_{\nu_{2}^{(1)}\nu_{2}^{(1)}}(0)}
\nonumber\\
&&
\times \frac{
\tN_{\emptyset\nu_{3}^{(1)}}\left(\ba_2+l_4+m_1-\frac{Q}{2}\right)
\tN_{\emptyset\nu_{2}^{(1)}}\left(\ba_1-\ba_2+m_2-\frac{Q}{2}\right)\tN_{\emptyset\nu_{3}^{(1)}}\left(\ba_2+l_4+m_2-\frac{Q}{2}\right)}{ \tN_{\nu_{2}^{(2)}\nu_{2}^{(2)}}(0) 
\tN_{\nu_{3}^{(1)}\nu_{3}^{(1)}}(0) 
\tN_{\nu_{2}^{(1)}\nu_{3}^{(1)}}(-\ba_1+2 \ba_2+l_4)}
\nonumber\\
&&
\times \frac{
\tN_{\nu_{2}^{(1)}\emptyset}\left(-\ba_1+\ba_2-m_3-\frac{Q}{2}\right)
\tN_{\emptyset\nu_{3}^{(1)}}\left(\ba_2+l_4+m_3-\frac{Q}{2}\right) 
\tN_{\nu_{2}^{(1)}\emptyset}\left(-\ba_1+\ba_2-m_4-\frac{Q}{2}\right)}{\tN_{\nu_{2}^{(1)}\nu_{3}^{(1)}}(-\ba_1+2 \ba_2-Q+l_4)  
\tN_{\nu_{1}^{(1)}\nu_{2}^{(1)}}(2 \ba_1-\ba_2-n_1) 
\tN_{\nu_{1}^{(1)}\nu_{2}^{(1)}}(2 \ba_1-\ba_2-Q-n_1)}
\nonumber\\
&&
\times \frac{
 \tN_{\nu_{3}^{(1)}\emptyset}\left(-\ba_2-l_4-m_4-\frac{Q}{2}\right)
 \tN_{\nu_{1}^{(1)}\nu_{2}^{(2)}}\left(\ba_1+\ba_3-l_1-l_2-n_1-\frac{Q}{2}\right)
 \tN_{\nu_{1}^{(1)}\emptyset}\left(\ba_1-m_2-n_1-\frac{Q}{2}\right)}{\tN_{\nu_{1}^{(1)}\nu_{3}^{(1)}}(\ba_1+\ba_2+l_4-n_1) \tN_{\nu_{1}^{(1)}\nu_{3}^{(1)}}(\ba_1+\ba_2-Q+l_4-n_1)}
 \nonumber\\
&&\times \frac{
\tN_{\nu_{1}^{(1)}\emptyset}\left(\ba_1-m_3-n_1-\frac{Q}{2}\right) \tN_{\nu_{1}^{(1)}\emptyset}\left(\ba_1-m_4-n_1-\frac{Q}{2}\right) \tN_{\emptyset\nu_{1}^{(1)}}\left(-\ba_1+m_1+n_1-\frac{Q}{2}\right) }{\tN_{\nu_{1}^{(2)},\nu_{2}^{(2)}}(2 \ba_3-l_1-l_2-n_1-n_2) \tN_{\nu_{1}^{(2)},\nu_{2}^{(2)}}(2 \ba_3-Q-l_1-l_2-n_1-n_2)}
 \nonumber\\
&&\times\, \tN_{\nu_{1}^{(2)}\nu_{2}^{(1)}}\left(\ba_1-\ba_2+\ba_3-n_1-n_2-\frac{Q}{2}\right) 
\tN_{\nu_{1}^{(2)}\nu_{3}^{(1)}}\left(\ba_2+\ba_3+l_4-n_1-n_2-\frac{Q}{2}\right)
\nonumber\\
&&\times\,
\tN_{\nu_{1}^{(1)}\nu_{1}^{(2)}}\left(\ba_1-\ba_3+n_2-\frac{Q}{2}\right)\tN_{\nu_{1}^{(3)}\nu_{2}^{(2)}}\left(\ba_3-l_2+n_4-\frac{Q}{2}\right) 
\tN_{\nu_{1}^{(2)}\nu_{1}^{(3)}}\left(\ba_3-l_1+n_3-\frac{Q}{2}\right),
\eeqa
where the summation goes over partitions $\boldsymbol{\nu}=\{\nu_1^{(1)},\nu_2^{(1)},\nu_3^{(1)},\nu_1^{(2)},\nu_2^{(2)},\nu_1^{(3)}\}$.
Let us perform the contour integrals over the Coulomb moduli $\bA_i$'s. As demonstrated in \ref{sec:pinching}, there are multiple ways to choose the contour in such  a way that the contours gets pinched in the semi-degeneration limit. We will in this appendix just show the computation for a contour that leads to a single residue contributing. We have also performed the computation for other contours and, up to an irrelevant multiplicity, have obtained the same results.
 
Let us start by looking at the mass parameters.
Using the $T_4$ parametrization of \eqref{eq:PQR}, we find the expressions for the K\"ahler parameters $Q_{m;i}^{(j)}$ and $Q_{l;i}^{(j)}$. 
The mass parameters for the 5-branes on the left side of the $T_4$ junction are parametrized as follows
\beq
\label{eq:regularizationT4}
\tilde{M}_1=\left(\frac{\ft}{\fq}\right)^{\frac{3}{2}}\tilde{K}d_1\,,\qquad \tilde{M}_2=\left(\frac{\ft}{\fq}\right)^{\frac{1}{2}}\tilde{K}d_2\,,\qquad \tilde{M}_3=\left(\frac{\ft}{\fq}\right)^{-\frac{1}{2}}\tilde{K}d_3\,,\qquad \tilde{M}_4=\left(\frac{\ft}{\fq}\right)^{-\frac{3}{2}}\tilde{K}^{-3}\,,
\eeq
with $\prod_{i=1}^3d_i=1$. We set $d_i=e^{-\beta \delta_i}$ with $\sum_{i=1}^3\delta_i=0$.  We will compute the integrals in the order $\bA_1$, $\bA_2$ and $\bA_3$ and are interested in the result in the limit $\delta_a\rightarrow 0$.  Thus, in the calculation of the contour integrals, we will only keep the residues that will diverge when the regulators $\delta_i$ are finally all set to zero. Their divergences will be canceled in the limit by the zeroes coming from the $\big|\calM\big(\tilde{M}_i\tilde{M}_j^{-1}\big)\big|^2$ in the numerator.

\begin{figure}[ht]
 \centering
  \includegraphics[height=6cm]{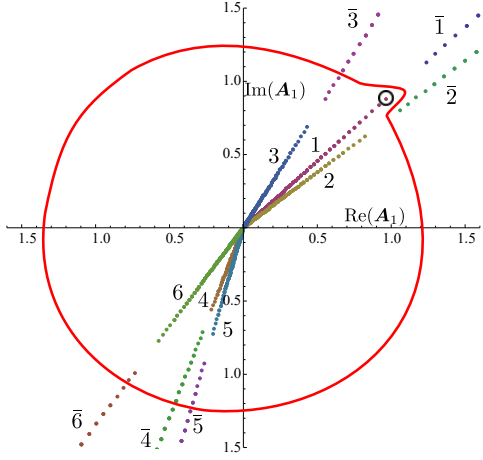}
  \caption{\it The figure presents our choice of the integration contour for \textbf{A}$_1$. 
  As the regulators $\delta_a$ are taken to zero, the integral is given by just one residue whose position is indicated by a small circle.}
  \label{fig:ContourT4A1}
\end{figure}

Let us now consider the contour integral over $\bA_1$. The possible contributing poles come from the following terms in the denominator of \eqref{eq:perturbativepartT4} 
\beq
\left|\prod_{j=1}^3 \calM\left(\sqrt{\frac{\ft}{\fq}}\frac{\bA_1 }{\tilde{M}_j \tilde{N}_1}\right)\prod_{k=1}^3\calM\left(\sqrt{\frac{\ft}{\fq}}\frac{\bA_1\tilde{M}_k }{\bA_2}\right)\right|^2.
\eeq
We number the terms with $j=1,2, 3$ as $1$ to $3$ and those with $k=1,2,3$ as $4$ to $6$ and we need to investigate which of them might pinch the integral contour.  
The situation for imaginary $\delta_a$ is depicted in figure~\ref{fig:ContourT4A1}. 
We see that for $|\tilde{K}|>1$ and imaginary masses $n_i$ and $l_i$
the contour for $\bA_1$ can be chosen in such a way that in the limit $\delta_a\rightarrow 0$ only one residue contributes, namely the one for
\beq
\label{eq:T4resforA1}
\bA_1=\frac{\ft}{\fq}\tilde{K}\tilde{N}_1d_1\, .
\eeq
Thus, we can compute the integral over $\bA_1$ just as in the $T_3$ case and, after some simplifications, obtain the integral expression
\begin{align}
\label{eq:T4afteroneintegral}
&\lim_{\delta_a\rightarrow 0}\ointctrclockwise\prod_{k=1}^3\left[\frac{d\bA_k}{2\pi i \bA_k}|M(\ft,\fq)|^2\right]\left|\calZ_4^{\text{top}}\right|^2=\lim_{\delta_a\rightarrow 0}\ointctrclockwise\prod_{k=2}^3\left[\frac{d\bA_k}{2\pi i \bA_k}|M(\ft,\fq)|^2\right]\text{Res}\left(\left|\calZ_4^{\text{top}}\right|^2,\bA_1=\tilde{K}\tilde{N}_1d_1\frac{\ft}{\fq}\right)\nonumber\\
&=
\lim_{\delta_a\rightarrow 0}\ointctrclockwise\prod_{k=2}^3\left[\frac{d\bA_k}{2\pi i \bA_k}|M(\ft,\fq)|^2\right] \frac{\left|\calM\left(\frac{\ft}{\fq}\frac{d_2}{d_3}\right)\calM\left(\frac{\ft^2}{\fq^2}\tilde{K}^4d_2\right)\calM\left(\frac{\ft}{\fq}\tilde{K}^4 d_3\right)\right|^2}{\left|\calM\left(\frac{\bA_2d_2}{\tilde{K}^2 \tilde{N}_1}\right)\calM\left(\frac{\fq}{\ft}\frac{\bA_2 d_3}{\tilde{K}^2 \tilde{N}_1 }\right)\calM\left(\frac{\ft}{\fq}\bA_2 \tilde{K} \tilde{L}_4 d_2 \right)\calM\left(\bA_2 \tilde{K} \tilde{L}_4 d_3\right)\right|^2}
\nonumber\\
&\times\frac{\left|\calM\left(\frac{\fq}{\ft}\frac{\bA_2 }{\tilde{K}^2 \tilde{N}_1 d_1^2}\right)\calM\left(\frac{\ft}{\fq}\bA_2 \tilde{K} \tilde{L}_4d_1\right)\calM\left(\frac{\fq}{\ft}\frac{\bA_2^2\tilde{L}_4}{\tilde{K}\tilde{N}_1d_1}\right)\calM\left(\frac{\bA_2^2\tilde{L}_4}{\tilde{K} \tilde{N}_1 d_1}\right)\right|^2}{\left|\calM\left(\frac{\fq}{\ft}\frac{\bA_2\tilde{L}_4}{\tilde{K}^3}\right)\calM\left(\frac{\ft}{\fq}\frac{\bA_2 \tilde{K}^2 }{\tilde{N}_1d_1}\right)\calM\left(\sqrt{\frac{\ft}{\fq}}\frac{\bA_2 }{\bA_3 \tilde{L}_3 }\right)\calM\left(\sqrt{\frac{\fq}{\ft}}\frac{\bA_2 \tilde{N}_2 }{\bA_3 \tilde{K}  d_1}\right)\calM\left(\sqrt{\frac{\fq}{\ft}}\frac{\bA_2 \bA_3 }{\tilde{K} \tilde{N}_1 \tilde{L}_1 \tilde{L}_2  d_1}\right)\calM\left(\sqrt{\frac{\ft}{\fq}}\frac{\bA_2 \bA_3\tilde{L_4}}{ \tilde{N}_1 \tilde{N}_2}\right)\right|^2}\nonumber\\
&\times \frac{\left|\calM\left(\frac{\bA_3^2}{ \tilde{N}_1 \tilde{N}_2 \tilde{L}_1 \tilde{L}_2}\right)\calM\left(\frac{\ft}{\fq}\frac{\bA_3^2}{ \tilde{N}_1 \tilde{N}_2\tilde{L}_1 \tilde{L}_2}\right)\right|^2}{\left|\calM\left(\sqrt{\frac{\ft}{\fq}}\frac{\bA_3 \tilde{N}_4}{\tilde{L}_2 }\right)\calM\left(\sqrt{\frac{\ft}{\fq}}\frac{\bA_3 \tilde{N}_3}{\tilde{L}_1 }\right)\calM\left(\left(\frac{\ft}{\fq}\right)^{\frac{3}{2}}\frac{\bA_3 \tilde{K}  d_1}{\tilde{L}_1 \tilde{L}_2 }\right)\calM\left(\left(\frac{\ft}{\fq}\right)^{\frac{3}{2}}\frac{\tilde{K} \tilde{N}_1 \tilde{N}_2 d_1}{\bA_3 }\right)\right|^2}\left|\calZ_4^{\text{inst}}\right|^2_{\big|\bA_1=\frac{\ft}{\fq}\tilde{K}\tilde{N}_1d_1}
\end{align}
where we have used \eqref{eq:residueformula}.

We must now perform the integration over $\bA_2$. We find that the relevant terms in the denominator of the integrand in \eqref{eq:T4afteroneintegral} are
\beq
\left|\calM\left(\frac{\bA_2d_2}{\tilde{K}^2\tilde{N}_1}\right)\calM\left(\frac{\fq}{\ft}\frac{\bA_2d_3}{\tilde{K}^2\tilde{N}_1}\right)\calM\left(\frac{\ft}{\fq}\bA_2\tilde{K}\tilde{L}_4 d_2\right)\calM\left(\bA_2\tilde{K}\tilde{L}_4 d_3\right)\right|^2\,.
\eeq
From the above, we read that there are two poles that are \textit{potentially} relevant for the semi-degenerate limit, namely those for
\beq
\label{eq:residues11and12}
\bA_2=\frac{\ft}{\fq}\tilde{K}^2 \tilde{N}_1d_3^{-1},\qquad \bA_2=\tilde{K}^{-1} \tilde{L}_4^{-1}d_3^{-1}\,.
\eeq
These are the two residues that \textit{could} contribute due to pinching. We need now to set the exact integral contour for $\bA_2$ to see which one of them actually contributes. 
The contour can be chosen in such a way as to have the residue at $\bA_2=\frac{\ft}{\fq}\tilde{K}^2 \tilde{N}_1d_3^{-1}$, but not the one at $\bA_2=\tilde{K}^{-1} \tilde{L}_4^{-1}d_3^{-1}$. 
Finally, we have to compute the integral over $\bA_3$. Arguments similar to the ones used for $\bA_2$ tell us that the contour can be chosen such as to have a pinching when the regulators are removed at the pole
\beq
\label{eq:T4A3contribution}
\bA_3=\sqrt{\frac{\ft}{\fq}}\tilde{K} \tilde{N}_1\tilde{N}_2d_1.
\eeq
Performing the same kind of computation that led to \eqref{eq:T4afteroneintegral}, we obtain the integral in the semi-degenerate limit
\begin{multline}
\label{eq:T4degpartitionfunction1}
\lim_{\delta_a\rightarrow 0}\ointctrclockwise\prod_{k=1}^3\left[\frac{d\bA_k}{2\pi i \bA_k}|M(\ft,\fq)|^2\right]\left|\calZ_4^{\text{top}}\right|^2=\\= \left|\frac{\calM\left(\tilde{K}^{-4}\right)}{\prod_{i=1}^4\calM\left(\frac{\tilde{N}_{5-i}\tilde{L}_i }{\tilde{K} }\right)}\right|^2\left|\calZ_4^{\text{inst}}\right|^2_{\big|\tilde{A}_i^{(j)}\rightarrow \left(\frac{\ft}{\fq}\right)^{\frac{i(4-i-j)}{2}}\tilde{K}^i\prod_{k=1}^{j}\tilde{N}_k}\, .
\end{multline}
Computing the ``instanton'' contribution to residues, we find that inserting the values of he Coulomb moduli, namely \eqref{eq:T4resforA1}, the left part of \eqref{eq:residues11and12} as well as \eqref{eq:T4A3contribution} into \eqref{eq:instantonpartT4} immediately gets rid of the sums over $\nu_1^{(1)}$, $\nu_1^{(2)}$ and $\nu_2^{(1)}$ due to \eqref{eq:tNdelta}. Thus, we obtain the ``instanton'' contribution to the contour integral in the semi-degenerate limit:
\begin{align}
\label{eq:instantonpartT4semidegenerate}
&\left(\calZ_4^{\text{inst}}\right)_{\big|\bA_1=\frac{\ft}{\fq}\tilde{K}\tilde{N}_1d_1,\bA_2=\frac{\ft}{\fq}\tilde{K}^2 \tilde{N}_1,\bA_3=\sqrt{\frac{\ft}{\fq}}\tilde{K} \tilde{N}_1\tilde{N}_2}=\sum_{\nu_1^{(3)},\nu_2^{(2)},\nu_3^{(1)}} 
\left(\frac{ \tilde{N}_3\tilde{L}_1}{\tilde{N}_4\tilde{L}_2}\right)^{\frac{\left|\nu_1^{(3)}\right|}{2}}\left(\frac{\tilde{N}_2\tilde{L}_2}{ \tilde{N}_3\tilde{L}_3}\right)^{\frac{\left|\nu_2^{(2)}\right|}{2} }
\left(\frac{\tilde{N}_1\tilde{L}_3 }{ \tilde{N}_2\tilde{L}_4}\right)^{\frac{\left|\nu_3^{(1)}\right|}{2} }
\nonumber \\&\times 
\frac{\tN_{\nu_1^{(3)}\emptyset}(n_4+l_1-\varkappa)
\tN_{\nu_2^{(2)}\nu_1^{(3)}}(n_3+l_2-\varkappa)
 \tN_{\nu_3^{(1)}\nu_2^{(2)}}( n_2+l_3-\varkappa)  
\tN_{\emptyset\nu_3^{(1)}}(n_1+l_4-\varkappa) }{\tN_{\nu_1^{(3)}\nu_1^{(3)}}(0) \tN_{\nu_2^{(2)}\nu_2^{(2)}}(0) \tN_{\nu_3^{(1)}\nu_3^{(1)}}(0)}\,.
\end{align} 
We can now  plug the summation formula \eqref{T_N-identity-main} in  \eqref{eq:instantonpartT4semidegenerate} and inserting the result in \eqref{eq:T4degpartitionfunction1} we get the final result:
\beq
\lim_{\delta_a\rightarrow 0}\ointctrclockwise\prod_{k=1}^3\left[\frac{d\bA_k}{2\pi i \bA_k}|M(\ft,\fq)|^2\right]\left|\calZ_4^{\text{top}}\right|^2
=\frac{\left|\calM(\tilde{K}^{-4})\prod_{1\leq i<j\leq 4}\calM\left(\nicefrac{\tilde{N}_j}{\tilde{N}_i}\right)\calM\left(\nicefrac{\tilde{L}_i}{\tilde{L}_j}\right)\right|^2}{\left|\prod_{i,j=1}^4\calM(\tilde{N}_i\tilde{L}_j\tilde{K}^{-1})\right|^2}\, . 
\eeq
Thus, we obtain our general formula \eqref{eq:calZNtopdegeneratefinalintegral}, specialized for $N=4$.


\providecommand{\href}[2]{#2}\begingroup\raggedright\endgroup

\end{document}